\documentclass[aps,prfluids,superscriptaddress,amsmath,amsfonts,preprintnumbers,floatfix]{revtex4-2}

\usepackage{graphicx} 
\usepackage{physics}
\usepackage{longtable}

\usepackage{xcolor}
\usepackage{placeins}

\DeclareMathOperator{\Div}{div}
\DeclareMathOperator{\Grad}{{\bf grad}}
\DeclareMathOperator{\Curl}{{\bf curl}}
\DeclareMathOperator{\Deltav}{{\bf \Delta}}

\newcommand{\x}{\mathbf{x}}
\usepackage{hyperref}

\begin{document}
\preprint{Published in Physical Review Fluids}

\title{Chaotic advection in a steady three-dimensional MHD flow}
   

\author{Julien Fontchastagner}
\email[]{julien.fontchastagner@univ-lorraine.fr} 
\affiliation{Université de Lorraine, GREEN, F-54000 Nancy, France}

\author{Jean-François Scheid}
\email[]{jean-francois.scheid@univ-lorraine.fr}
\affiliation{Université de Lorraine, CNRS, Inria, IECL, F-54000 Nancy, France}

\author{Jean-Régis Angilella}
\email[]{jean-regis.angilella@unicaen.fr}
\affiliation{Normandie Université, UNICAEN, UNIROUEN, ABTE, F-14000 Caen, France}

\author{Jean-Pierre Brancher}
\email[]{jean-pierre.brancher@univ-lorraine.fr}
\affiliation{Université de Lorraine, CNRS, IECL, F-54000 Nancy, France}

 

\begin{abstract}
We investigate the 3D stationary flow of a weakly conducting fluid in a cubic cavity, driven by the Lorentz force created by  two  permanent magnets and a weak constant current. Our goal is to determine the conditions leading to efficient mixing within the cavity. 
The flow is composed of a large recirculation cell created by one side magnet, superposed to two recirculation cells created by a central magnet perpendicular to the first one.  The overall structure of this flow, obtained here by solving the Stokes equations with Lorentz forcing, is similar to the tri-cellular model flow studied by Toussaint {\it et. al.} (Phys. Fluids. 7, 1995).
Chaotic advection in this flow is  analyzed by means of Poincaré sections, Lyapunov exponents and expansion entropies. In addition, we quantify the quality of mixing by computing contamination rates, homogeneity, as well as mixing times. 
Though individual vortices have poor mixing properties, the superposition of both flows creates chaotic streamlines and efficient mixing.
\end{abstract}


\maketitle

\section{\label{section:one}Introduction and motivation}

\subsection{General considerations}

Mixing devices at small Reynolds number have been studied during the past fifty years. These are challenging situations where turbulence cannot be used to produce random motion \cite{Ottino1989,younes}. The low Reynolds number condition is generally met when viscosity is large, or dimensions are small (e.g. microfluidic devices), or  when fluids are fragile and require moderate stirring. 
Nevertheless, high quality mixing can be achieved even for such fluids, provided the flow exhibits chaotic trajectories corresponding to efficient stretching and folding  mechanisms \cite{AB1986,ProceedingsISFMSM,Ottino1989}.

Chaotic advection has been studied and exploited in two-dimensional time-dependent flows, such as flows between eccentric cylinders \cite{Ottino1989,KW1993}, flows induced by moving walls in a cavity \cite{LO1989}, or electromagnetically driven flows \cite{Figueroa2017}.  
In three dimensions, the pioneering works of Dombre {\it et al.} \cite{dombre1986} have shown that the flow does not need to be time-dependent to produce chaotic advection. Indeed, streamlines in steady 3D flows can be chaotic, so that fluid points trajectories, which are identical to streamlines in steady flows, are extremely sensitive to initial positions.  This property was exploited in a steady open pipe flow  with electrically conducting fluids driven from outside by high frequency helical inductors \cite{JPJR2001,JP2004}, as well as in steady flows in a cubic periodic box \cite{theseVToussaint,articleVToussaint}. The authors of the latter reference considered a simple arbitrary analytic flow composed of a steady vortex superposed to a vortex pair perpendicular to the first one.
In the present paper, a flow with a similar tri-cellular structure is considered. However, instead of using a simple analytical model, our flow is obtained by means of magnetohydrodynamic (MHD) forcing, involving two magnets to create the  system of vortices, and will be calculated by solving the Stokes equation with Lorentz forcing. This is a first step towards an experimental facility.

\subsection{Qualitative description of the  flow: side magnet and central magnet}

\begin{figure} 
  \centering
  \begin{tabular}{c c c}
    \includegraphics[width=6.0cm]{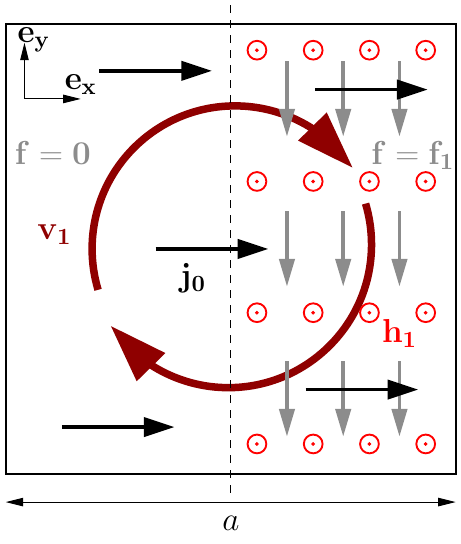} & &
    \includegraphics[width=6.0cm]{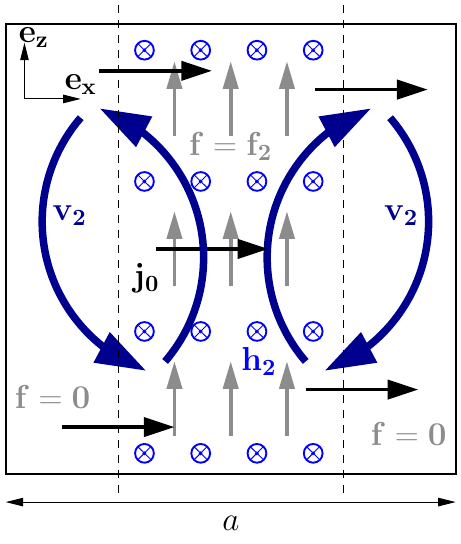} \\
    (a) & & (b)  \\
    \multicolumn{3}{c}{\includegraphics[width=12cm]{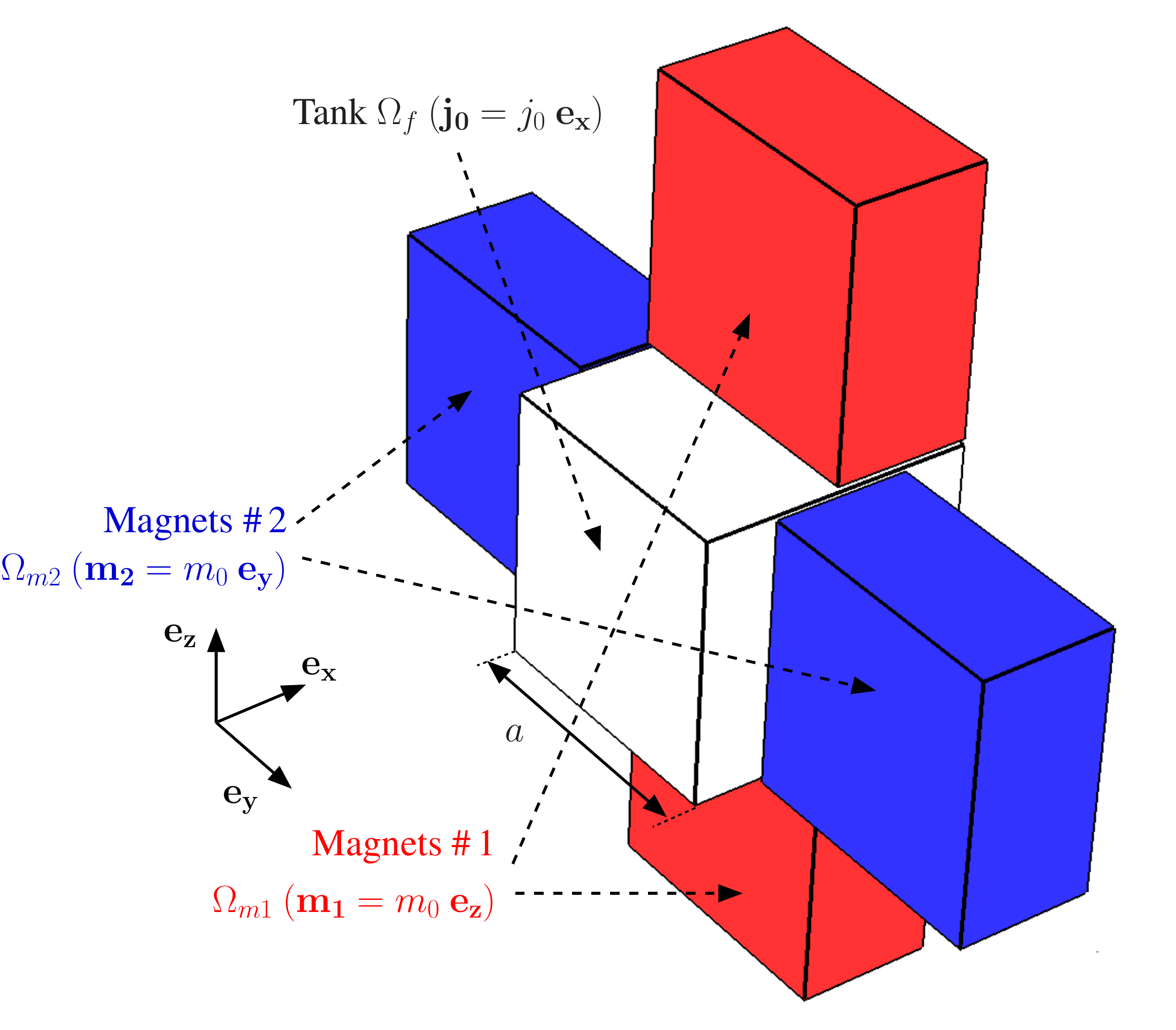}}\\
    \multicolumn{3}{c}{(c)} \\
  \end{tabular}
  \caption{\label{fig:intro}Upper graphs: sketches of the flows when a single magnet is present. (a): single-cell in the $(x,y)$ plane induced by a side magnet.   (b): double-cell in the $(x,z)$ plane created by a central magnet. Current density is shown in black, force fields are shown in gray. The corresponding magnetic fields $\mathbf{h_1}$ (configuration (a)) and $\mathbf{h_2}$ (configuration (b)) are shown respectively in red and blue. Velocities $\mathbf{v_1}$ and $\mathbf{v_2}$ are in dark red and dark blue. Lower graph (c): Setup of the mixing MHD device creating these configurations.}
\end{figure}

To create such a tri-cellular flow in a realistic way, we consider a viscous fluid with  low electric conductivity (e.g. salted water), confined within a cubic cavity. It is submitted to a magnetic field induced by two magnets located outside the box, and to a  uniform horizontal current with density $\mathbf{j_0}$ passing through the cavity (\figurename\,\ref{fig:intro}(c)). The axis $z$ will be denoted as "vertical" in the following.

In the case where a uniform vertical magnetic field $\mathbf{h_1}$ is applied over  the right-hand-side of the cavity, and set to zero in the other half (\figurename\,\ref{fig:intro}\,(a)), one can check that the liquid cannot remain at rest, and that a flow  will appear \cite{JPJR2001}.
Indeed, if the fluid is at rest, the force balance imposes $\mathbf{f} = \mathbf{grad}\,p$ everywhere, where  $\mathbf{f}$ is the Lorentz force and $p$ is the pressure, so that the curl of the Lorentz force must be zero everywhere. This condition cannot be fulfilled when the magnetic field is present only over  a portion of the volume of the cavity: here the curl of the Lorentz force is non-zero along the surface of discontinuity of the magnetic field, the hydrostatic condition is not fulfilled, and motion will appear. Under the combined effect of the Lorentz force and of the walls, a large recirculation cell as the one sketched in \figurename\,\ref{fig:intro}\,(a) will appear. This situation corresponds to the case where a {\sl side magnet} like the blue one of   \figurename\,\ref{fig:intro}\,(c) is placed near the box.  The corresponding Lorentz force will be denoted $\mathbf{f}_1$ in the following. 
Similarly, a vertical force $\mathbf{f_2}$ can be created near the mid-plane of the cubic box by means of a magnetic field $\mathbf{h_2}$ centered around this mid plane.  It will create a double recirculation cell, as sketched in  \figurename\,\ref{fig:intro}\,(b). This can be achieved by placing a centered magnet like the red one of \figurename\,\ref{fig:intro}\,(c).

\medskip
The two forces $\mathbf{f_1}$ and $\mathbf{f_2}$ can be superimposed to generate a complex flow, in view of creating a simple MHD mixing device involving static sources only. As discussed above, the structure of this flow is very close to one investigated in Refs.  \cite{theseVToussaint} and \cite{articleVToussaint}, which involved a simple analytic velocity field. Note however that our flow has no-slip boundary conditions at the walls of the cubic box, as expected for viscous flows, in contrast with Refs.  \cite{theseVToussaint} and \cite{articleVToussaint} where fluid is allowed to slip along the faces of the cubic box.

In this article, we will make use of the two configurations described above, except that the magnetic fields will not be taken  piecewise constant, but will be issued from a real physical device and computed from Maxwell's equations.
Such a device, based solely on the use of two pairs of permanent magnets, is illustrated in \figurename\,\ref{fig:intro}\,(c), and described in the next section.
The velocity field in the fluid will then be  computed numerically with these magnetic forces by means of a vorticity -- streamfunction formulation of the Stokes equations.

The equations of the coupled system under study for the magnetic and velocity fields are presented in the next chapter. Then, numerical simulations will be performed for various parameters of interest. The magnetic force that will be imposed to the fluid is a weighted average of the two forces $\mathbf{f_1}$ (the Lorentz force due to the side magnet) and $\mathbf{f_2}$ (due to the central magnet), and the ratio of the two will be the control parameter of the device. It will be shown that chaotic advection is obtained for specific choices of this ratio. The computation of Poincaré sections, Lyapunov exponents and expansion entropies will provide qualitative and quantitative information about the chaotic streamlines we obtain.  
A brief study of mixing efficiency is presented in section \ref{sec:mixing}.

\section{Motion equations and calculation of the flow} 

\subsection{General considerations}

A cubic tank of side length $a$ containing an incompressible Newtonian fluid with kinematic viscosity $\nu$ is considered. A uniform current density ${\bf j_0}$ (directed along the unit vector ${\bf e_x}$) is imposed inside the tank by means of a constant voltage drop. As described in the previous section,  the two vortex configurations are obtained by means of two pairs of magnets placed  in two orthogonal directions (\figurename\,\ref{fig:intro}(c)). 
The side magnet (magnet \#1, in red) is magnetized along ${\bf e_z}$  and is used to create the force   ${\bf f_1}$. The central magnet (magnet \#2, in blue) is magnetised along ${\bf e_y}$ and produces the force  ${\bf f_2}$. 
The length and the depth of the magnets are equal to $a$ whereas their width is equal to $a/2$. As explained in the introduction, the force ${\bf f_1}$ alone would produce a single recirculation occupying the whole domain, and  the force ${\bf f_2}$ alone would produce a double-cell. It is the combination of both forces that will produce a complex flow.

Throughout the paper, the Reynolds number based on the size of the box and on the typical fluid velocity is assumed to be small. The value of the current density is chosen to be sufficiently low compared to the magnetization of the permanent magnets, so that its impact on the total magnetic field, and therefore on the total corresponding force, can be neglected. This assumption was verified by \textit{a posteriori} simulations with and without current density contribution. The $\text{L}_2$ norm of the relative difference was less than 0.1\% for the typical values of the source terms that were considered.

\subsection{Governing equations}
 
Following to the previous assumptions, the flow  corresponds to a Stokes flow, governed by
\begin{equation}\label{eq:Stokes1}
  \left\{
    \begin{array}{rcl}
      \nu \Deltav {\bf v} - \Grad p + {\bf f} &=& {\bf 0} \\
      \Div {\bf v} &=& 0  \\
    \end{array}
  \right. ~,~\text{in}~\Omega_f
\end{equation}
where ${\bf v}$ is the velocity field, $p$ is the pressure field (including gravity effects if any),  ${\bf f}$ is the Lorentz force density, and $\Omega_f$ denotes the fluid domain.  
The fluid being viscous,  no-slip conditions are imposed on the tank boundary $\Gamma_f$ for the velocity field:
\begin{equation}
  {\bf v} = {\bf 0} ~\text{on}~ \Gamma_f  .
\end{equation}
The contribution of each pair of magnets can be treated separately according to the superposition principle:
\begin{equation}
  \displaystyle   {\bf f} = \mu_0 \, {\bf j_0} \times {\bf h} = \underbrace{\mu_0 \, {\bf j_0} \times {\bf h_1}}_{\bf f_1} + \underbrace{\mu_0 \, {\bf j_0} \times {\bf h_2}}_{\bf f_2}
\end{equation}
where ${\bf h_i},~i = \{1,2\}$ is the field produced by magnet $\# i$. 
In this way, due to the linearity of the Stokes equations, the velocity field $\bf v$ in \eqref{eq:Stokes1} can be decomposed into 
\begin{equation}
  {\bf v}={\bf v_1}+{\bf v_2}
\end{equation}
where each part ${\bf v_1}$ and ${\bf v_2}$ satisfies:
\begin{equation}
  \left\{
    \begin{array}{cl}
      \nu \Deltav {\bf v_i} - \Grad p_i + \mu_0 \, {\bf j_0}\times{\bf h_i} = {\bf 0} & \text{in}~\Omega_f \\
      \Div {\bf v_i} = 0 & \text{in}~\Omega_f  \\
      {\bf v_i} = {\bf 0} & \text{on}~ \Gamma_f  
    \end{array}
  \right.,~\text{for}~i = \{1,2\}
\end{equation}
The magnetic field ${\bf h_i}$ and flux density ${\bf b_i}$ in the overall domain $\Omega$ ($=\mathbb{R}^3$) are governed by the Maxwell-Ampère and Maxwell-Thomson equations: 
\begin{equation}
  \Curl {\bf h_i} = {\bf 0}~, ~~~ \Div {\bf b_i} = 0,
\end{equation}
and both fields vanish far from the magnets ($\norm{\bf x} \to\infty$). 
The overall domain $\Omega$  
corresponds to the air containing the fluid domain (the tank) and the pairs of magnets (see \hyperref[myappendix]{\appendixname} and \figurename \ref{fig:mesh} for the precise overall domain used for computations).

The source term at the origin of the magnetic flux density is due to the magnetization of the magnets, ${\bf m_1}=m_0\,{\bf e_z}$ for   magnet $\#1$ and ${\bf m_2}=m_0\,{\bf e_x}$ for   magnet $\#2$ (see \figurename\,\ref{fig:intro}(c)). Then, magnetization vectors have the same norm, $m_0$. 
The magnetic behaviour law linking the flux densities   and the magnetic fields   is then 
\begin{equation}
  \left\{
    \begin{array}{ll}
      {\bf b_i} = \mu_0\,({\bf h_i}+{\bf m_i}) & \text{in}~\Omega_{mi} \\
      {\bf b_i} = \mu_0\,{\bf h_i} & \text{elsewhere} 
    \end{array}\right.,~\text{for}~i = \{1,2\}
\end{equation}
Extending magnet magnetizations as piecewise functions on the whole domain:
\begin{equation}
  {\bf m_i} = \left\{ 
    \begin{array}{cl}
      {\bf m_i} & \text{in}~\Omega_{mi} \\
      {\bf 0} & \text{in}~\Omega\backslash\Omega_{mi}
    \end{array}
  \right.,
\end{equation}
the magnetic fields ${\bf h_1}$ and ${\bf h_2}$ are finally governed by
\begin{equation}\label{eq:mag1}
  \displaystyle \left\{
    \begin{array}{c l}
      \Curl {\bf h_i} = {\bf 0} & \text{in}~\Omega  \\
      \Div ({\bf h_i} + {\bf m_i}) = 0 &  \text{in}~\Omega  \\
      \displaystyle \lim_{\norm{\bf x}\to\infty} {\bf h_i} = {\bf 0}
    \end{array}
  \right.,~\text{for}~i = \{1,2\}
\end{equation}
These equations will be set non-dimensional as follows. The side length of the cubic tank $a$ is chosen as the characteristic length of the problem, so that the  non-dimensional position vector is defined as: $\widehat{\bf x}= {\bf x}/a$. The norm ${j_0}$ of the current density can be expressed from the total current $I_0$ flowing through the system so that the non-dimensional current density can be defined by
\begin{equation}
  \widehat{\bf j_0}(\widehat{\bf x}) = \frac{{\bf j_0}({\bf x})}{j_0} = \frac{a^2}{I_0}\,{\bf j_0}({\bf x}).
\end{equation}
In the same way, a non-dimensional magnetization vector $\widehat{\bf m_i}$, and the corresponding $\widehat{\bf h_i}$ field are introduced as follows:
\begin{equation}
  \widehat{\bf m_i}(\widehat{\bf x}) = \frac{\beta}{m_0}\,{\bf m_i}({\bf x}) ~,~~~
  \widehat{\bf h_i}(\widehat{\bf x}) = \frac{\beta}{m_0}\,{\bf h_i}({\bf x}) ~,~~~\text{for}~i = \{1,2\}
\end{equation}  
where $\beta$ is an arbitrary scaling factor designed to avoid too low values of magnetic fields (and therefore forces and velocities) when solving the problem. In terms of these non-dimensional quantities, Eq.\  (\ref{eq:mag1}) now reads  
\begin{equation}\label{eq:strongformmag}
  \renewcommand{\arraystretch}{1.25}  
  \displaystyle 
  \left\{
    \begin{array}{c l}
      \Curl \widehat{\bf h_i} = {\bf 0} & \text{in}~\widehat{\Omega}~~\left(\Omega = a\,\widehat{\Omega}\right)  \\
      \Div \left(\widehat{\bf h_i} + \widehat{\bf m_i}\right) = 0 &  \text{in}~\widehat{\Omega}  \\
      \displaystyle \lim_{\norm{\widehat{\bf x}}\to\infty} \widehat{\bf h_i} = {\bf 0}
    \end{array}
  \right.,~\text{for}~i = \{1,2\}
\end{equation}
Velocity and pressure are set non-dimensional as: 
\begin{equation}
  \widehat{\bf v_i}(\widehat{\bf x}) = \frac{\bf v_i ({\bf x})}{v_0} ~,~~~
  \widehat{p_i}(\widehat{\bf x}) = \frac{p_i({\bf x})}{p_0} ~,~~~\text{for}~i = \{1,2\}
\end{equation}
where $v_0$ and $p_0$ are characteristic velocity and pressure that will be defined below.
The fluid equations becomes
\begin{equation}
  \Deltav \widehat{\bf v_i} - \frac{a\,p_0}{\nu\,v_0}\,\Grad \widehat{p_i} + \frac{\mu_0\,I_0\,m_0}{\nu\,v_0\,\beta} ~\widehat{\bf j_0}\times\widehat{\bf h_i} = {\bf 0} ~,~~~\text{for}~i = \{1,2\} .
\end{equation}
By choosing 
\begin{equation}
  v_0 = \frac{\mu_0\,I_0\,m_0}{\nu\,\beta} ~~\text{and}~~~
  p_0 = \frac{\mu_0\,I_0\,m_0}{a\,\beta}~, 
\end{equation}
the motion equation finally reads: 
\begin{equation}\label{strongformfluid}
  \renewcommand{\arraystretch}{1.25}  
 \left\{
    \begin{array}{cl}
      \Deltav \widehat{\bf v_i} - \Grad \widehat{p_i} + \widehat{\bf j_0}\times\widehat{\bf h_i} = {\bf 0} & \text{in}~\widehat{\Omega}_f ~~\left(\Omega_f = a\,\widehat{\Omega}_f\right) \\
      \Div \widehat{\bf v_i} = 0 & \text{in}~\widehat{\Omega}_f \\
      \widehat{\bf v_i} = {\bf 0} & \text{on}~\widehat{\,\Gamma}_f 
    \end{array}
  \right.,~\text{for}~i = \{1,2\} .
\end{equation}
Finally, by sequentially solving Eqs.\ (\ref{eq:strongformmag}) and (\ref{strongformfluid}), for $i$=1,2,  
we obtain the two velocity fields $\widehat{{\bf v_1}}$ and $\widehat{{\bf v_2}}$. 

In the remainder of this article, we will be concerned only with non-dimensional quantities, and for convenience the hat symbols ($\widehat{\,.\,}$)  will be omitted.

\section{\label{model_num}Numerical computations of fields}

To avoid projection or interpolation errors due to the use of different meshes or function spaces, we have chosen to solve the governing equations by means of the same computational tool. Mixed finite elements methods, based on both nodal and edge basis functions, provide an efficient and convenient framework for solving problems of low-frequency computational electromagnetism \cite{Nedelec1980,ren, ThesisGeuzaine}.  However, they can also be used to correctly approximate the Stokes system \cite{Nedelec1980,Nedelec1986} as described in section\,\ref{stokes_form}.    


\subsection{Magnetic fields: scalar magnetic potential formulation} 
First of all, we should point out that the magnetic fields produced by the magnets (and only these fields) can be calculated purely analytically in our particular case. Indeed, analytical formulas exist for rectangular magnets in air, see \cite{yonnet} or more recently \cite{Furlani2001}. However, the authors preferred to use a numerical model, as this will be necessary in our future work in order to design an experimental test bench. To reduce the total volume of magnets required, ferromagnetic yokes will be added to each pair of magnets to loop their magnetic flux and reinforce the field inside the tank. In this case, an analytical model is no longer possible.
However, we compared both approaches on our study case, and the results from the analytical and numerical models were perfectly consistent. No significant difference were observed on the final velocities. This allows us to validate the numerical calculation of the magnetic fields detailed below.

\begin{figure}
  \centering
  \includegraphics[width=12.9cm]{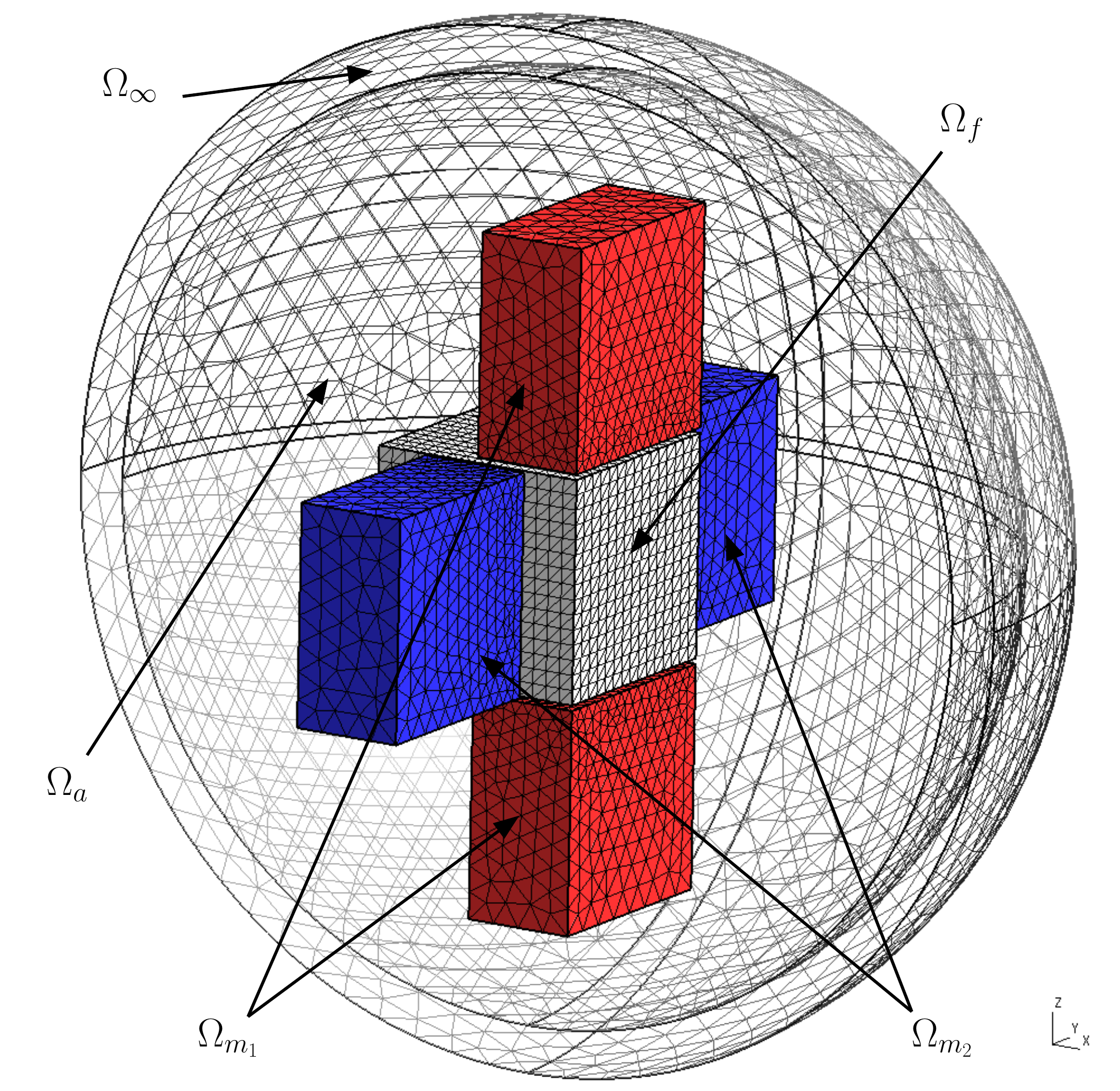}
  \caption{\label{fig:mesh}Representation of the total computational domain, and the corresponding mesh used for the numerical computations (control parameter $ne = 15$).}
\end{figure}

\medskip 
The overall domain of computation $\Omega$ can be reduced to $\Omega = \Omega_f\cup \Omega_{m1}\cup \Omega_{m2}\cup \Omega_{a}\cup \Omega_{\infty}$ which is represented in \figurename\,\ref{fig:mesh}. 
$\Omega_f$ corresponds to the tank, $\Omega_{m1}$ and $\Omega_{m2}$ to the pairs of magnets, and $\Omega_a$ to the air region around them. Their union forms the domain of interest $\Omega_i= \Omega_f\cup \Omega_{m1}\cup \Omega_{m2}\cup \Omega_{a}$ in which the magnetic fields are wanted. The external boundary of $\Omega_i$ is that of a sphere of radius $R_i$. This region is surrounded by a spherical finite shell $\Omega_{\infty}$ with inner and outer radii respectively $R_i$ and $R_o$. This will be useful to deal with the open boundary condition on fields as explained below.

The spaces $\text{L}^2(\Omega)$ and $\textbf{L}^2(\Omega)$ can be defined by:
\begin{multline}
  \text{L}^2(\Omega) = \{ u:\Omega\longrightarrow\mathbb{R} ~ \big| \int_{\Omega} u({\bf x})^2 \text{d}{\bf x} < \infty \}~,~~\text{and}~
  \textbf{L}^2(\Omega) = \{ {\bf u}:\Omega\longrightarrow\mathbb{R}^3 ~\big| \int_{\Omega} \norm{{\bf u}({\bf x})}^2 \text{d}{\bf x} < \infty \}
\end{multline}
They permit to define the Sobolev space $\text{H}(\Grad,\Omega)$, more commonly known as $\text{H}^1(\Omega)$:
\begin{equation}
  \begin{split}
    \text{H}(\Grad,\Omega) &= \{u \in \text{L}^2(\Omega) \mid \Grad u \in \textbf{L}^2(\Omega) \} \\
                           &= \{u \in \text{L}^2(\Omega) \mid \partial_x u, \partial_y u, \partial_z u \in \text{L}^2(\Omega) \} = \text{H}^1(\Omega) 
  \end{split}
\end{equation}
The Maxwell-Ampère equation $\Curl {\bf h_i} = {\bf 0}$ implies that there exists a magnetic scalar potential ${\phi_i} \in \text{H}(\Grad,\Omega)$ such that:
\begin{equation}
  {\bf h_i} = -\Grad {\phi_i}
\end{equation}
To deal with the ${\bf h_i}\xrightarrow[\norm{\bf x}\to \infty]{}{\bf 0}$ condition, a bijective transformation that maps all the region $\mathbb{R}^3\backslash\Omega_i$ onto $\Omega_{\infty}$ is constructed thanks to the relation:
\begin{equation}
  \norm{\bf x} = R_i \frac{R_o-R_i}{R_o-\norm{\bf X}}
\end{equation}
where ${\bf X}$ is the position vector in the shell $\Omega_{\infty}$ \cite{ShellTransform}. The considered transformation $\boldsymbol{\mathcal{F}} : \mathbb{R}^3\backslash\Omega_i \to\Omega_{\infty}$ is therefore defined by: 
\begin{equation}
  {\bf X} = \boldsymbol{\mathcal{F}}({\bf x}) =\frac{\norm{\bf X}}{\norm{\bf x}}\,{\bf x}
  =\left[\frac{R_o}{\norm{\bf x}} - \frac{R_i (R_o-R_i)}{\norm{\bf x}^2} \right]\,{\bf x}
\end{equation}
The open-boundary conditions on ${\bf h_1}$ and ${\bf h_2}$ can be replaced by imposing homogeneous Dirichlet conditions on $\phi_1$ and $\phi_2$ on $\Gamma_{\infty} = \partial\Omega$ (the sphere of radius $R_o$). We will also use the transformation $\boldsymbol{\mathcal{F}}$ and its jacobian matrix ${\bf J_{\mathcal{F}}}$ to bring back all integrals posed on $\mathbb{R}^3\setminus \Omega_i$ to $\Omega_{\infty}$.

The second equation of $\left(S_{m,i}\right)$ becomes: 
\begin{equation}\label{eq:maxwell2}
  \Div (\Grad \phi_i - {\bf m_i}) = 0 \ \ \text{in } \Omega
\end{equation} 
Multiplying \eqref{eq:maxwell2} by a test function $\xi_i \in \text{H}_0(\Grad,\Omega) = \{u \in \text{H}(\Grad,\Omega) : \left.u\right|_{\Gamma_{\infty}} = 0\}$, then using the transformation $\boldsymbol{\mathcal{F}}$ to write down the integrals terms over $\mathbb{R}^3\setminus \Omega_i$ to $\Omega_{\infty}$ and finally integrating by parts, we obtain the weak form  $\left(\Sigma_{m,i}\right)$ of the system $\left(S_{m,i}\right)$ given by (\ref{eq:strongformmag}):
\begin{equation}\label{eq:wfmagnet2}
  \renewcommand{\arraystretch}{1.5}   
  \left(\Sigma_{m,i}\right) \left\{
    \begin{array}{l}
      \text{Find~} \phi_i \in \text{H}_0(\Grad,\Omega) \text{~such that:~} ~\forall \xi_i \in \text{H}_0(\Grad,\Omega),  \\
      \displaystyle \int_{\Omega_i} \Grad {\phi_i} \cdot \Grad {\xi_i}~\text{d}{\bf x} - \int_{\Omega_{mi}} {\bf {\bf m_i}} \cdot \Grad {\xi_i}~\text{d}{\bf x} \\
      \displaystyle \qquad\qquad + \int_{\Omega_{\infty}} {\bf J_{\mathcal{F}}^{-1}} \Grad {\phi_i} \cdot {\bf J_{\mathcal{F}}^{-1}} \Grad {\xi_i}~|\det \mathbf{J}_{\mathcal{F}}|~\text{d}{\bf X} = 0
    \end{array}\right. 
\end{equation}
Once $\phi_i$ is computed from \eqref{eq:wfmagnet2}, the force density involved in the non-dimensional Stokes flow system  in $\Omega_f$ will be: 
\begin{equation}
  {\bf f_i} = -{\bf j_0}\times \Grad \phi_i   
\end{equation}

\subsection{Flows velocities: vorticity -- streamfunction formulation}\label{stokes_form}

Let us begin by defining the spaces: 
\begin{align}
  \textbf{H}(\Curl,\Omega_f) &= \{ {\bf u} \in \textbf{L}^2(\Omega_f) \mid \Curl {\bf u} \in {\bf L}^2(\Omega_f)\}\\
  \textbf{H}_0(\Curl,\Omega_f) &= \{ {\bf u} \in \textbf{H}(\Curl,\Omega_f) \mid \left.{\bf u}\times{\bf n}\right|_{\Gamma_f} = {\bf 0} \}
\end{align}
where $\Gamma_f$ is the boundary of $\Omega_f$ ($\Gamma_f = \partial\Omega_f$). 
It is possible to rewrite the Stokes system (\ref{strongformfluid}) in terms of the vorticity fields $\boldsymbol{\omega_i} \in \textbf{H}(\Curl,\Omega_f)$ defined by $\boldsymbol{\omega_i} = \Curl {\bf v_i}$:
\begin{equation}
  \Curl \boldsymbol{\omega_i} + \Grad p_i - {\bf f_i} = {\bf 0}
\end{equation}
And the fact that ${\bf v_i} = {\bf 0}$ on $\Gamma_f$ directly implies:
\begin{equation}
  0 = \Curl {\bf v_i} \cdot {\bf n} = \boldsymbol{\omega_i}\cdot{\bf n} ~~~ \text{on}~\Gamma_f 
\end{equation}
Since ${\bf v_i}$ is divergence free, there exists a vector potential  $\boldsymbol{\psi_i} \in \textbf{H}_0(\Curl,\Omega_f)$ such that \cite{armouche1998}:
\begin{equation} 
  {\bf v_i} = \Curl \boldsymbol{\psi_i}~,~~\text{and}~~~ \Div \boldsymbol{\psi_i} = 0
\end{equation}
By using this potential and applying the $\Curl$ operator to the first equation of the system $\left(S_{f,i}\right)$, we obtain:
\begin{equation}
  \renewcommand{\arraystretch}{1.25}  
  \left\{
    \begin{array}{ccc}
      \Curl \Curl \boldsymbol{\omega_i} = \Curl {\bf f_i}~,~~ & \Div \boldsymbol{\omega_i} = 0  & ~~~\text{in}~\Omega_f \\
      \boldsymbol{\omega_i} = \Curl \Curl \boldsymbol{\psi_i}~,~~ & \Div \boldsymbol{\psi_i} = 0  & ~~~\text{in}~\Omega_f \\
      \boldsymbol{\omega_i}\cdot{\bf n} = 0~,~~ \boldsymbol{\psi_i}\times{\bf n} = {\bf 0}~, & \Curl \boldsymbol{\psi_i} \times {\bf n} = {\bf 0} & ~~~\text{on}~\Gamma_f  
    \end{array}
  \right.
\end{equation}
The vorticity $\boldsymbol{\omega_i}$ can be decomposed as $\boldsymbol{\omega_i} = \boldsymbol{\omega_i^0} + \boldsymbol{\omega_i^*}$
where $\boldsymbol{\omega_i^0}\in \textbf{H}_0(\Curl,\Omega_f)$ is a regular part and  $\boldsymbol{\omega_i^*}\in \textbf{H}(\Curl,\Omega_f)$ is the harmonic contribution \cite{Amara1999,amara-bernardi1999}.  We rewrite the global system in two weakly coupled subsystems as follows:
\begin{equation}
  \renewcommand{\arraystretch}{1.25}
  \begin{array}{ll}
    \left\{
    \begin{array}{c}
      -\Delta \boldsymbol{\omega_i^0} = \Curl \Curl \boldsymbol{\omega_i^0} = \Curl {\bf f_i}  \\
      \Div \boldsymbol{\omega_i^0} = 0 \\
      \left.\boldsymbol{\omega_i^0} \times{\bf n}\right|_{\Gamma_f} = {\bf 0}
    \end{array}
    \right.
    &~,~~~
      \left\{
      \begin{array}{c}
        -\Delta \boldsymbol{\psi_i} = \Curl \Curl \boldsymbol{\psi_i} = \boldsymbol{\omega_i^0} + \boldsymbol{\omega_i^*}\\
        -\Delta \boldsymbol{\omega_i^*} = \Curl \Curl \boldsymbol{\omega_i^*} = {\bf 0} \\
        \Div \boldsymbol{\omega_i^*} = 0 ~,~~ \Div \boldsymbol{\psi_i} = 0 \\
        \left.\left(\boldsymbol{\omega_i^0} + \boldsymbol{\omega_i^*}\right)\cdot{\bf n}\right|_{\Gamma_f} = {\bf 0} \\ 
        \left.\boldsymbol{\psi_i}\times{\bf n}\right|_{\Gamma_f} = {\bf 0} ~,~ \left.\Curl\boldsymbol{\psi_i}\times{\bf n}\right|_{\Gamma_f} = {\bf 0}
      \end{array}
      \right.          
  \end{array}
\end{equation}
For each subsystem, a mixed formulation can be deduced by introducing the scalar fields $\lambda_i^0$, $\lambda_i^*$ and $\eta_i$ which permit respectively to weakly impose the divergence free properties of $\boldsymbol{\omega_i^0}$, $\boldsymbol{\omega_i^*}$ and $\boldsymbol{\psi_i}$, by acting as Lagrange multipliers.
The systems $\left(S_{f,i}\right)$ are then equivalent to the following weak form $\left(\Sigma_{f,i}\right)$, given by (\ref{eq:systemfluidweak}) (see \cite{Amara1999}).
\begin{figure*}
\begin{equation}
  \label{eq:systemfluidweak}
  \renewcommand{\arraystretch}{1.5}  
  \left(\Sigma_{f,i}\right) \left\{
    \begin{array}{l}
      \left(\Sigma_{f,i}^{(1)}\right) \left\{
      \begin{array}{l}
        \text{Find~} \left(\boldsymbol{\omega_i^0},\lambda_i^0\right) \in \textbf{H}_0(\Curl,\Omega_f) \times \text{H}(\Grad,\Omega_f) \text{~such that:~}  \\
        \displaystyle \int_{\Omega_f} \Curl \boldsymbol{\omega_i^0}\cdot\Curl \boldsymbol{\varphi_i}~\text{d}{\bf x} + \int_{\Omega_f} \Grad \lambda_i^0 \cdot \boldsymbol{\varphi_i}~\text{d}{\bf x} = \\
        \displaystyle \hfill \int_{\Omega_f} {\bf f_i}\cdot\Curl \boldsymbol{\varphi_i}~\text{d}{\bf x} ~,~~ \forall \boldsymbol{\varphi_i} \in \textbf{H}_0(\Curl,\Omega_f) \\
        \displaystyle \int_{\Omega_f} \boldsymbol{\omega_i^0} \cdot \Grad \mu_i~\text{d}{\bf x} = 0 ~,~~ \forall \mu_i \in \text{H}_0(\Grad,\Omega_f)\\
      \end{array}
      \right.\\
      ~\\
      \left(\Sigma_{f,i}^{(2)}\right)\left\{
      \begin{array}{l}
        \text{Find~} \left(\boldsymbol{\omega_i^*},\lambda_i^*\right) \in \textbf{H}(\Curl,\Omega_f) \times \text{H}(\Grad,\Omega_f) ~\text{and} \\ \hfill
        \left(\boldsymbol{\psi_i},\eta_i\right) \in \textbf{H}_0(\Curl,\Omega_f) \times \text{H}(\Grad,\Omega_f)  \text{~such that:~}   \\
        \displaystyle \int_{\Omega_f} \boldsymbol{\omega_i^*} \cdot \boldsymbol{\varphi_i}~\text{d}{\bf x} - \int_{\Omega_f} \Curl \boldsymbol{\psi_i} \cdot \Curl \boldsymbol{\varphi_i}~\text{d}{\bf x} + \int_{\Omega_f} \Grad \eta_i \cdot \boldsymbol{\varphi_i}~\text{d}{\bf x} = \\
        \hfill \displaystyle - \int_{\Omega_f} \boldsymbol{\omega_i^0} \cdot \boldsymbol{\varphi_i}~\text{d}{\bf x}  ~,~~ \forall \boldsymbol{\varphi_i} \in \textbf{H}(\Curl,\Omega_f) \\
        \displaystyle \int_{\Omega_f} \Curl \boldsymbol{\omega_i^*} \cdot \Curl \boldsymbol{\xi_i}~\text{d}{\bf x} + \int_{\Omega_f} \Grad \lambda_i^* \cdot \boldsymbol{\xi_i}~\text{d}{\bf x} = 0 ~, 
        \forall \boldsymbol{\xi_i} \in \textbf{H}_0(\Curl,\Omega_f)\\
        \displaystyle \int_{\Omega_f} \boldsymbol{\psi_i} \cdot \Grad \chi_i~\text{d}{\bf x} = 0 ~,~~ \forall \chi_i \in \text{H}_0(\Grad,\Omega_f)\\
        \displaystyle \int_{\Omega_f} (\boldsymbol{\omega_i^0}+\boldsymbol{\omega_i^*}) \cdot \Grad \mu_i~\text{d}{\bf x} = 0 ~,~~ \forall \mu_i \in \text{H}(\Grad,\Omega_f)\\
      \end{array}
      \right.
    \end{array}
  \right.
\end{equation}
\end{figure*}

It is important to mention that the condition $(\boldsymbol{\omega_i^0}+\boldsymbol{\omega_i^*})\cdot{\bf n} = 0$ on $\Gamma_f$ is implicitely imposed by choosing the test functions $\mu_i$ in $\text{H}(\Grad,\Omega_f)$; and for the boundary condition $\Curl\boldsymbol{\psi_i}\times{\bf n} = 0$ on $\Gamma_f$, by choosing in $\left(\Sigma_{f,i}^{(2)}\right)$ the test functions $\boldsymbol{\varphi_i}$  in $\textbf{H}(\Curl,\Omega_f)$ and not in $\textbf{H}_0(\Curl,\Omega_f)$. 

\begin{figure}
  \centering
  \begin{tabular}{c c}
    ${\bf h_1} = -\Grad \phi_{h1}$ & ${\bf h_2} = -\Grad \phi_{h2}$ \\
    \includegraphics[width=0.45\textwidth]{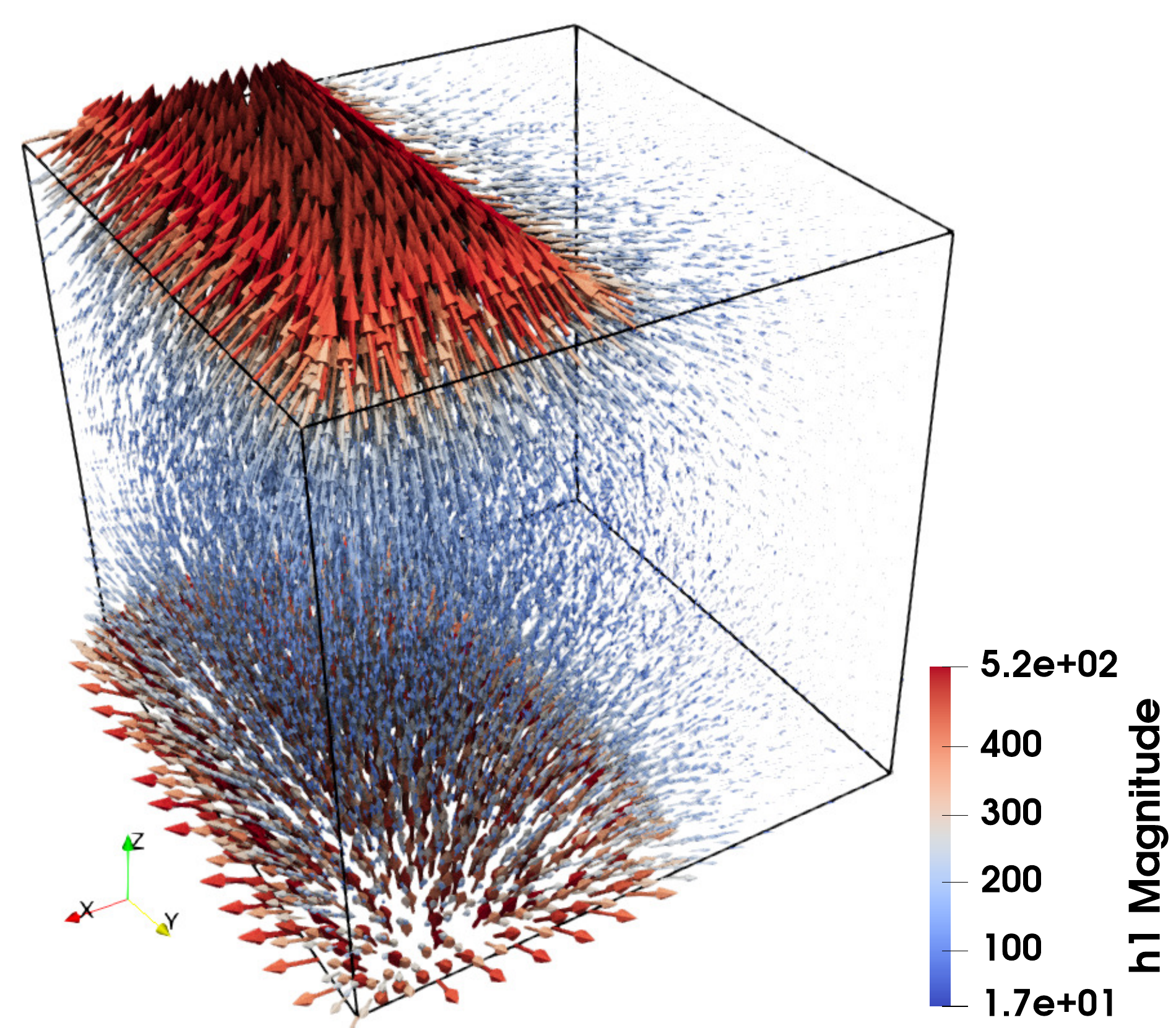} &\includegraphics[width=0.45\textwidth]{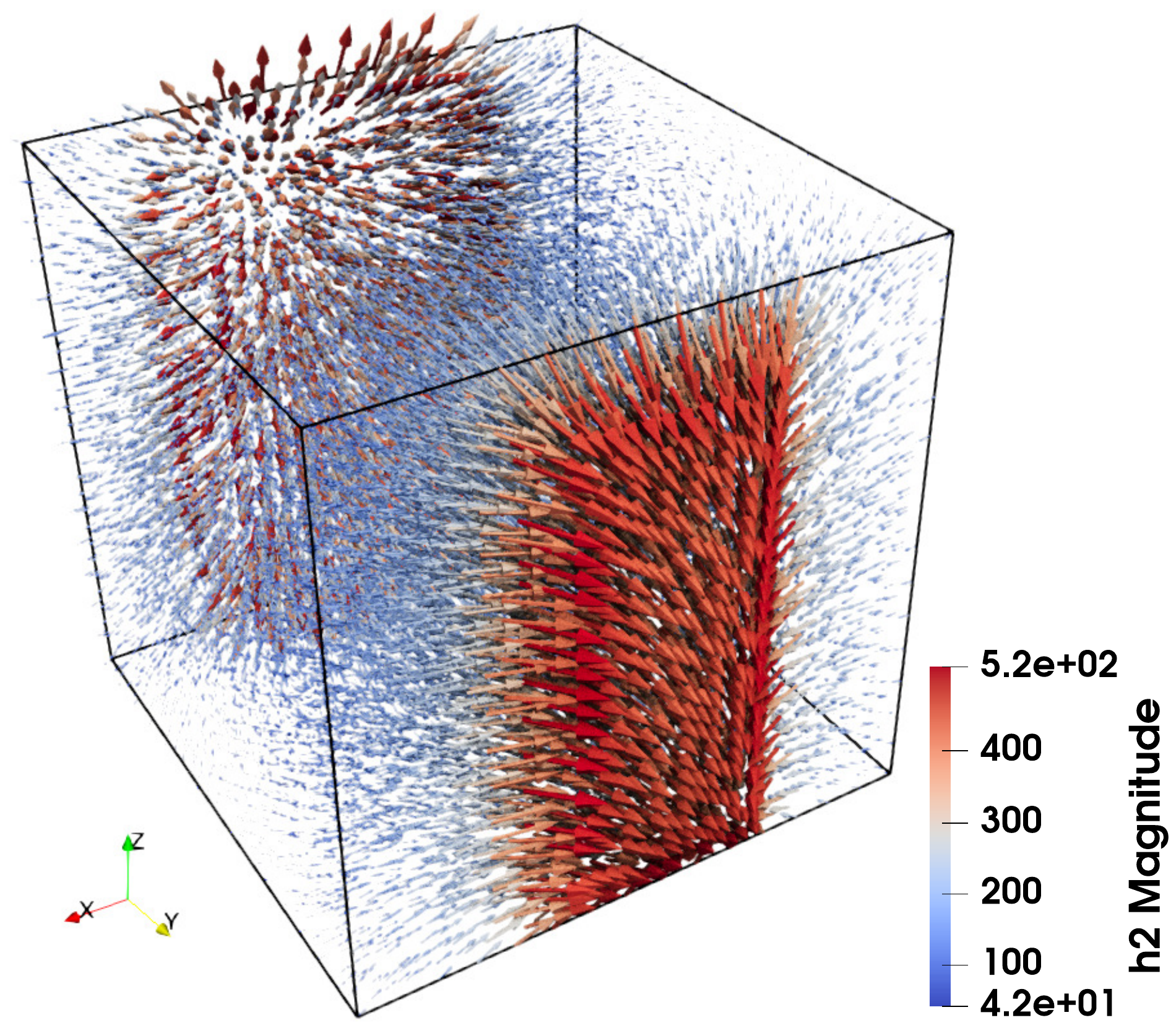}\\
    ${\bf f_1} = -{\bf j_0}\times\Grad \phi_{h1}$ & ${\bf f_2} = -{\bf j_0}\times\Grad \phi_{h2}$ \\
    \includegraphics[width=0.45\textwidth]{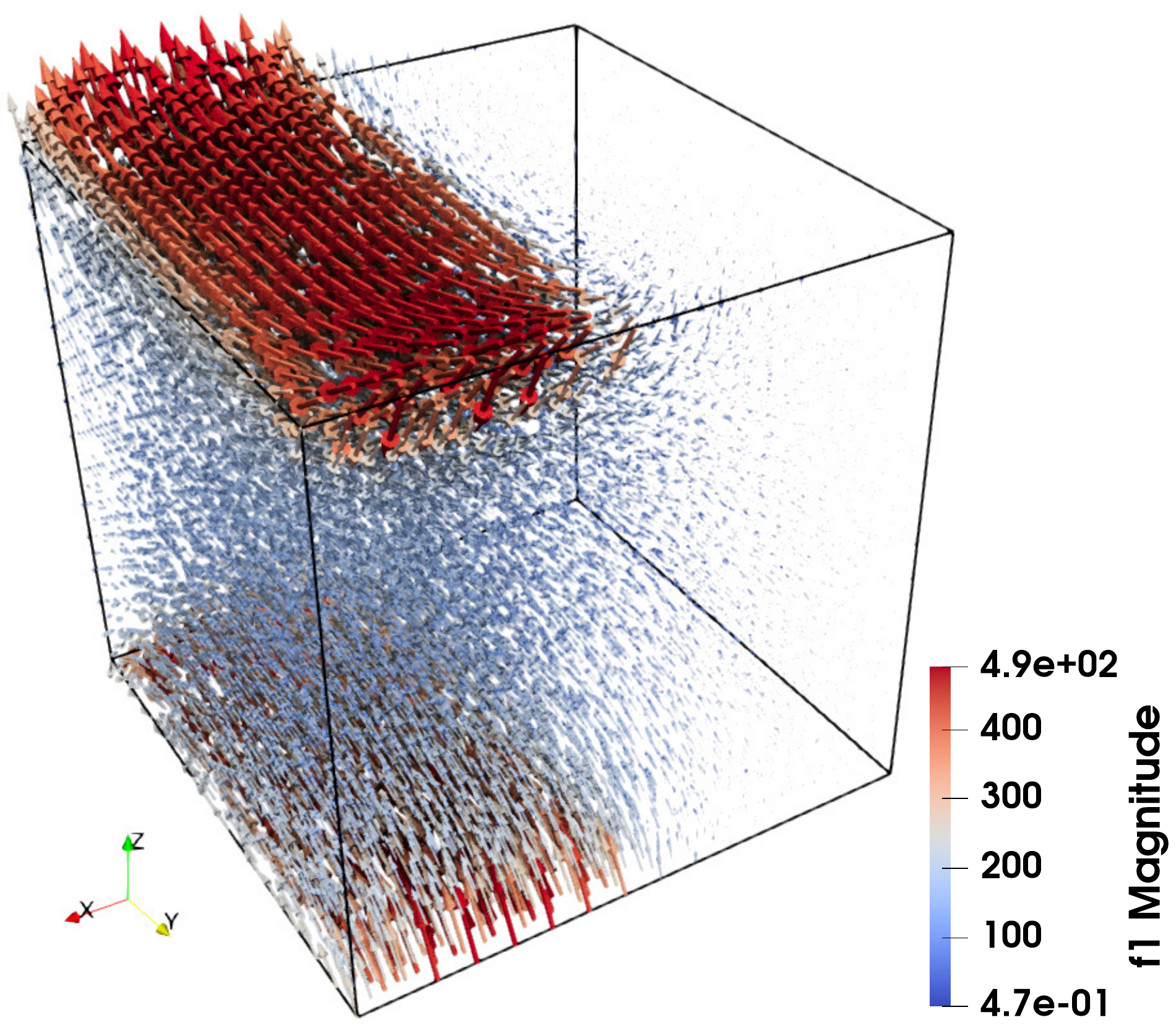} &\includegraphics[width=0.45\textwidth]{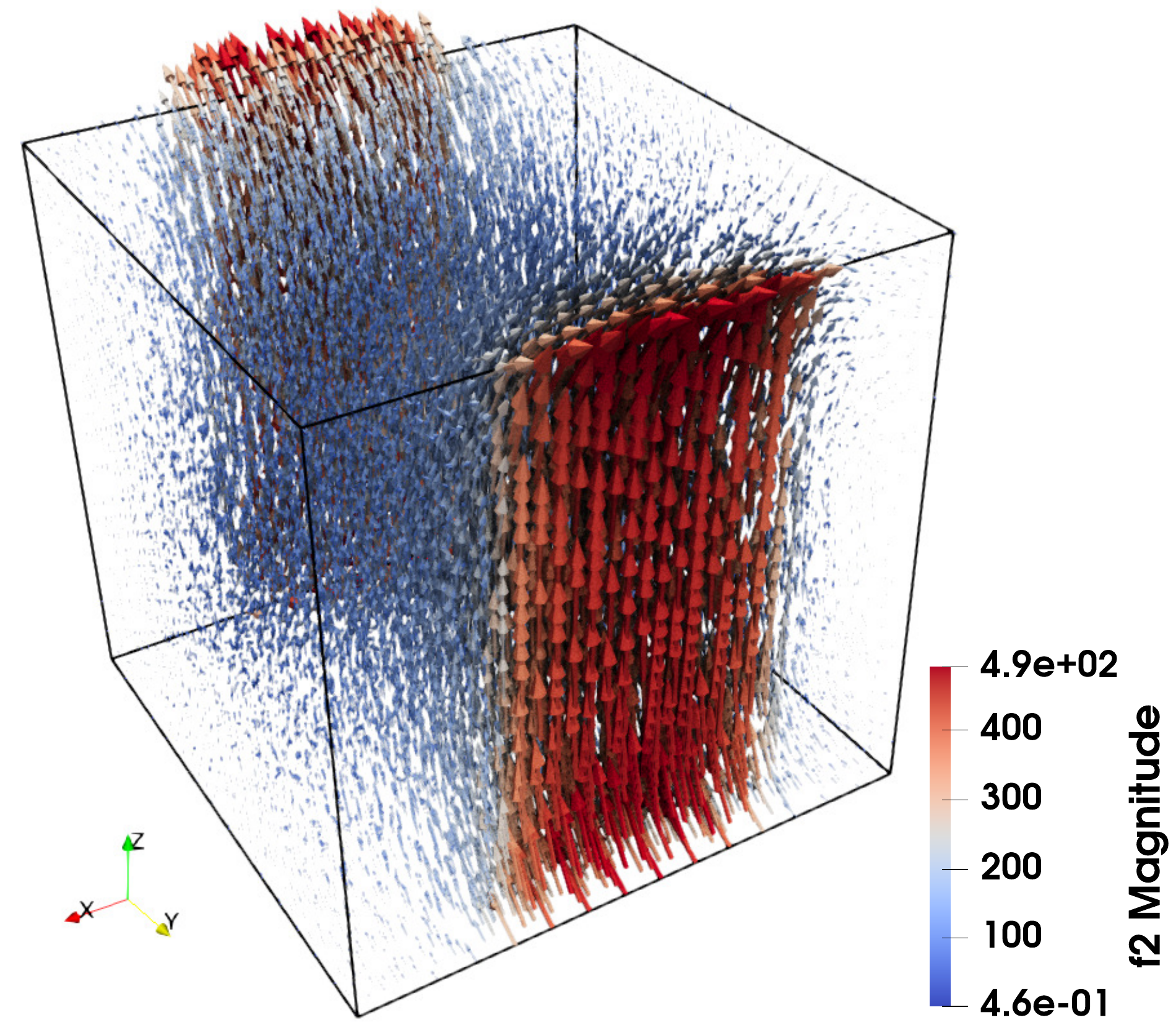}\\
    ${\bf v_1} = \Curl \boldsymbol{\psi_{h1}}$ & ${\bf v_2} = \Curl \boldsymbol{\psi_{h2}}$ \\
    \includegraphics[width=0.45\textwidth]{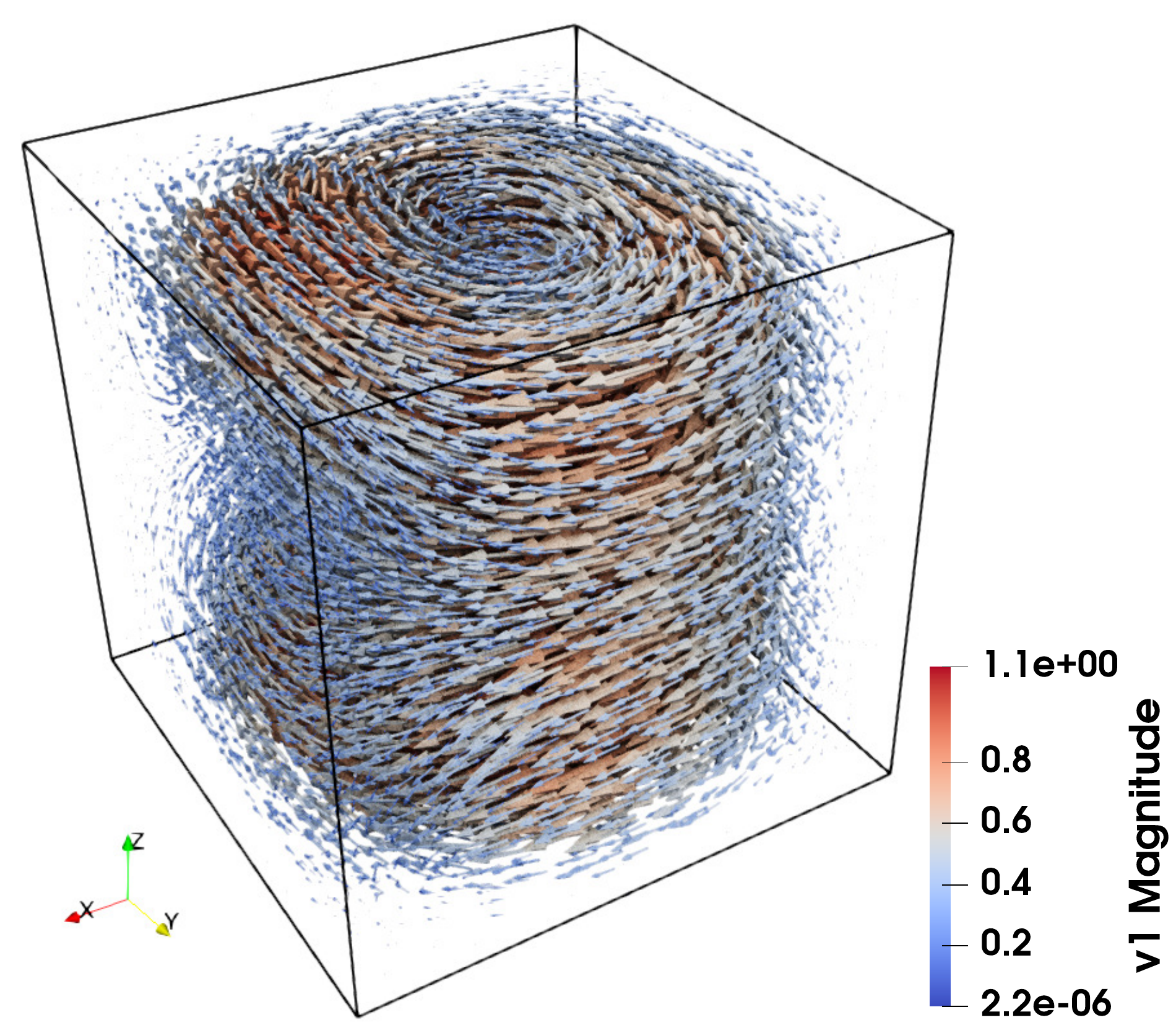} &\includegraphics[width=0.45\textwidth]{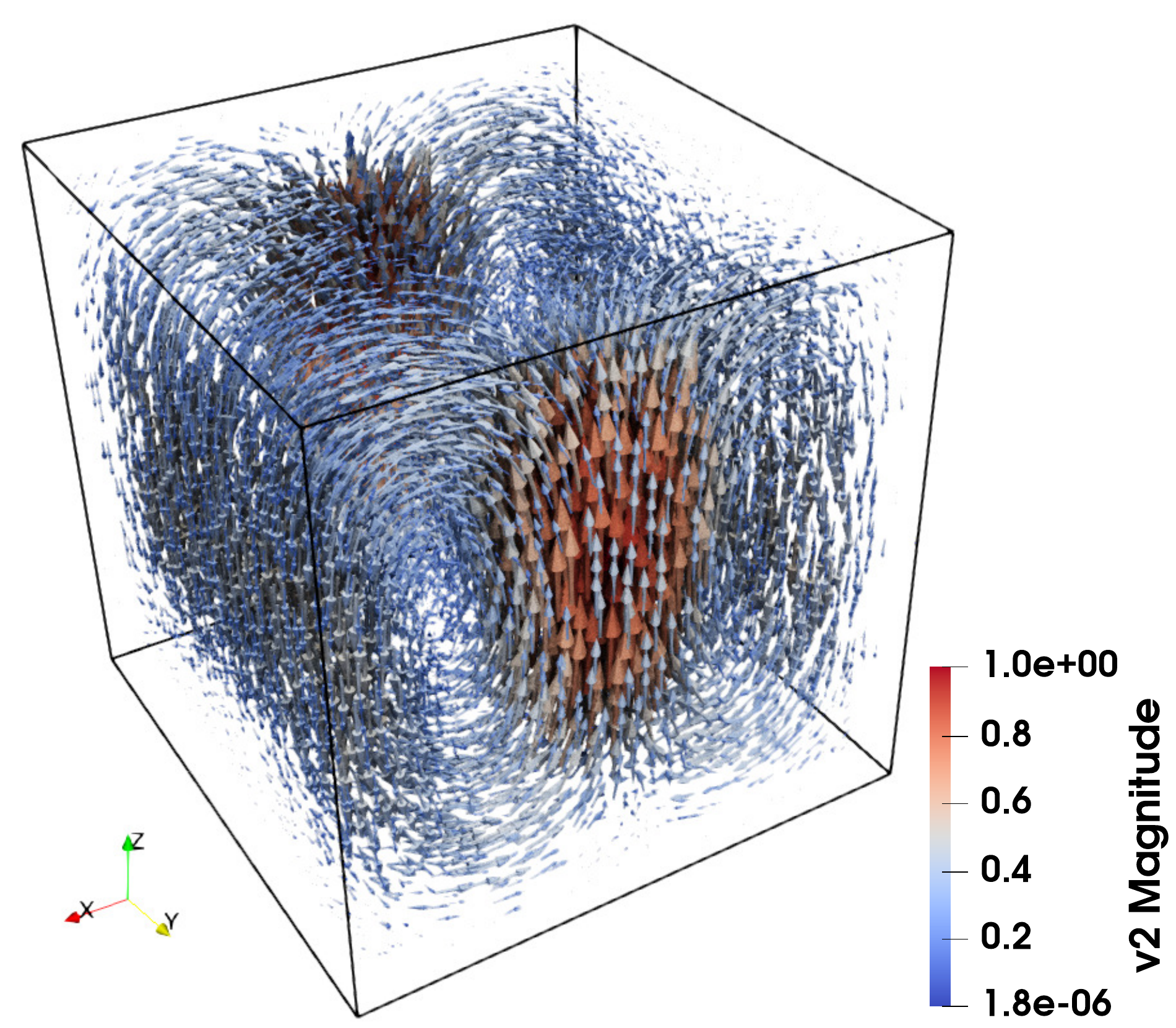}\\
  \end{tabular}
  \caption{\label{fig:numres}Plots of calculated non-dimensional fields, when a single magnet is present: ${\bf h_1}$ and ${\bf h_2}$ are the magnetic fields,  ${\bf f_1}$ and ${\bf f_2}$ are the corresponding Lorentz forces, ${\bf v_1}$ and ${\bf v_2}$ are the resulting  velocities (see \hyperref[myappendix]{\appendixname} for details).}
\end{figure}

By sequently solving $\left(\Sigma_{m,i}\right)$, then $\left(\Sigma_{f,i}^{(1)}\right)$, and lastly $\left(\Sigma_{f,i}^{(2)}\right)$ for $i=\{1,2\}$, the desired velocities are finally obtained by ${\bf v_1} = \Curl \boldsymbol{\psi_1}$ and  ${\bf v_2} = \Curl \boldsymbol{\psi_2}$.
The practical implementation with high order basis functions and discrete form of the full system solved are detailed in \hyperref[myappendix]{\appendixname}. The plots of the calculated magnetic fields, the corresponding forces and the resulting velocities (when a single magnet is present) are provided by \figurename\,\ref{fig:numres}. 

\begin{figure}
  \centering
  \begin{tabular}{c c}
    \includegraphics[width=6.45cm]{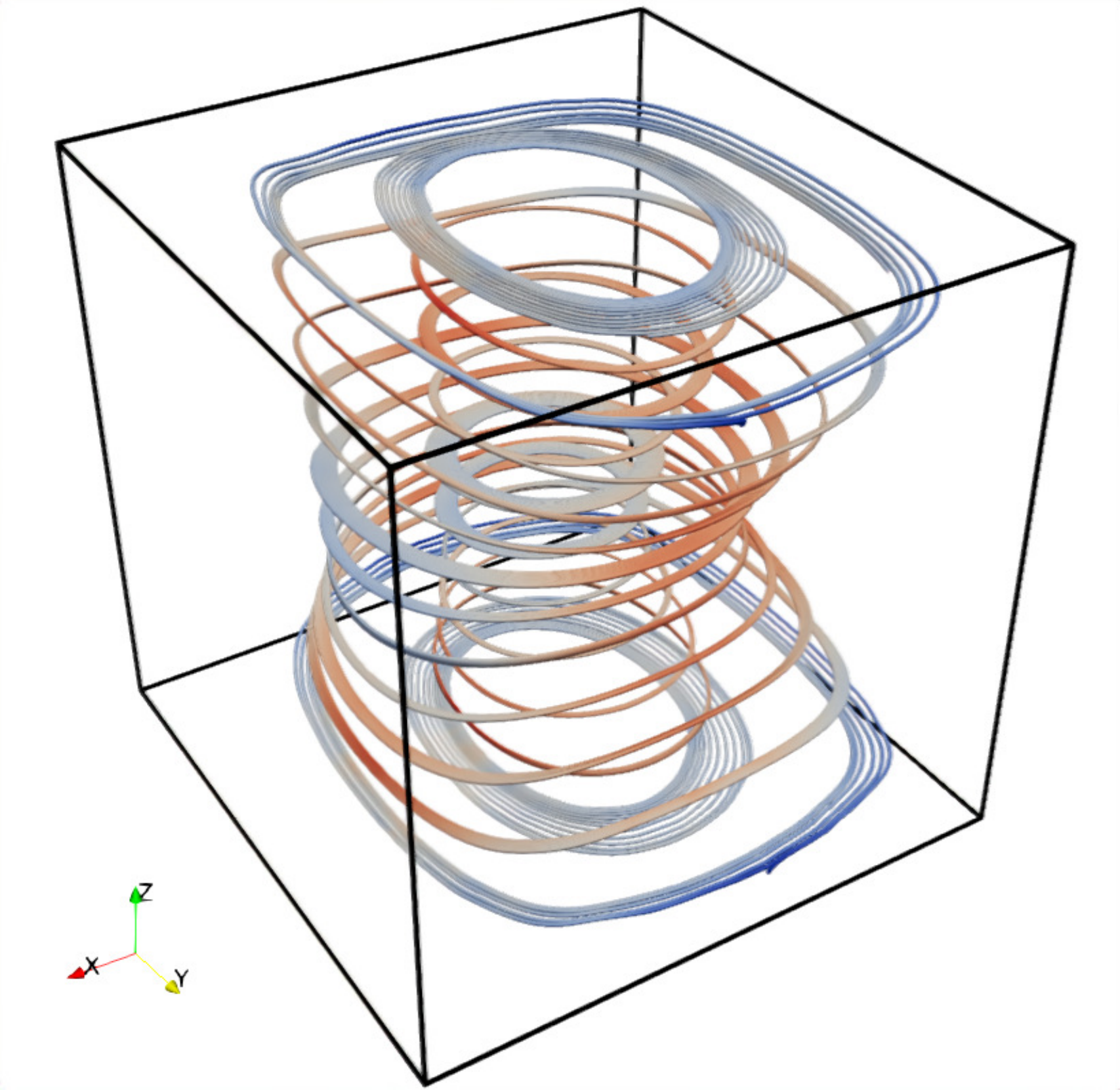} &\includegraphics[width=6.45cm]{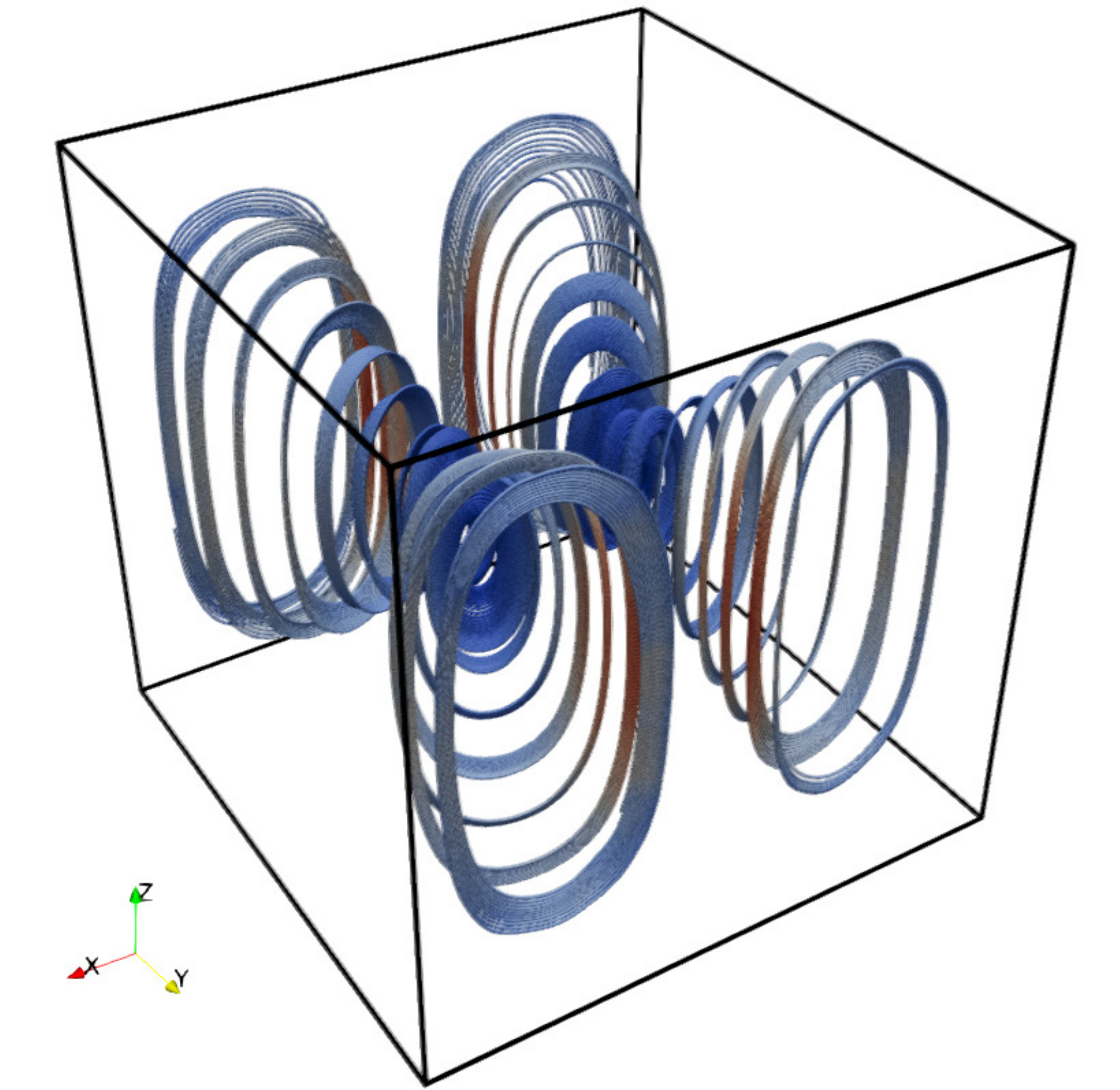}\\
  \end{tabular}
  \caption{\label{fig:streamlines}Examples of velocity streamlines when a single magnet is present. Left: ${\bf v_1}$ (side magnet). Right: ${\bf v_2}$ (central magnet).}
\end{figure}

As expected, the magnetic fields created by the magnets remain  concentrated in the area between each of them, and so does the electromagnetic forces. 
The structure of these velocity fields corresponds to the sketches of \figurename\,\ref{fig:intro}, with one  recirculation cell created by the side magnet \#1 and two cells created by the central magnet \#2. To make this clearer, a few streamlines of ${\bf v1}$ and ${\bf v_2}$ are also displayed in \figurename\,\ref{fig:streamlines}. The motion of tracers in the steady flow obtained by involving the two magnets, i.e. by superposing ${\bf v1}$ and ${\bf v_2}$, is studied in the next section.

Once the computation is complete, the two velocity fields are exported on a regular grid composed of $100\times 100\times 100$ points covering the whole tank domain $\Omega_f$. Values are computed thanks to our high order basis functions in order to keep the best possible accuracy. These export files constitute the final results of the numerical procedure described in this section, and are the starting point of the study of the chaotic behaviour detailed in the following.

\section{Lagrangian chaos and mixing analysis}
\subsection{Highlighting and quantifying chaos}

\subsubsection{Trajectories and sensitivity to initial conditions} 

In the following, we will assume that it will always be possible to modulate the velocity fields ${\bf v_1}$ and ${\bf v_2}$ (e.g. by adjusting the magnet length), so that the total velocity ${\bf v}$ may be written as :
\begin{equation}
  {\bf v} = \alpha\,{\bf v_1} + (1-\alpha)\,{\bf v_2} ~,~~ \text{for}~ \alpha \in [0,1]
  \label{defalpha}
\end{equation}
where  $\alpha$ is the convex combination coefficient of the two velocity fields. 
To find out the value of this velocity field at any point ${\bf x}=(x,y,z)$ within the tank domain $[-0.5, 0.5]^{\,3}$, we simply interpolate  the values calculated in the previous section by means of cubic splines. In the following, these interpolated functions will be still denoted ${\bf v_1}$ and ${\bf v_2}$.

The key point of this part is to find out the parameter $\alpha$ leading to the best possible mixing. To do this, we will study the nonlinear dynamical system corresponding to the trajectory of a tracer particle with position ${\bf x}(t)$, initially released at ${\bf x_0}$. In our case, this reads as:
\begin{equation}
  \renewcommand{\arraystretch}{1.5}
  \left\{
    \begin{array}{l}
      \displaystyle \frac{\text{d}\,{\bf x}}{\text{d}t} = {\bf \dot x}(t) = \alpha\,{\bf v_1}\left({\bf x}\left(t\right)\right) + (1-\alpha)\,{\bf v_2}\left({\bf x}\left(t\right)\right),  t\geq 0\\
      {\bf x}(0) = {\bf x_0} = (x_0, y_0, z_0).
    \end{array}
  \right.
  \label{dxdt} 
\end{equation}
{Here also, velocities, lengths and times have been set non-dimensional by means of $v_0$, $a$ and $a/v_0$ respectively}. 

The ordinary differential system (\ref{dxdt}) has been implemented with the Julia programming language using the {\tt DynamicalSystemsBase.jl} package (and its subset {\tt ChaosTool.jl}) \cite{Datseris2018}. 
We choose a solver based on an Adams-Bashforth explicit method: a fixed time step fifth-order Adams-Bashforth-Moulton method, where a 5th order Adams-Bashforth method is used as predictor and an Adams-Moulton 4-steps method is used for the corrector. {To calculate starting values}, we make use of a Runge-Kutta method of order 4. Throughout the rest of the paper,  
the non-dimensional time step is fixed to $5\cdot 10^{-4}$.

\begin{figure}
  \centering\renewcommand{\arraystretch}{1}
  \begin{tabular}{c c c}
    \includegraphics[width=0.3\textwidth]{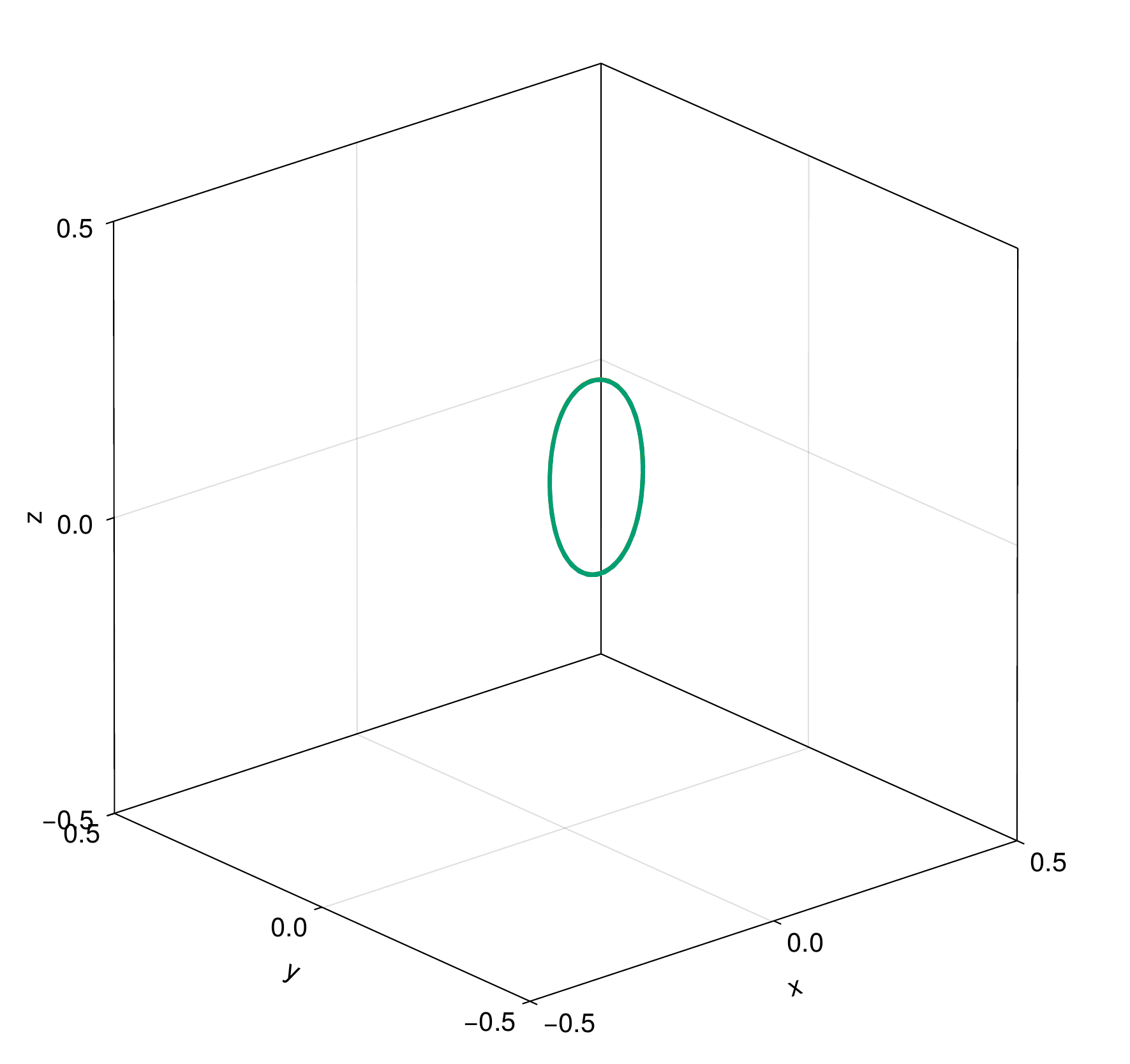} & \includegraphics[width=0.3\textwidth]{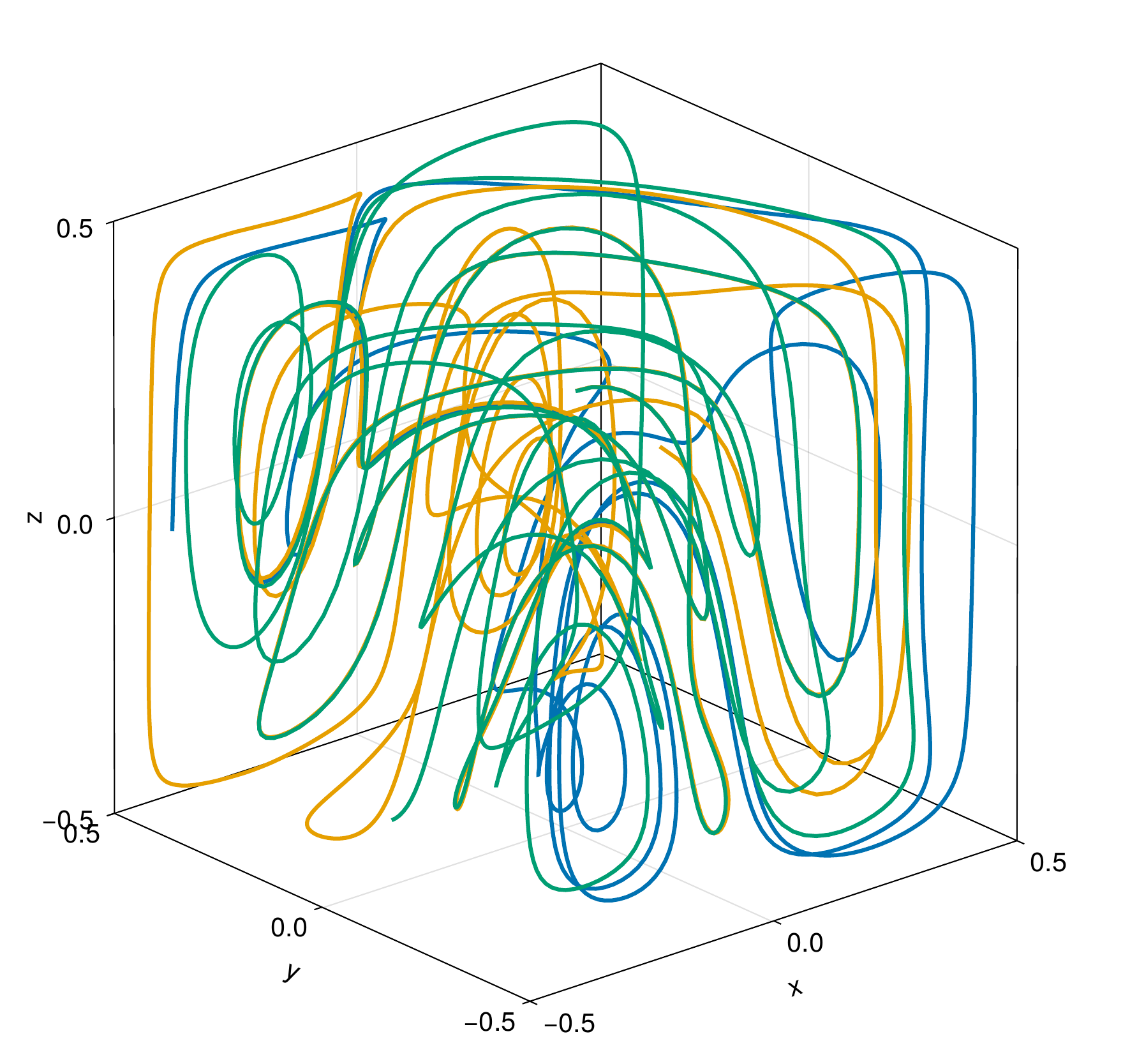} & \includegraphics[width=0.3\textwidth]{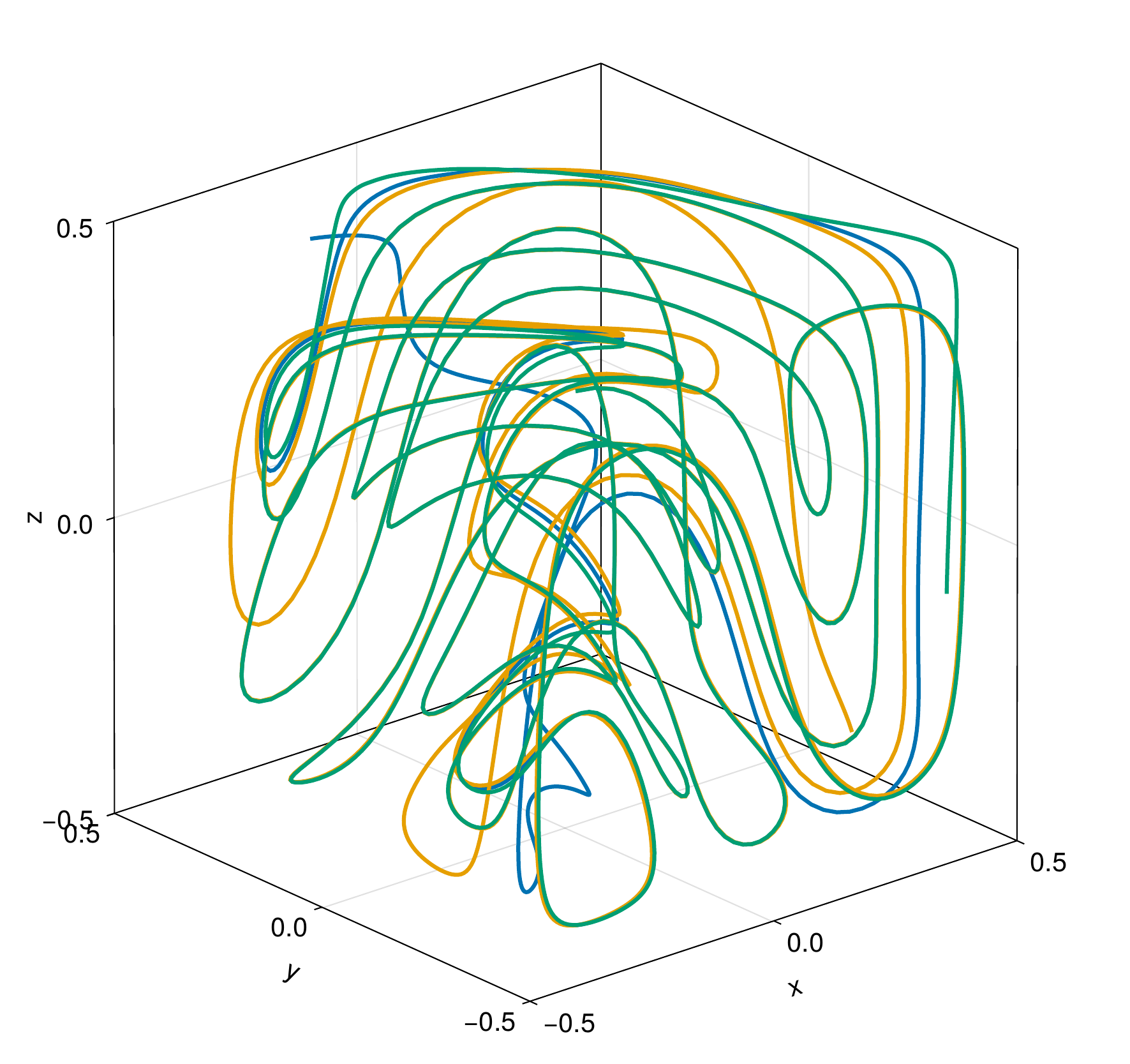}\\
    $\alpha =0$ & $\alpha = \frac{1}{4}$ & $\alpha = \frac{1}{3}$ \\
    \includegraphics[width=0.3\textwidth]{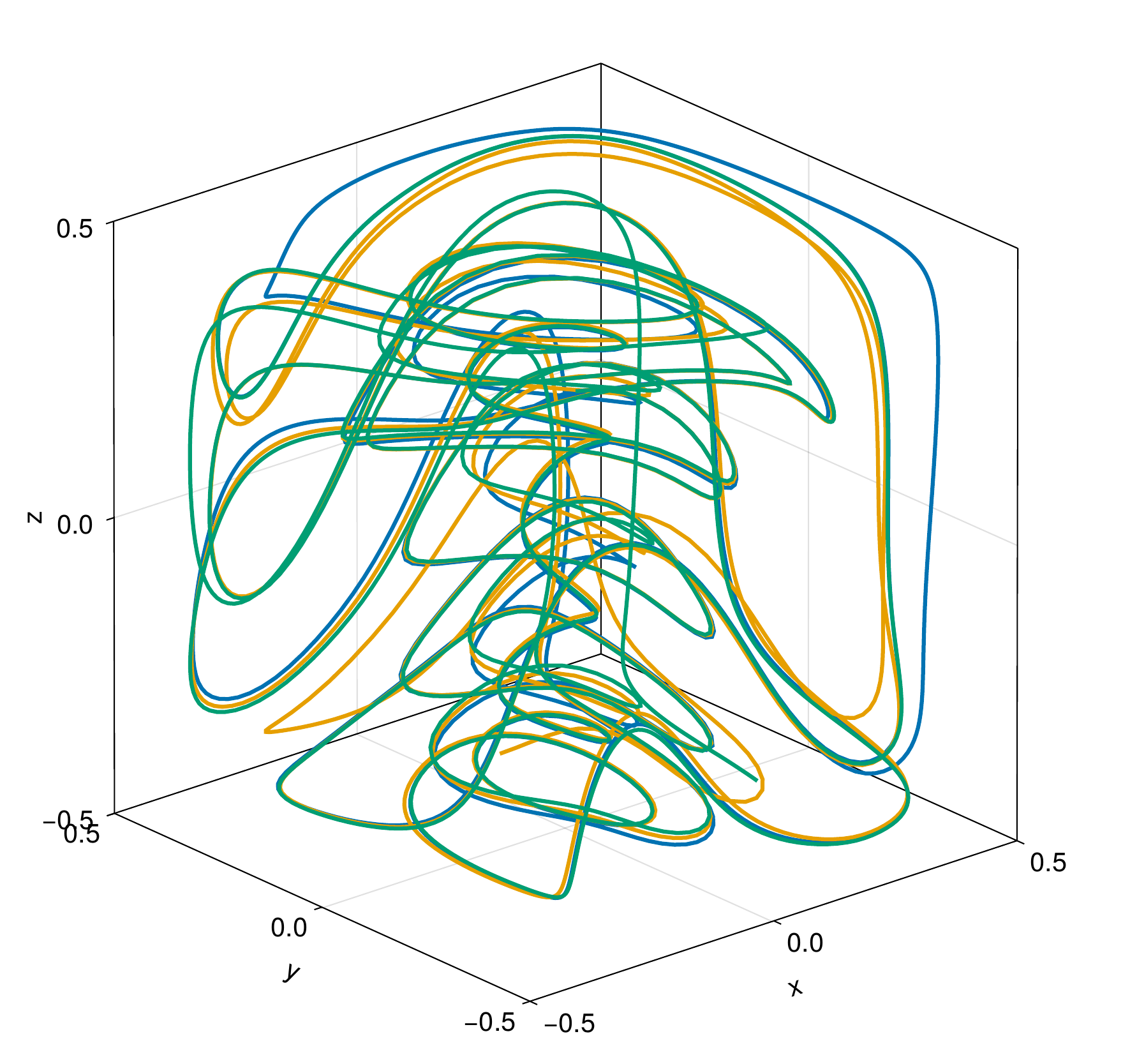}&
    \includegraphics[width=0.3\textwidth]{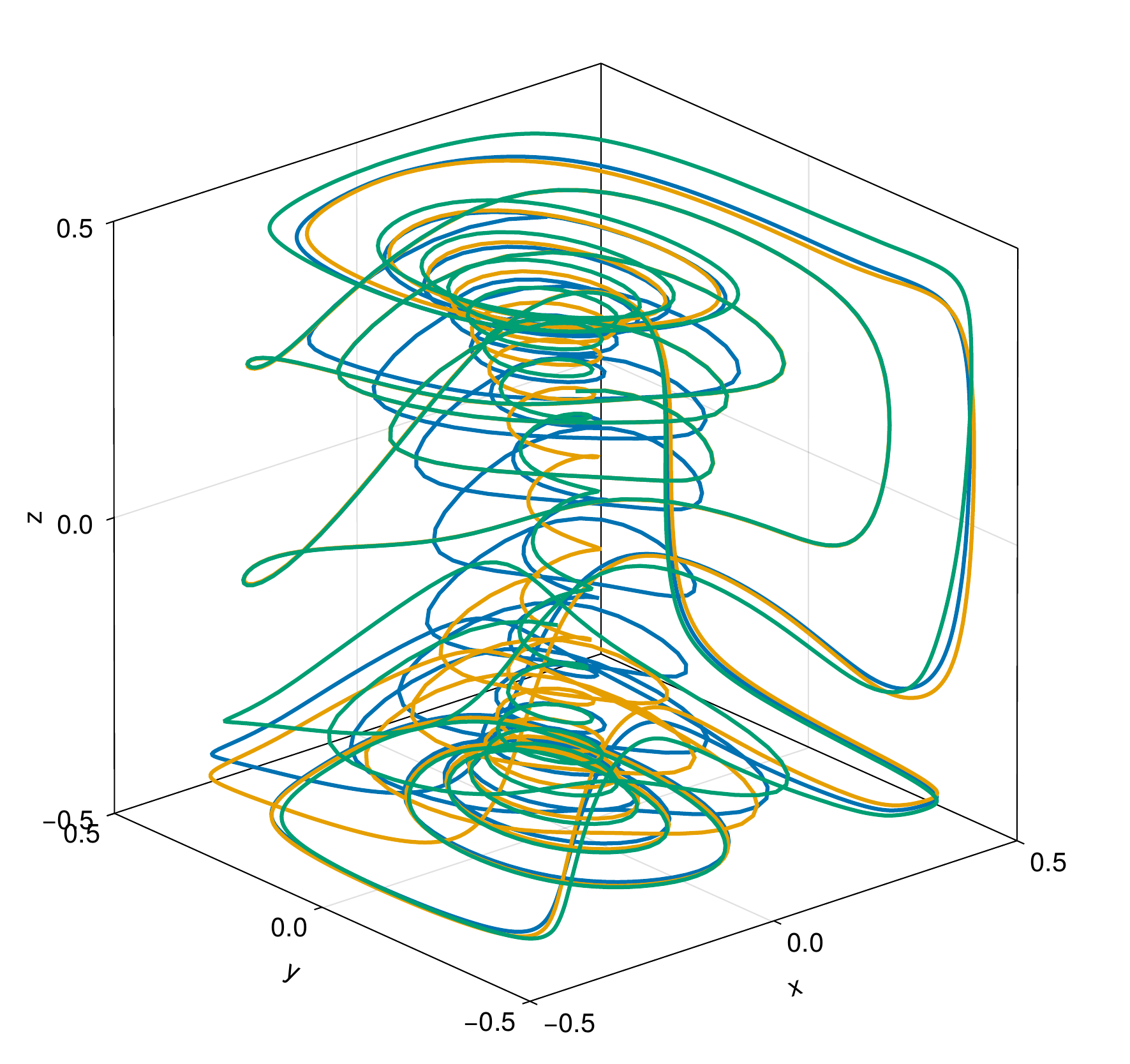}&
    \includegraphics[width=0.3\textwidth]{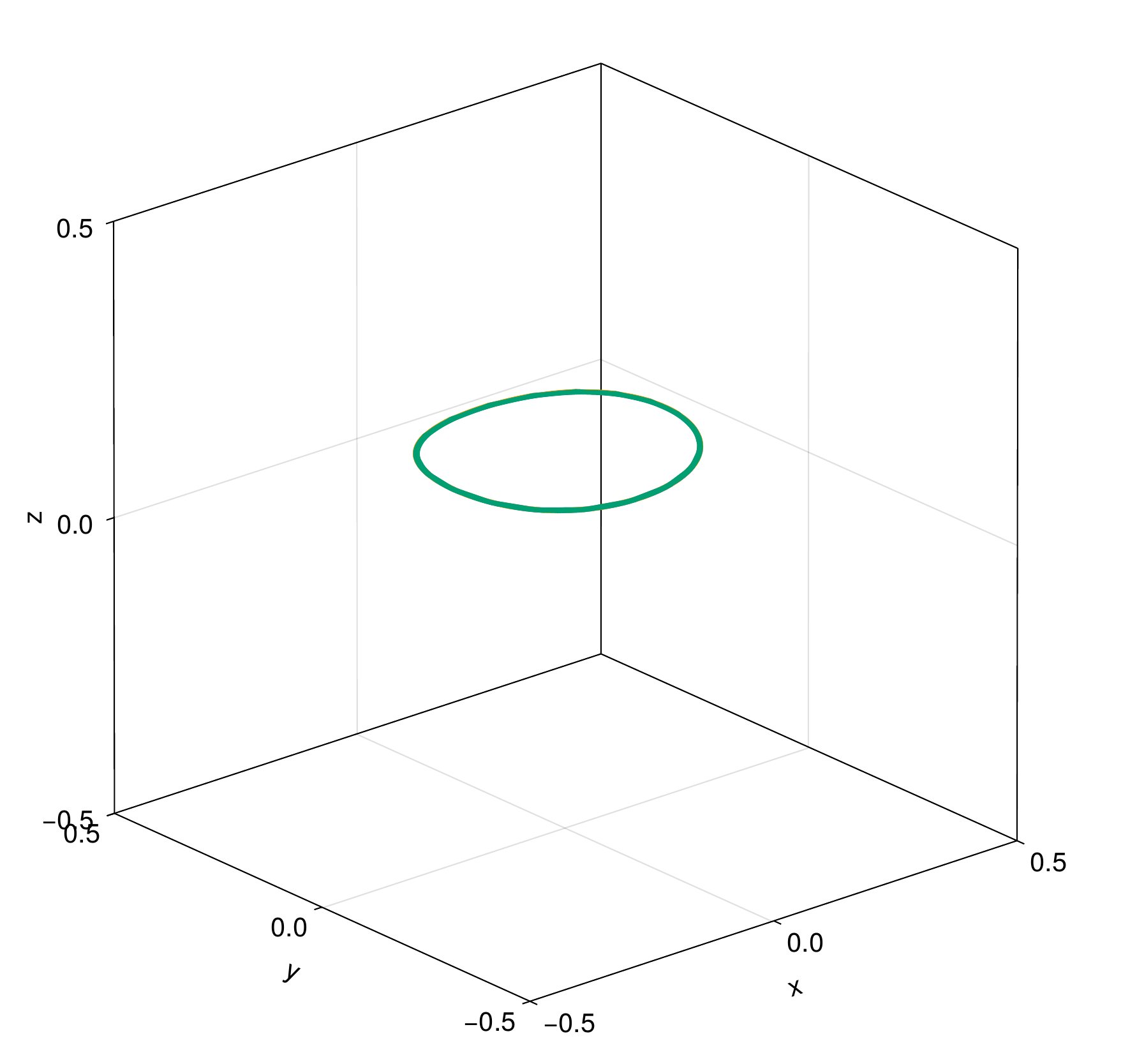}\\
    $\alpha = \frac{1}{2}$ & $\alpha = \frac{2}{3}$ & $\alpha = 1$ \\
  \end{tabular}
  \caption{\label{fig:trajectories}Short-term trajectories for various values of $\alpha$ and for 3 starting points very close to ${\bf x_0} = (0.15,0.15,0.15)$. For $\alpha \not = 0$ and $1$, trajectories are irregular and sensitive to initial conditions. }
\end{figure}

To get started, some trajectories are mapped out in order to study the sensitivity to initial conditions. Three very close starting points are chosen: ${\bf x_0} = (0.15,0.15,0.15)$, ${\bf x_{01}} = {\bf x_0} + \boldsymbol{\varepsilon}$ and ${\bf x_{02}} = {\bf x_0} - \boldsymbol{\varepsilon}$, with $\boldsymbol{\varepsilon} = (1,1,1)\cdot 10^{-3}$.  
The early times of the three corresponding trajectories, for six values of $\alpha$ arbitrarily chosen between 0 and 1, are plotted in  \figurename\,\ref{fig:trajectories}. 

As expected, trajectories correspond to the streamlines of the single-magnet flows (\figurename\,\ref{fig:streamlines})  when $\alpha = 0$ (${\bf v} = {\bf v_2}$) and $\alpha = 1$ (${\bf v} = {\bf v_1}$).  No sensitivity to initial conditions is observed there.

In contrast,  other values of $\alpha$  between these two extremes lead to  much less regular trajectories
(see \figurename\,\ref{fig:trajectories}) and to sensitivity to initial conditions. This property, which characterizes chaos, is particularly pronounced for $\alpha \simeq 0.25$, but is always present for other values of $\alpha$ (except, of course, for $\alpha = 0$ and 1).

It is therefore necessary to study deeper the impact of the coefficient $\alpha$, using other tools such as Poincaré sections and Lyapunov exponents.   

\subsubsection{Poincaré sections}
\label{subsec:Poinca}

\begin{figure}
  \centering
  \centering\renewcommand{\arraystretch}{1.5}
  \begin{tabular}{ c | c | c | c }
    $\alpha = 0.2$ & $\alpha = 0.25$ & $\alpha = 0.3$ & $\alpha = 0.35$\\
    \includegraphics[width=0.225\textwidth]{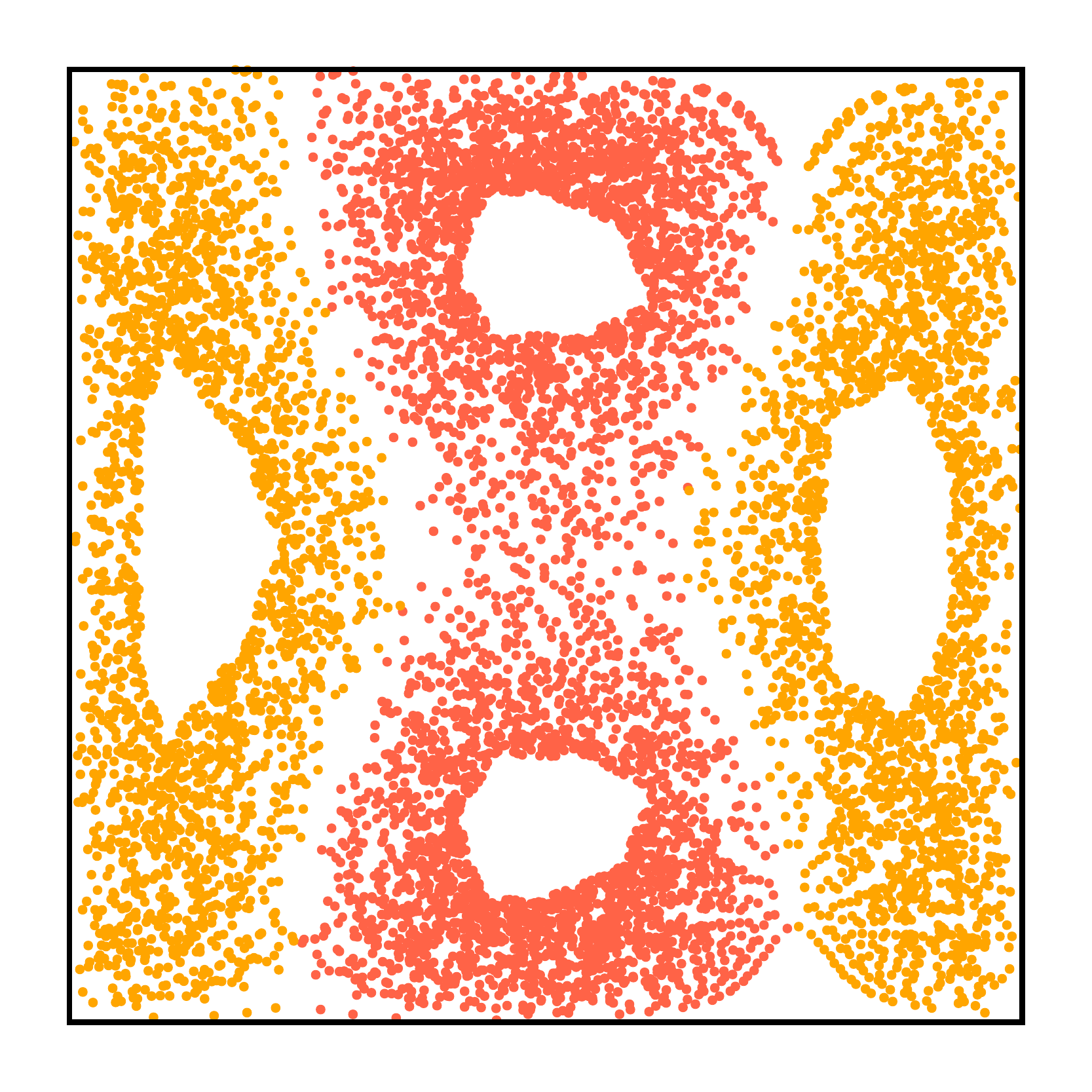}&
    \includegraphics[width=0.225\textwidth]{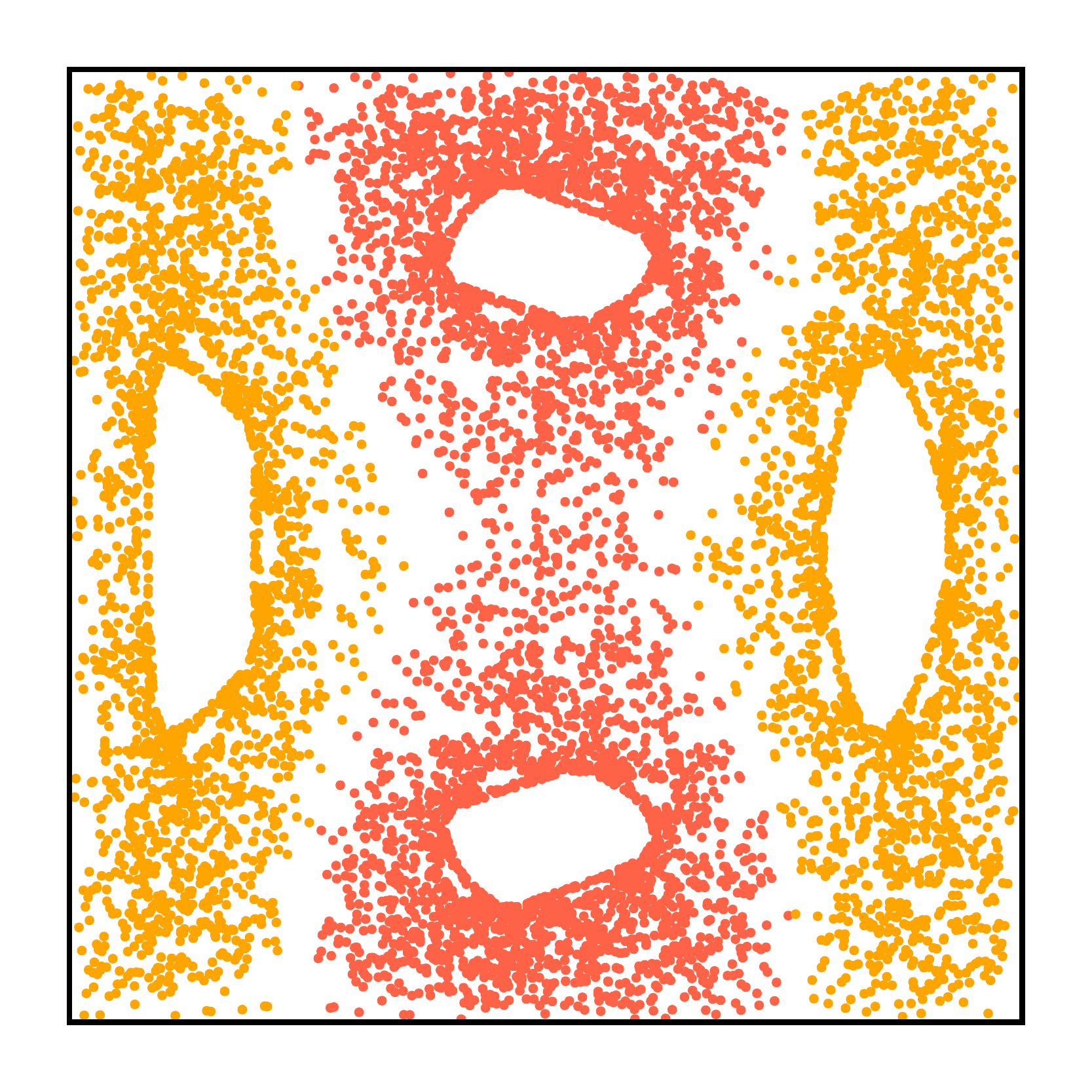}&
    \includegraphics[width=0.225\textwidth]{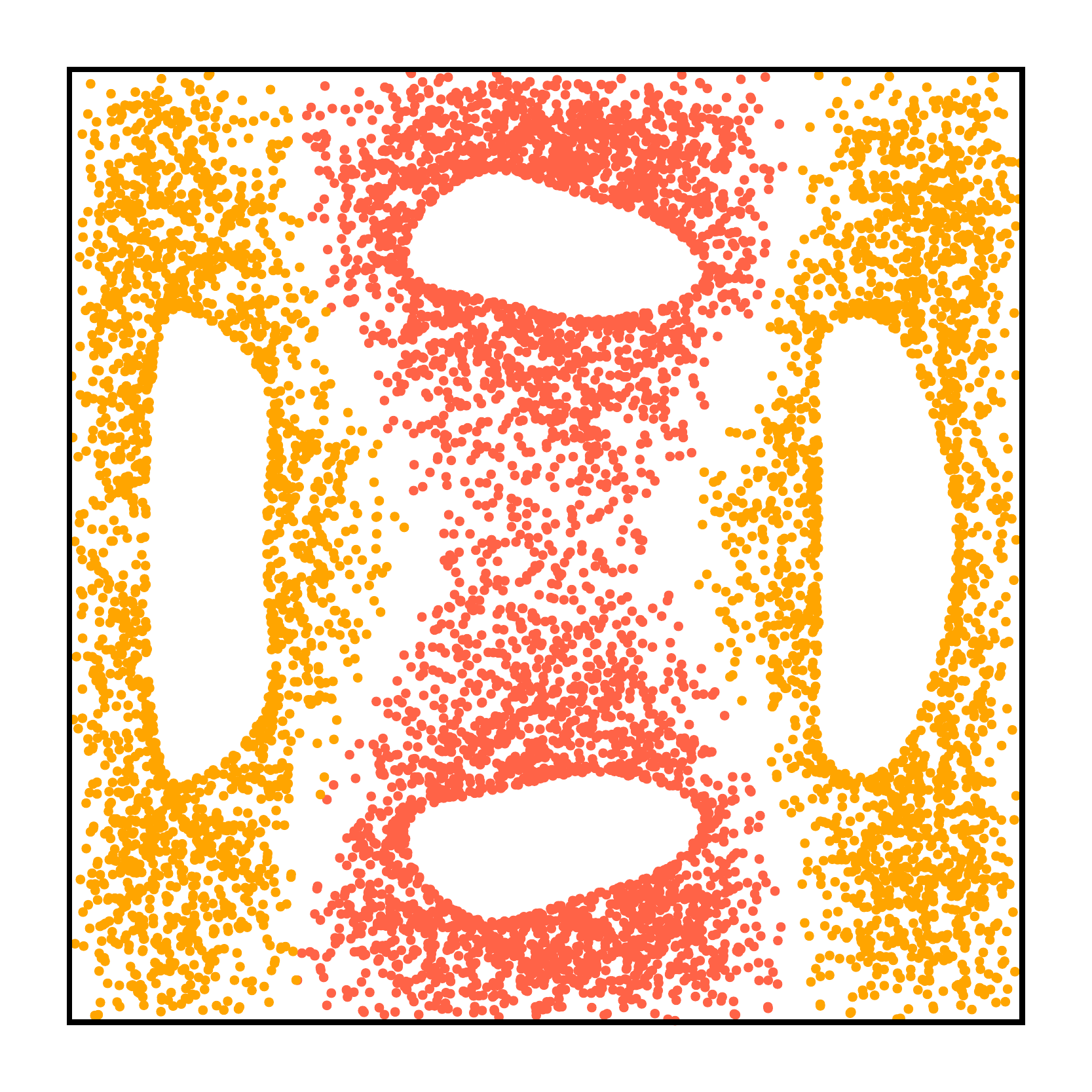}&
    \includegraphics[width=0.225\textwidth]{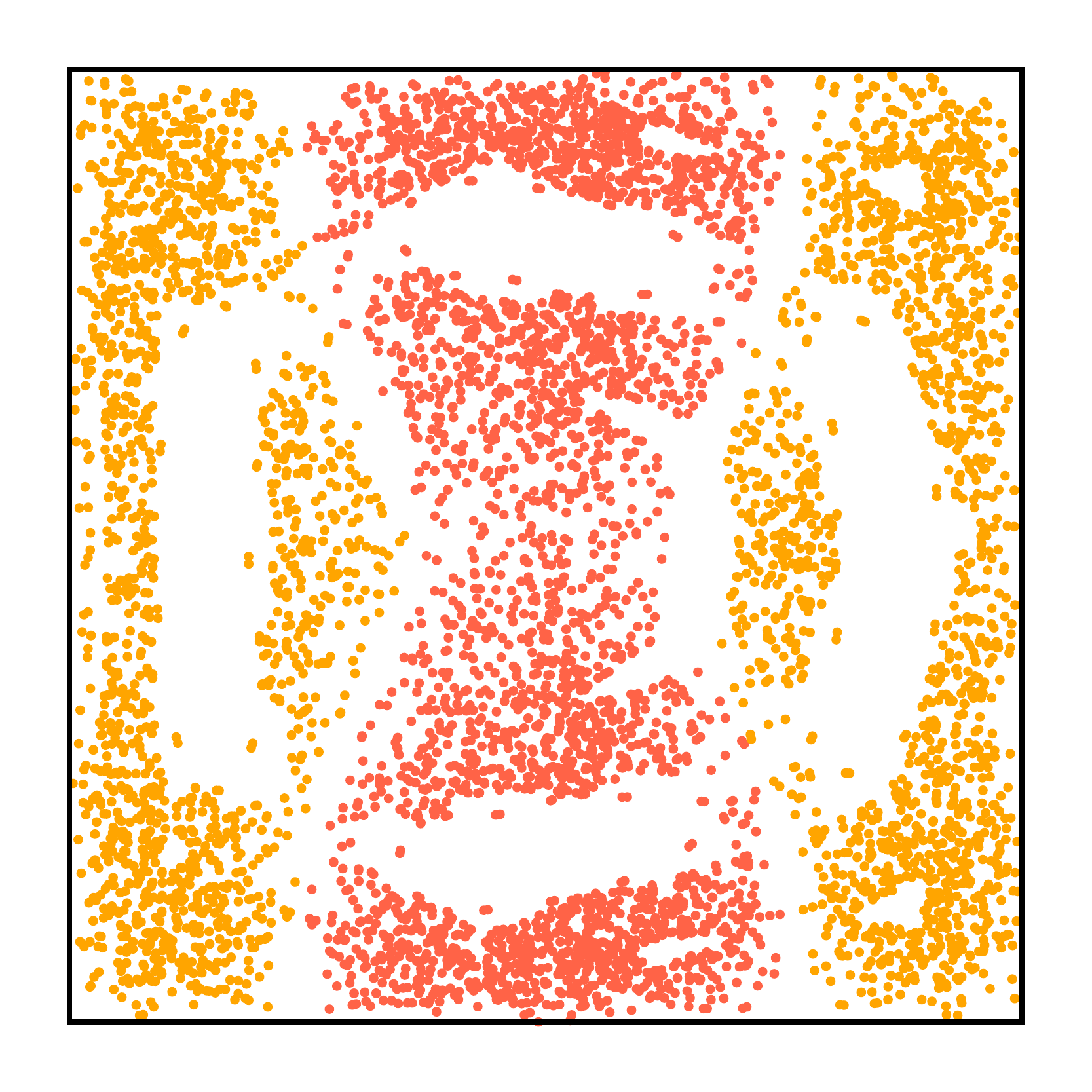}\\
    \hline
    $\alpha = 0.40$ & $\alpha = 0.45$ & $\alpha = 0.5$ & $\alpha = 0.6$\\
    \includegraphics[width=0.225\textwidth]{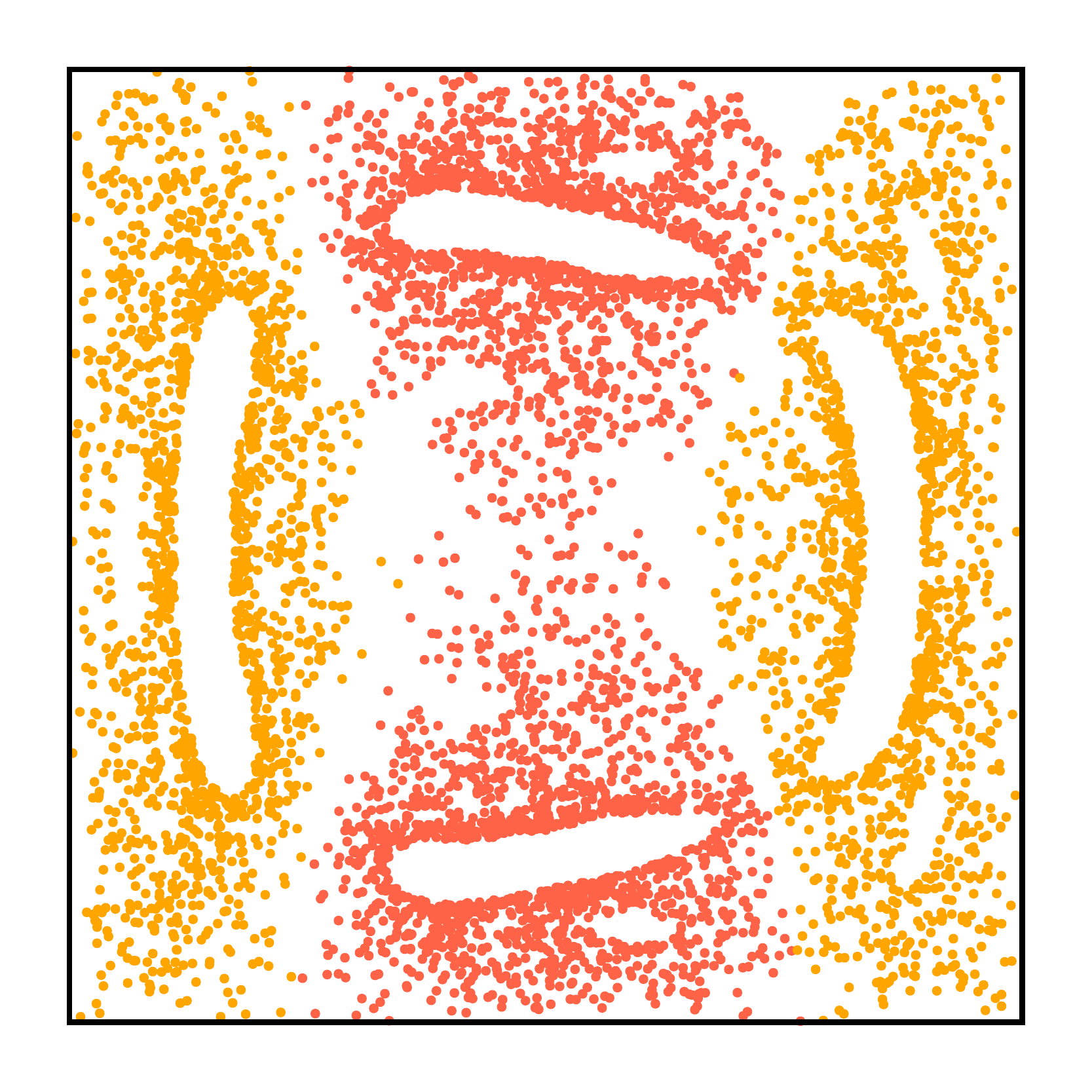}&
    \includegraphics[width=0.225\textwidth]{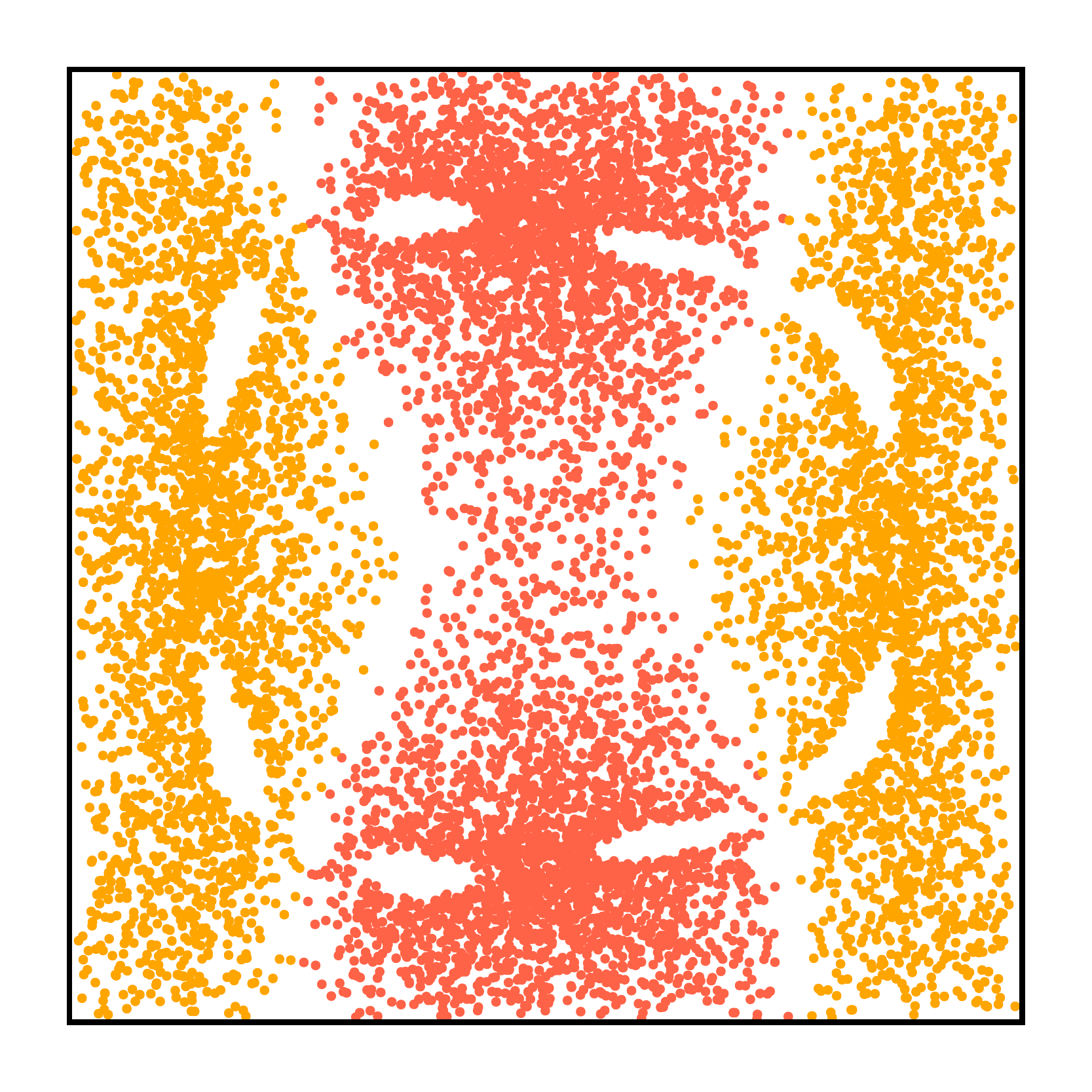}&
    \includegraphics[width=0.225\textwidth]{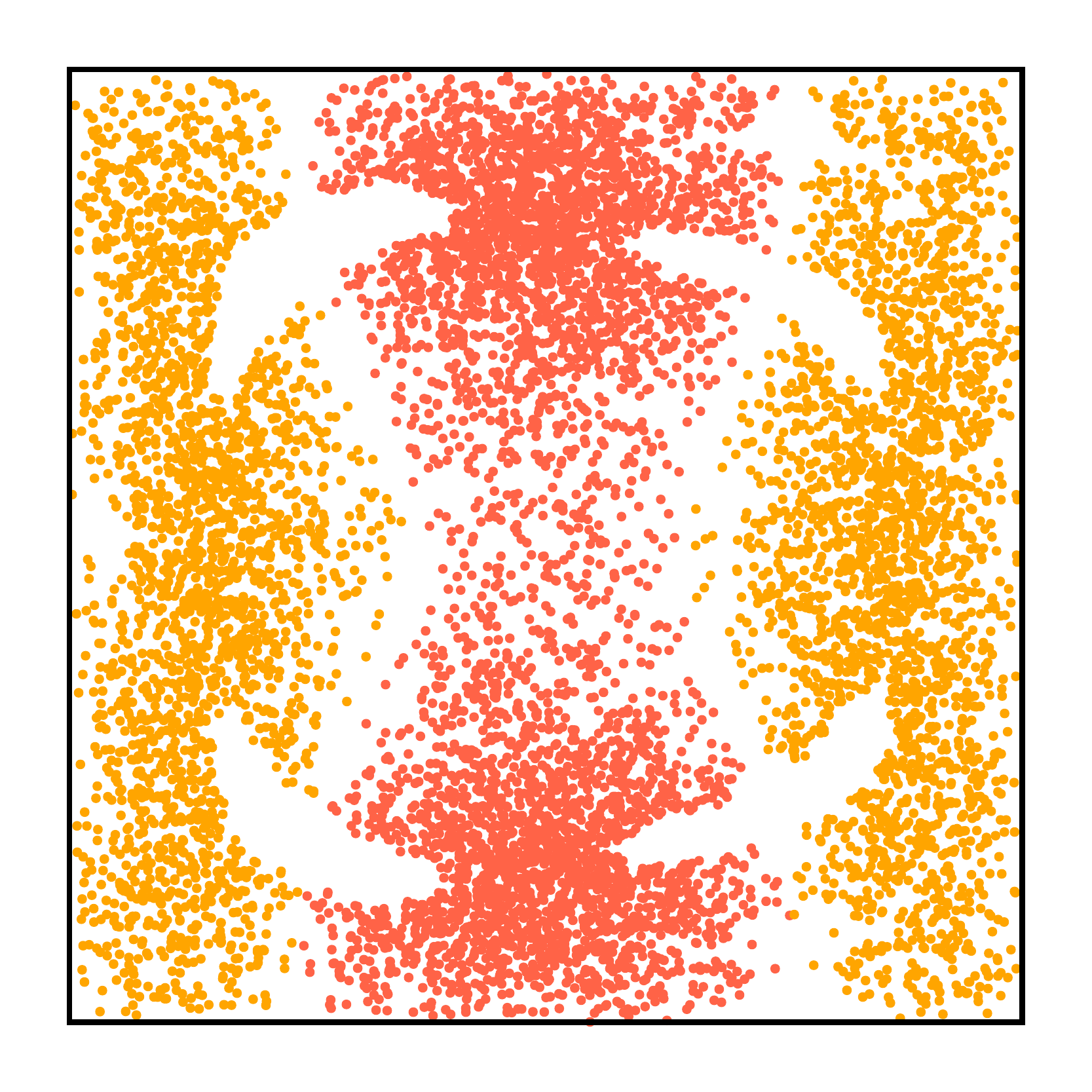}&
    \includegraphics[width=0.225\textwidth]{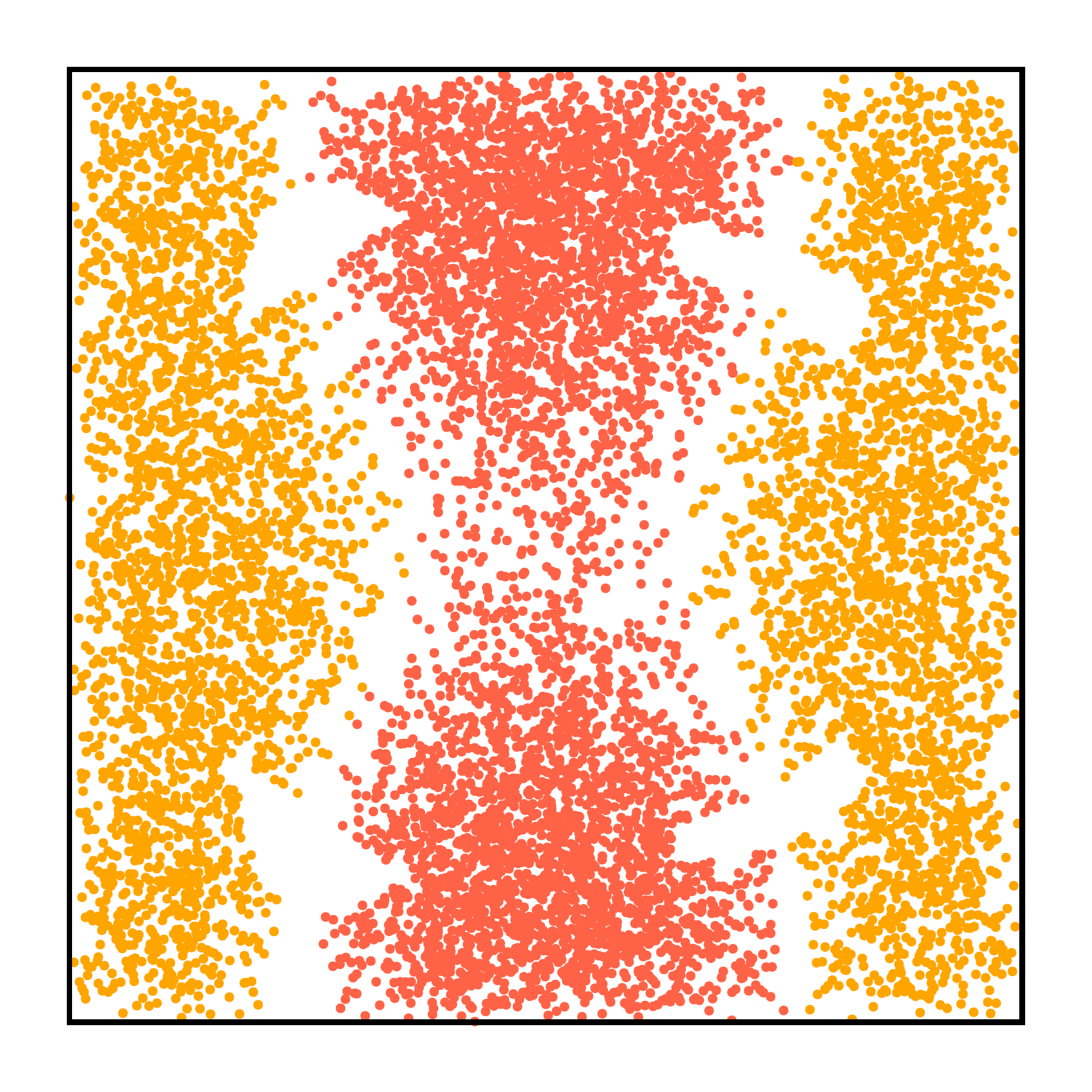}\\
  \end{tabular}
  \caption{\label{fig:sosz0vsalpha}Influence of the coefficient $\alpha$ on the size and location of tori of periodic trajectories: dead zones represented by a hole in the Poincaré sections ($z=0$,  ${\bf x_0} = (0.15,0.15,0.15)$), the darkest color corresponds to outgoing points and the lightest color to incoming points.}
\end{figure}

For the single starting point ${\bf x_0} = (0.15,0.15,0.15)$  we draw the corresponding Poincaré sections in the plane $z=0$. They are shown  in \figurename\,\ref{fig:sosz0vsalpha}, for eight values of $\alpha$  equally distributed between 0.2 and 0.6. It is important to point out that these plots have been obtained for a unique starting point, and not for a set of several points tightened around the initial condition. 
In the three cases $\alpha = 0.3$, 0.4 and 0.55, just a second starting point very close to ${\bf x_0}$ has been added to obtain a similar number of impact points for all sections. The high densities of the plots show that the problem is solved with a very good accuracy.  

First, we observe in \figurename\,\ref{fig:sosz0vsalpha} that most of the Poincaré sections are covered, which means that the entire tank is explored by each trajectory. This is reassuring in terms of mixing efficiency. 
Nonetheless, we note the presence of holes in each case. This seems to indicate the presence of regions of regular trajectories which must take the form of torii of periodic trajectories. Depending on the values of the coefficient $\alpha$, the position of these holes varies, with a $+$-shaped placement for low values of $\alpha$  and a shift towards an $\times$-shape for higher values. The changeover seems to take place around $\alpha=0.45$. 

\begin{figure}
  \centering\renewcommand{\arraystretch}{2}
  \begin{tabular}{| c c |}
    \hline
    \multicolumn{2}{|c|}{$\alpha = 0.25$ / ${\bf x_0} = (0.15,0.15,0.15)$:\,orange, $(0.35,0,0)$:\,blue} \\
    \includegraphics[width=0.5\textwidth]{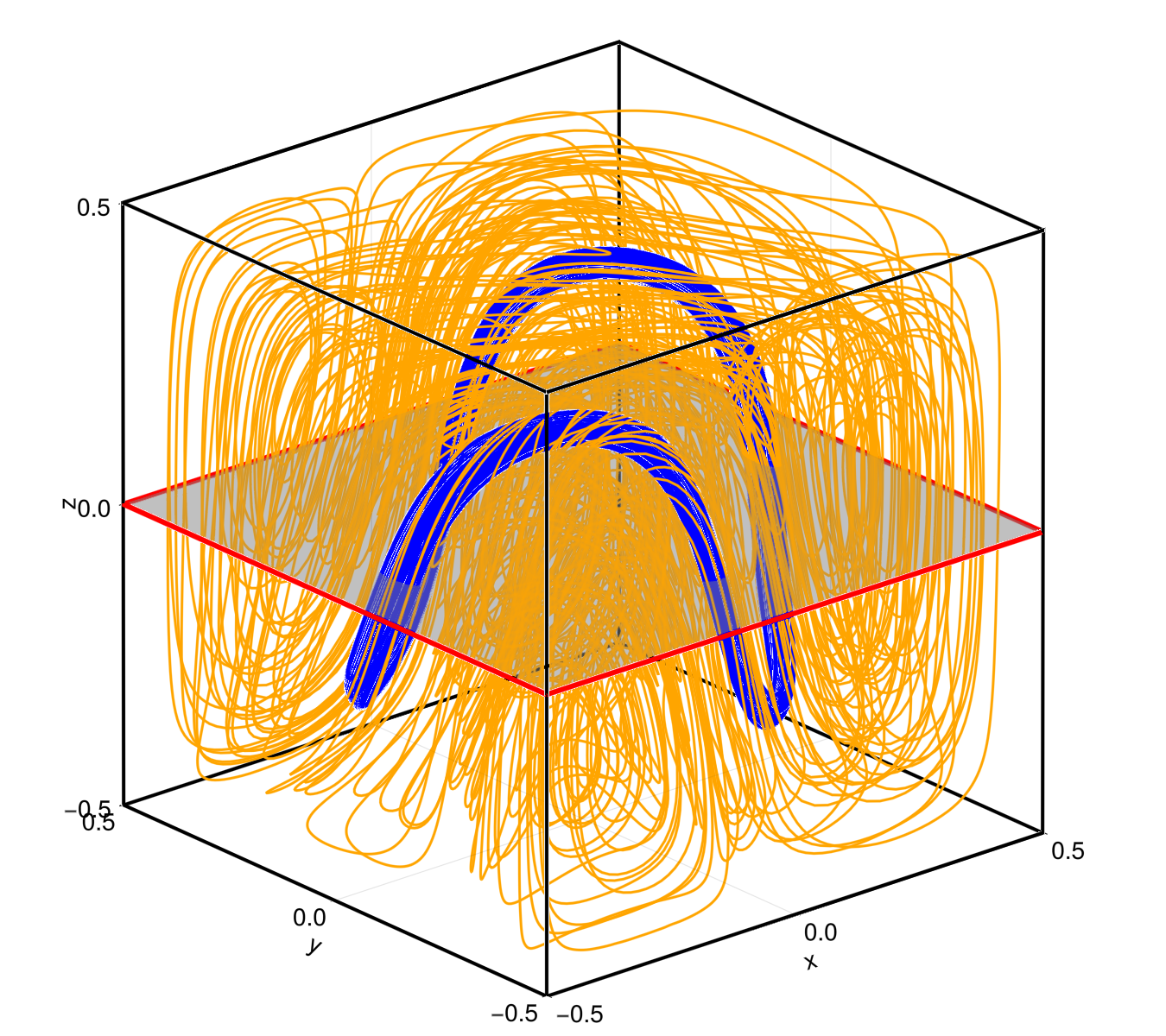} & \includegraphics[width=0.45\textwidth]{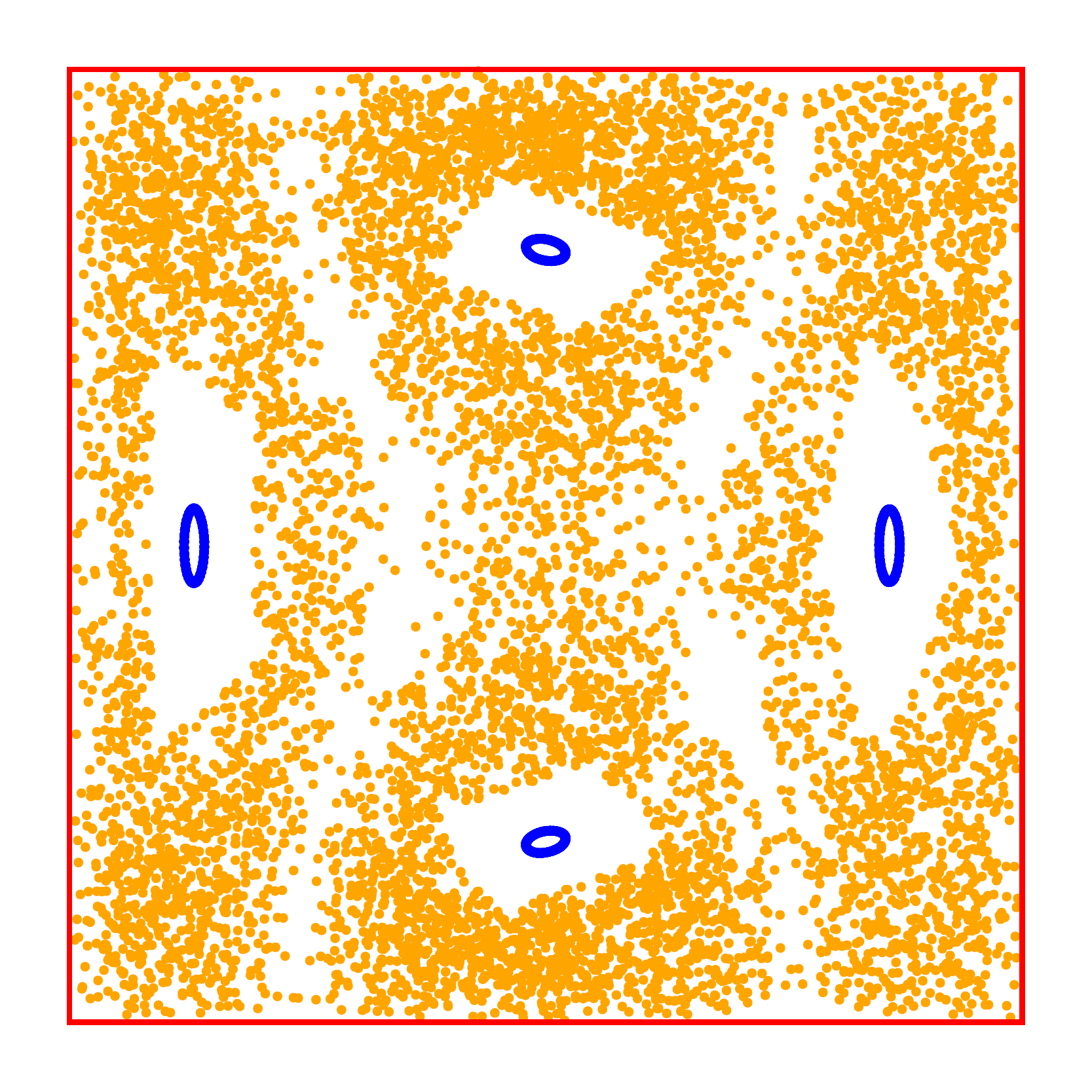} \\
    \hline
    \multicolumn{2}{|c|}{~$\alpha = 0.5$ / ${\bf x_0} = (0.15,0.15,0.15)$:\,orange, $(0.25,0.25,0)$:\,blue,
    $(-0.25,0.3,0)$:\,green~}\\
    \includegraphics[width=0.5\textwidth]{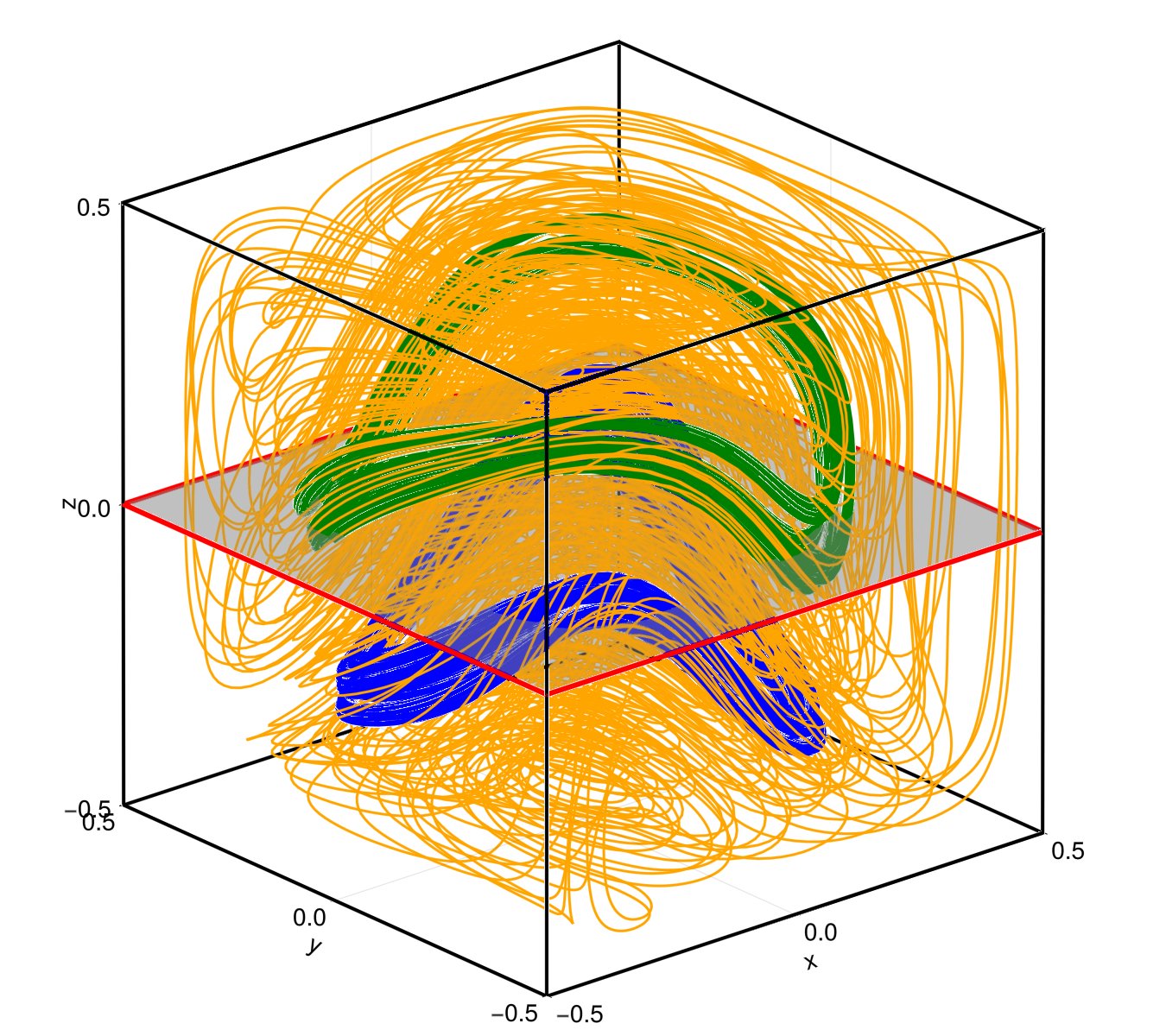} & \includegraphics[width=0.45\textwidth]{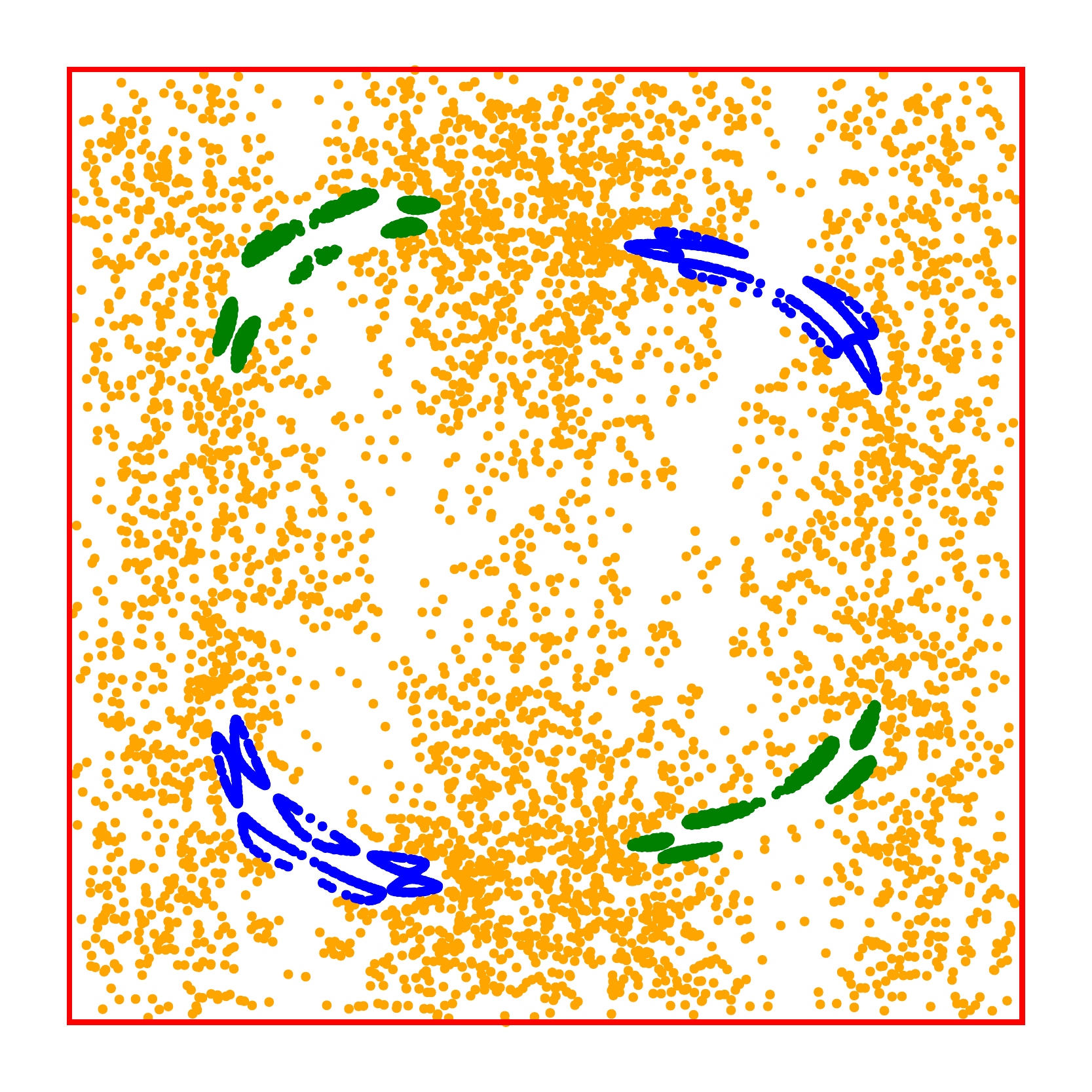}\\
    \hline
  \end{tabular}
  \caption{\label{fig:tori}Trajectories and corresponding $z=0$ Poincaré sections with highlight of one torus or two tori of periodic trajectories.}
\end{figure}

To better visualize these torii of periodic trajectories, a 3D plot is shown in \figurename\,\ref{fig:tori}. For two values of $\alpha$  (0.25 and 0.5), the beginning of trajectories is plotted for the previous starting point (${\bf x_0}=(0.15,0.15,0.15)$) and for points taken in one or two holes. 
The corresponding Poincaré sections $z=0$ have been added alongside.   
The plots clearly show the periodic trajectories inside the orange chaotic trajectory. We can see one torus for the case $\alpha = 0.25$, corresponding to the $+$-shaped holes. For the case $\alpha=0.5$,  where the holes follow an $\times$-shape, two tori are present. 
However, even in the presence of these regular trajectories, the flow may be chaotic from a global point of view. To characterize this, we will calculate some Lyapunov exponents in the next section. 
 
\begin{figure}
  \centering\renewcommand{\arraystretch}{1.5}
  \begin{tabular}{c | c | c}
    $x = -0.25$ &  $x = 0$ &  $x = 0.25$\\[-0.25em]
    \includegraphics[width=0.3\textwidth]{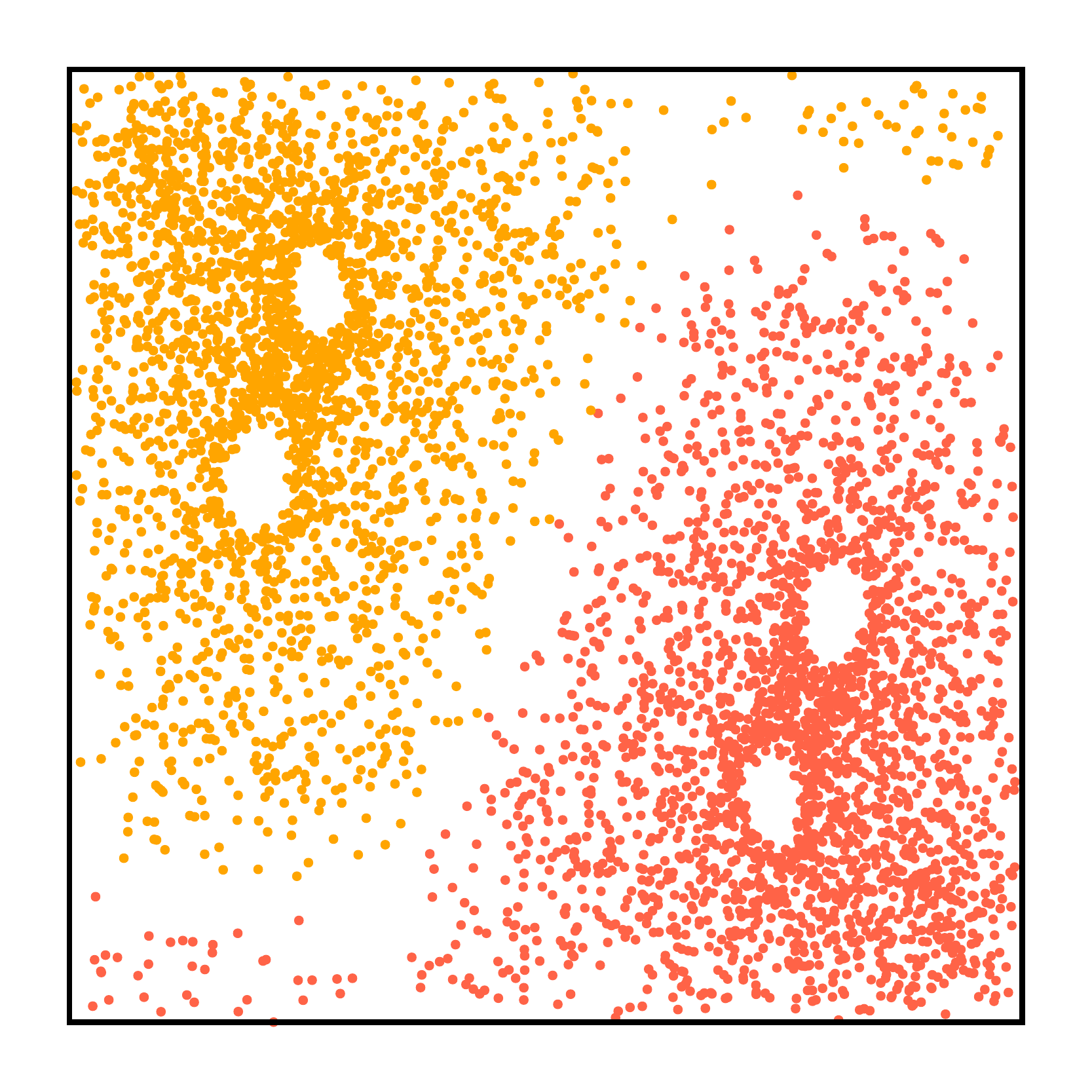} & \includegraphics[width=0.3\textwidth]{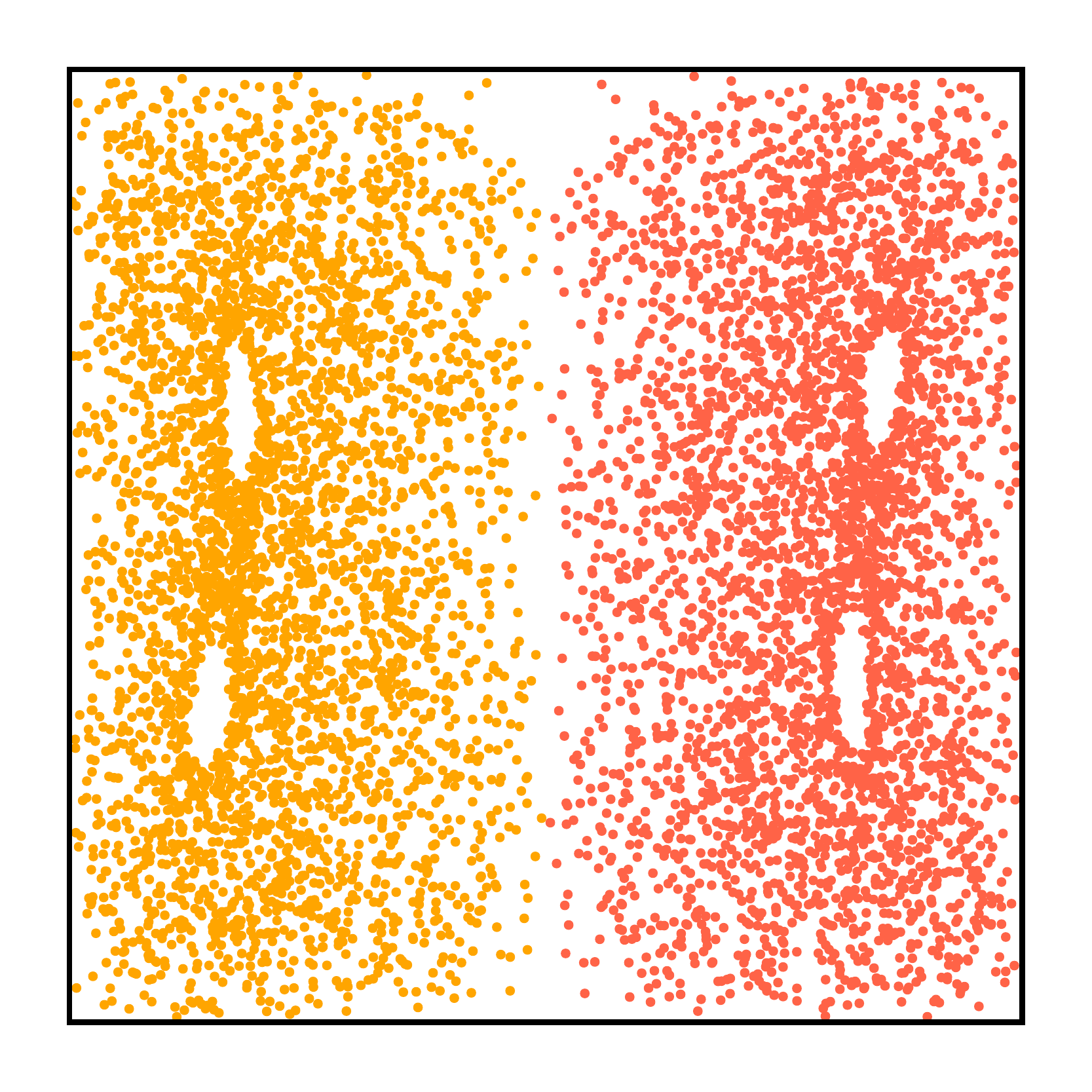} & \includegraphics[width=0.3\textwidth]{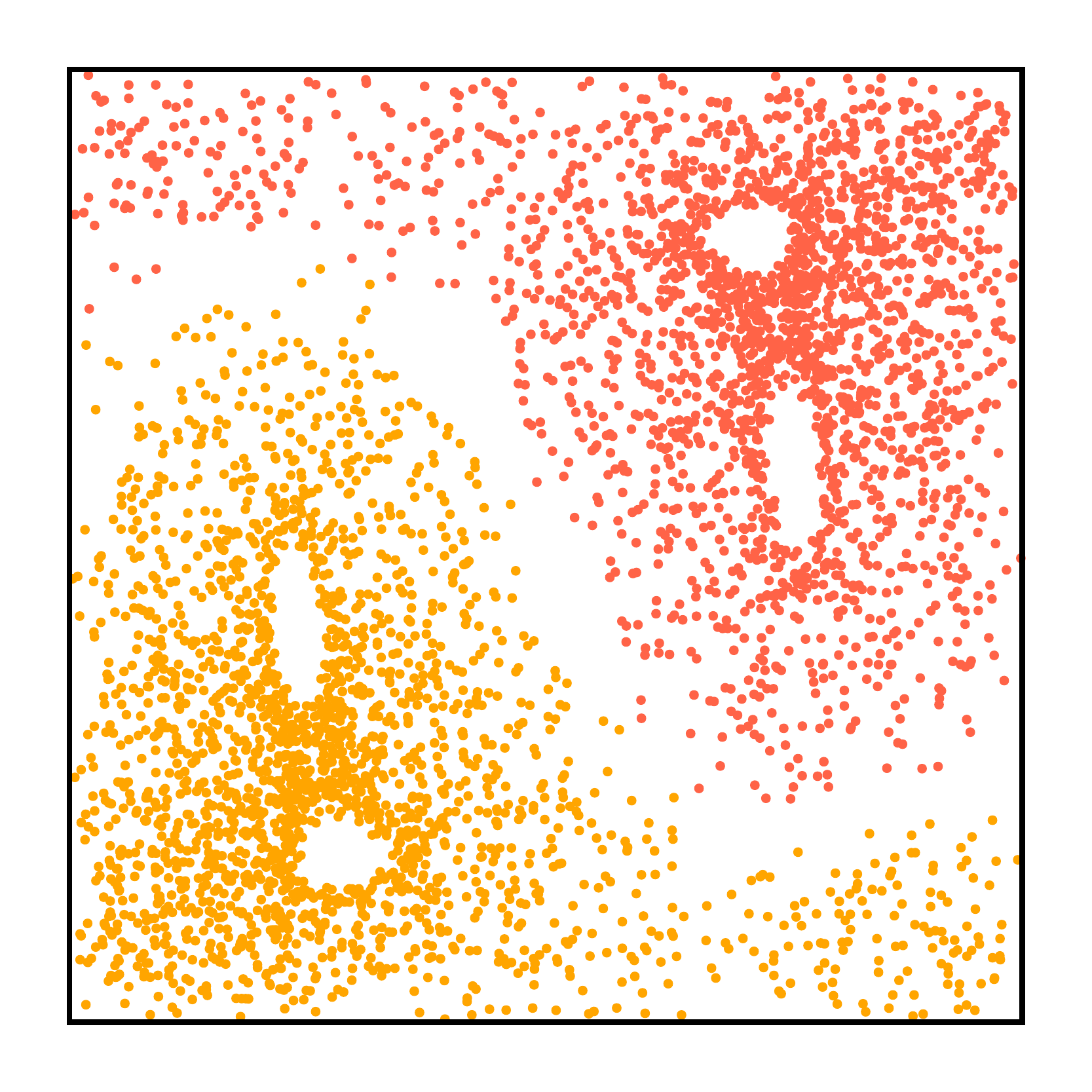}    \\
    \hline
    $y = -0.25$ &  $y = 0$ &  $y = +0.25$\\[-0.25em]                                         
    \includegraphics[width=0.3\textwidth]{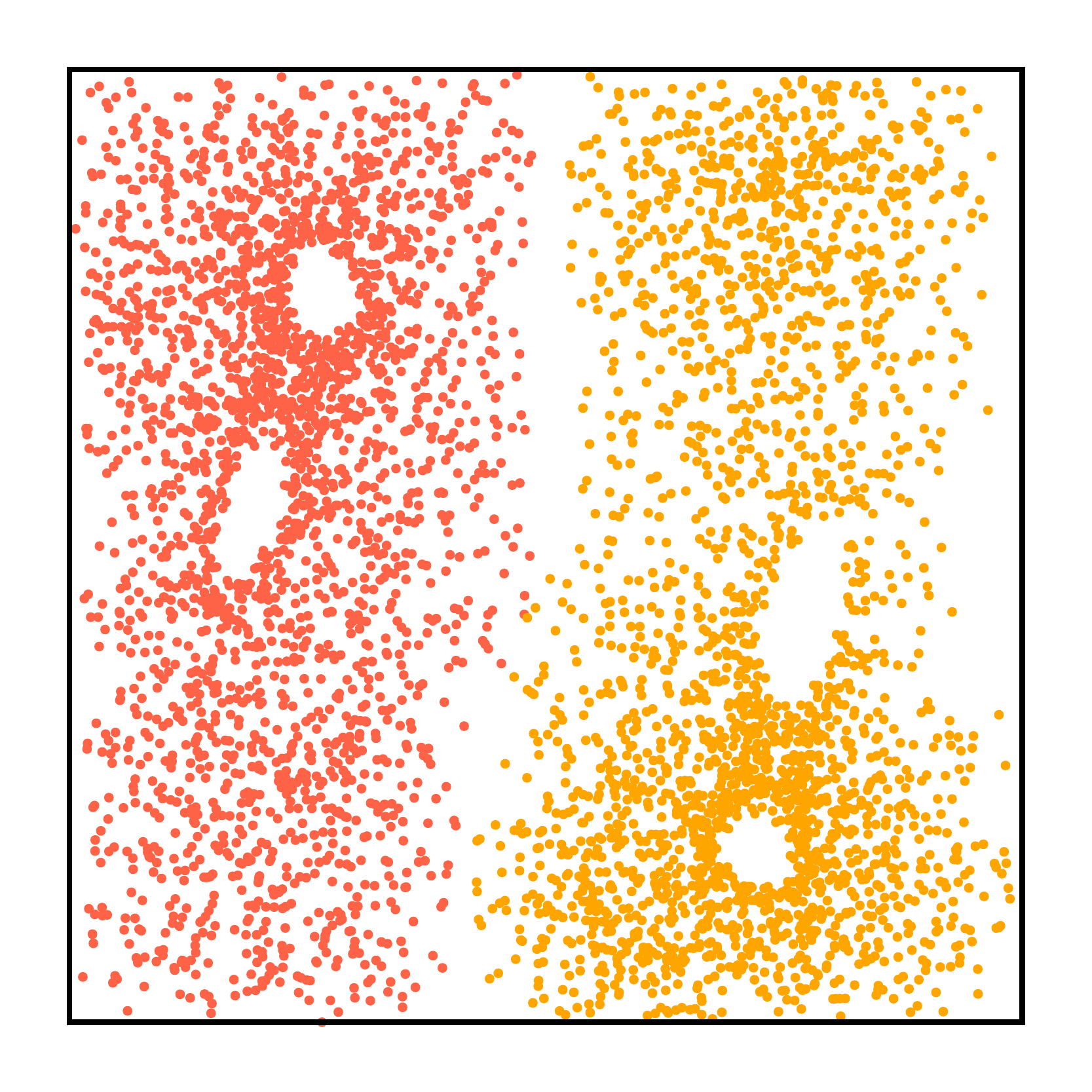} & \includegraphics[width=0.3\textwidth]{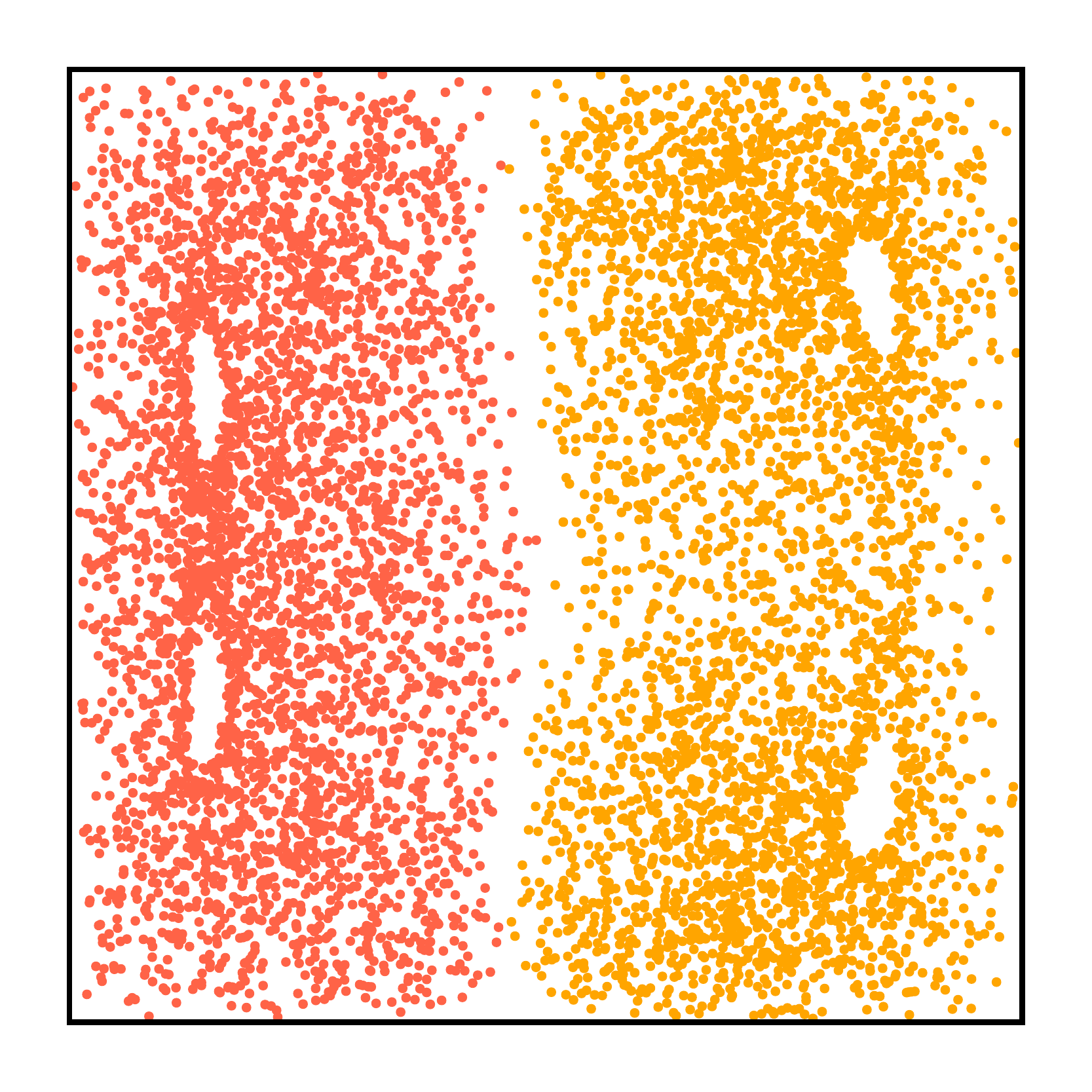} & \includegraphics[width=0.3\textwidth]{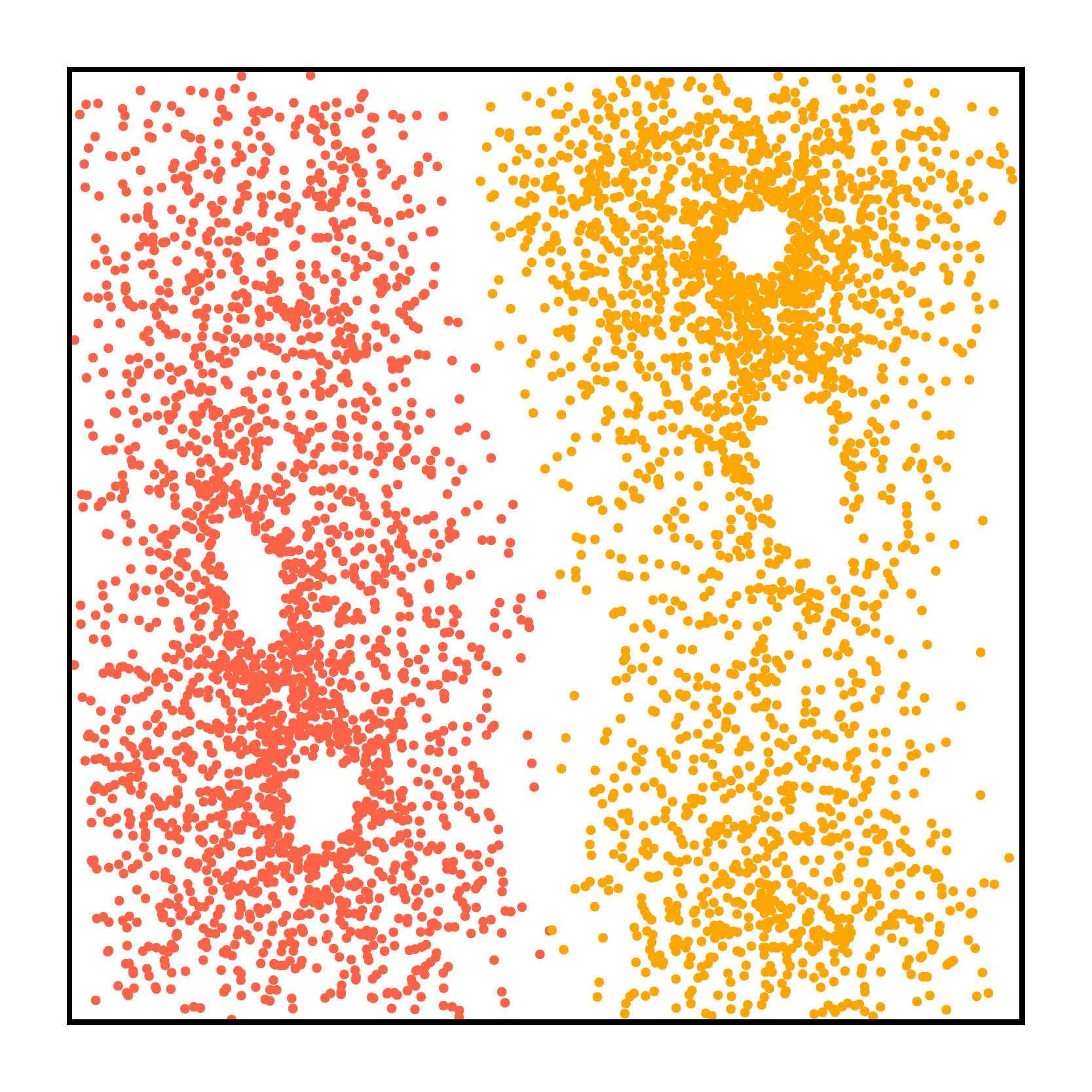} \\
    \hline
    $z = -0.25$ &  $z = 0$ &  $z = -0.25$\\[-0.25em]  
    \includegraphics[width=0.3\textwidth]{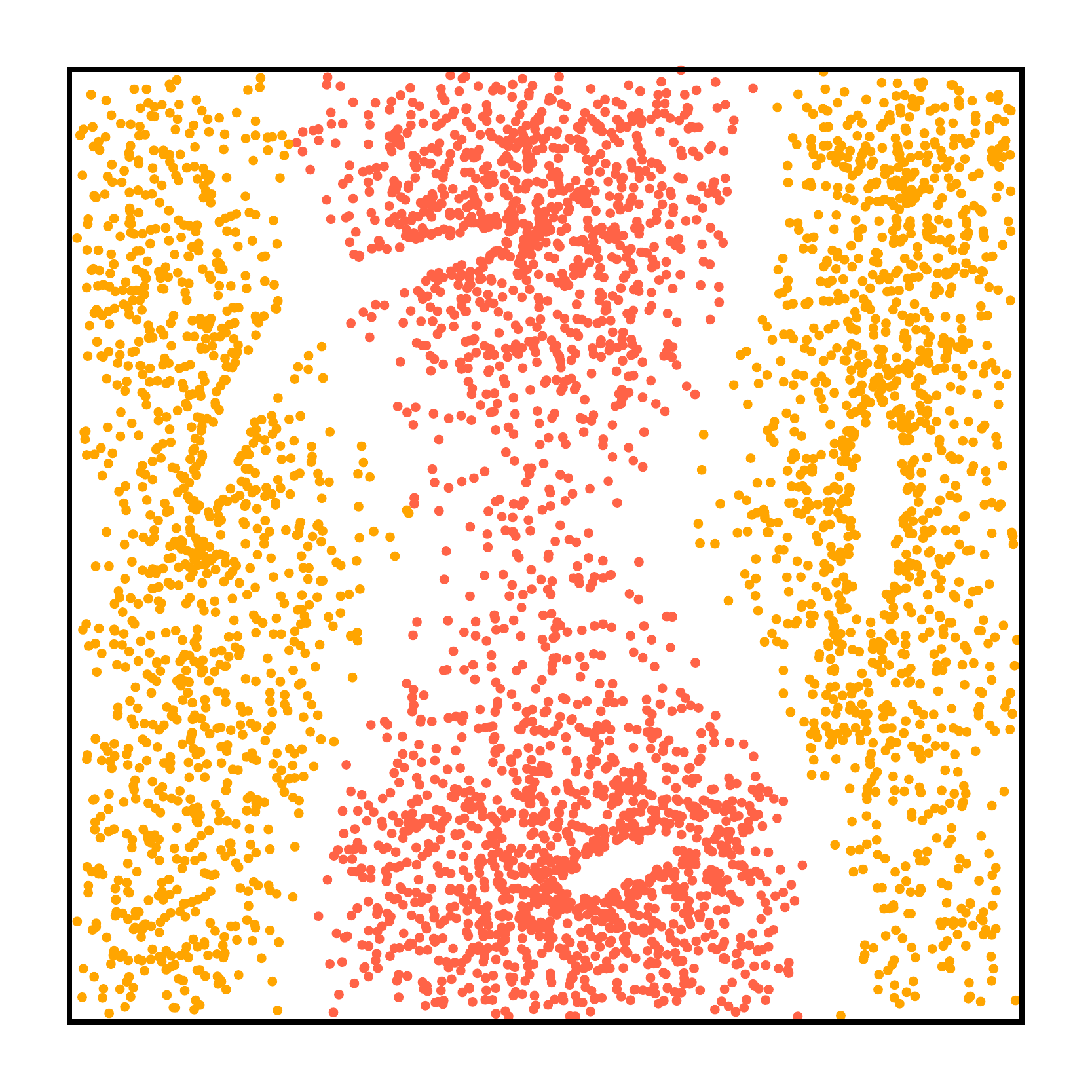} &
    \includegraphics[width=0.3\textwidth]{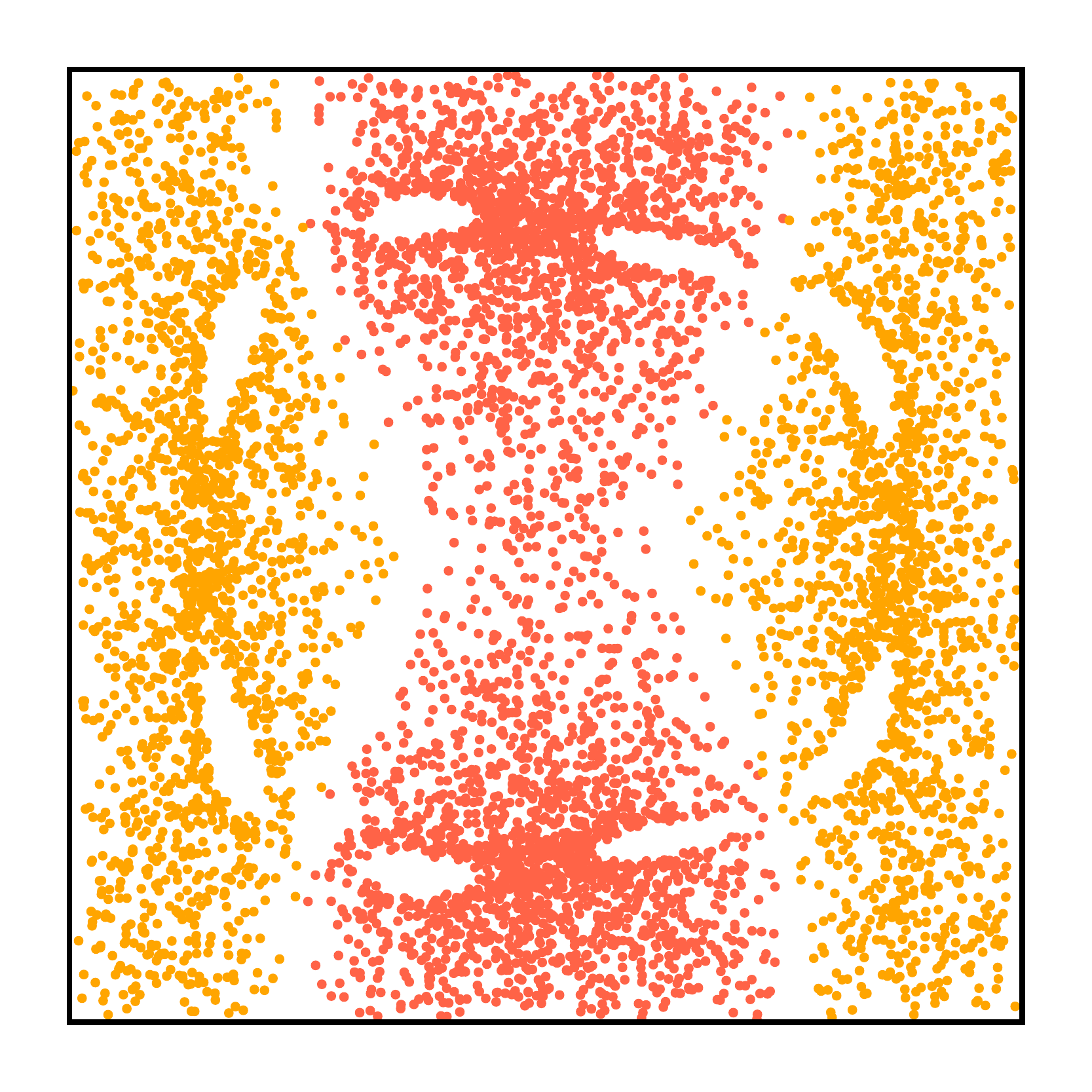} &
    \includegraphics[width=0.3\textwidth]{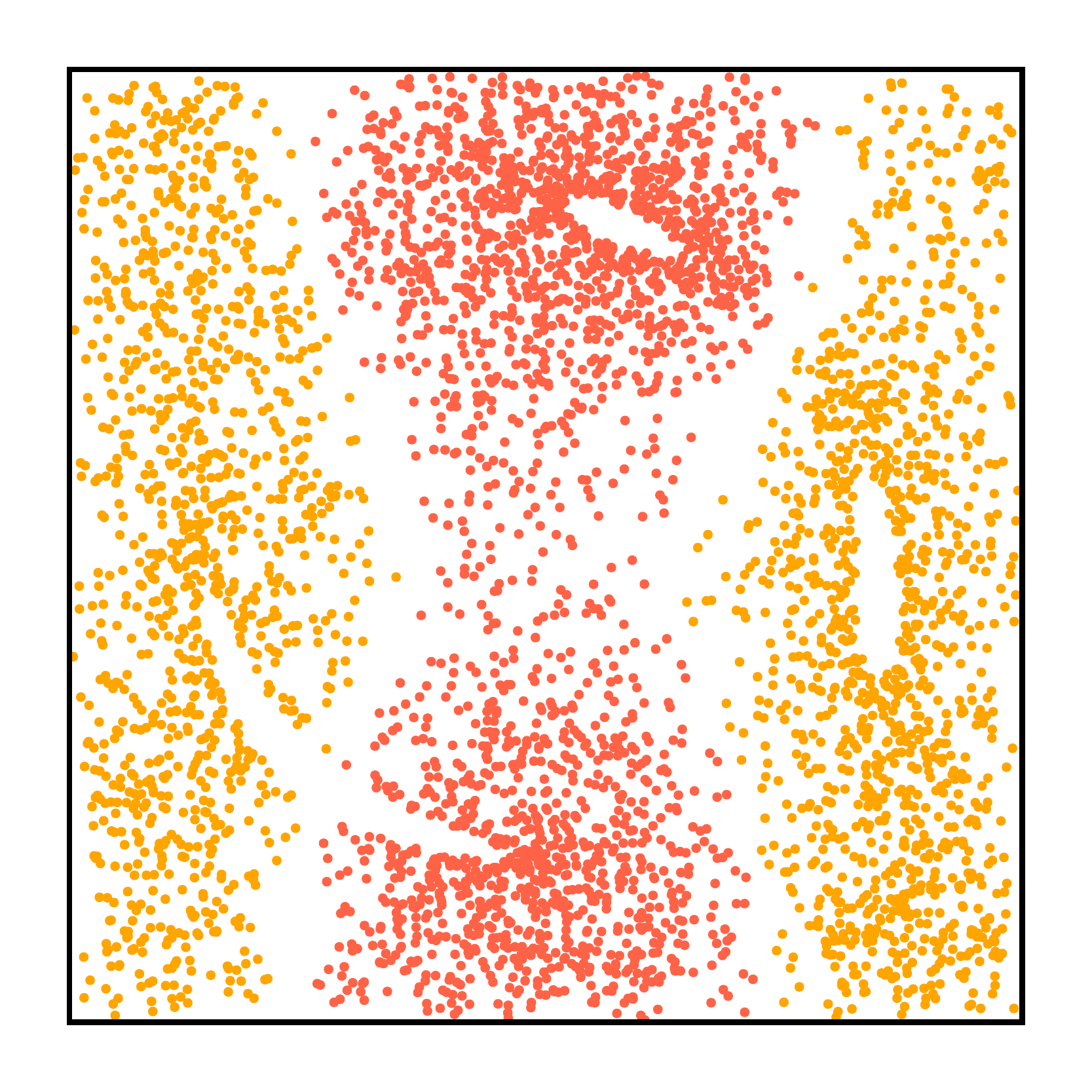} \\
  \end{tabular}
  \caption{\label{fig:sos}  Poincaré sections for one single starting point $(0.15,0.15,0.15)$ in the case $\alpha = 0.45$. The darkest color corresponds to outgoing points and the lightest color to incoming points.}
\end{figure}

Before that, and to conclude this paragraph, we plot in \figurename\,\ref{fig:sos} a set of Poincaré sections (in three planes in each direction of space), still for the same starting point and in the case $\alpha = 0.45$, which seems to be the one where periodic trajectories are less prominent. Indeed, the plots clearly show that the entire space is almost visited by the particle, suggesting that mixing is efficient.

\subsubsection{Lyapunov exponents}

Lyapunov exponents are often used to quantify deterministic chaos. 
For a given trajectory ${\bf x}(t)$ of a chaotic dynamical system, we perturb the initial position ${\bf x}(0)={\bf x}_0$ at time $t=0$. The perturbed trajectory, denoted ${\bf y}(t)$ in the following, originates at ${\bf y}(0) = {\bf y}_0$. 
As time $t$ goes on, ${\bf y}(t)$ moves further away from ${\bf x}(t)$ with a distance $ \delta (t) \equiv \|{\bf x} - {\bf y}\|  \simeq \delta_0 \exp(\lambda t)$, where   $\delta_0 = \|{\bf x}_0 - {\bf y}_0\|$ and the real number $\lambda$, called Lyapunov exponent, is positive in the case of a chaotic system.
Looking for perturbations along the three axes of the frame $(O,x,y,z)$, leads to three Lyapunov exponents $\lambda_1, \lambda_2$ and $\lambda_3$.
For the dynamical system \eqref{dxdt}, we consider a perturbation matrix $Y$ according to the three cardinal axes, which satisfies the linearized dynamical system $\dot {Y}(t) = \nabla {\bf v} ({\bf x}(t))\,  Y(t)$ where ${\bf v}$ is given by \eqref{defalpha}. The Lyapunov exponents can be computed from a {\it QR}-decomposition of the perturbation matrix $Y(t) = Q(t)R(t)$ with $Q$ being an orthogonal matrix and $R$ an upper triangular matrix with positive diagonal coefficients $R_{ii} >0$ (see \cite[Appendix A]{DatserisParlitz2022} and \cite{EckmannRuelle1985}): 
\begin{equation}
  \lambda_i = \lim_{t\to +\infty} \frac{1}{t} \ln R_{ii}(t), \quad \text{ for } i=1,2,3.
\end{equation}
For practical numerical computations of the Lyapunov exponents, we perform an iterated {\it QR}-decomposition of $Y$ at discrete times $t_k$ with a time step $\Delta t$. We have $Y(t_k) = Q_kR_k$ with $Y$ satisfying 
\begin{equation}
  \left\{
    \begin{array}{l}
      \dot Y((t) = \nabla {\bf v} ({\bf x}(t))\,  Y(t), \quad t\in [t_{k-1}, t_k]
      \\
      Y(t_{k-1}) = Q_{k-1} .
    \end{array}
  \right.
\end{equation}
For a final time $T = N \Delta t$, we obtain the   iterated {\it QR}-decomposition with $Y(T) = Q_N R_NR_{N-1}\dots R_1$ and we compute the Lyapunov exponents according to
\begin{equation}
  \lambda_i \simeq \frac{1}{N\Delta t} \sum_{j=1}^N \ln {(R_j)}_{ii}.
\end{equation}

The dynamical system \eqref{dxdt} will present a chaotic behavior if at least one Lyapunov exponent is positive. Since the velocity $\bf v$ in \eqref{dxdt} is divergence-free, the folllowing property holds:
\begin{equation}\label{lambdasum}
  \lambda_1+\lambda_2+\lambda_3 = 0.
\end{equation}
We emphasize that the values of the Lyapunov exponents depend on the initial position ${\bf x}(0)={\bf x}_0$ of the trajectory. It is known that autonomous dynamical systems always possess a zero
Lyapunov exponent provided that the trajectory of the flow does not
converge to a steady state. We refer to \cite{articleVToussaint} and
\cite{Pikovsky_Politi_2016} \S 2.5.6 where a sketch of a simple proof is
given. In a chaotic region (where the initial position ${\bf x}_0$ has been chosen), we then necessarily get from \eqref{lambdasum} 
\begin{equation}\label{eq:lambda_chaos}
  \lambda_1 >0,\  \lambda_2 =0,\ \lambda_3 = -\lambda_1 <0,
\end{equation}
while in a smooth or regular region, we expect to have
\begin{equation}
  \lambda_1  = \lambda_2 =\lambda_3 = 0.
\end{equation}

\begin{figure}
  \centering 
  \includegraphics[width=\columnwidth]{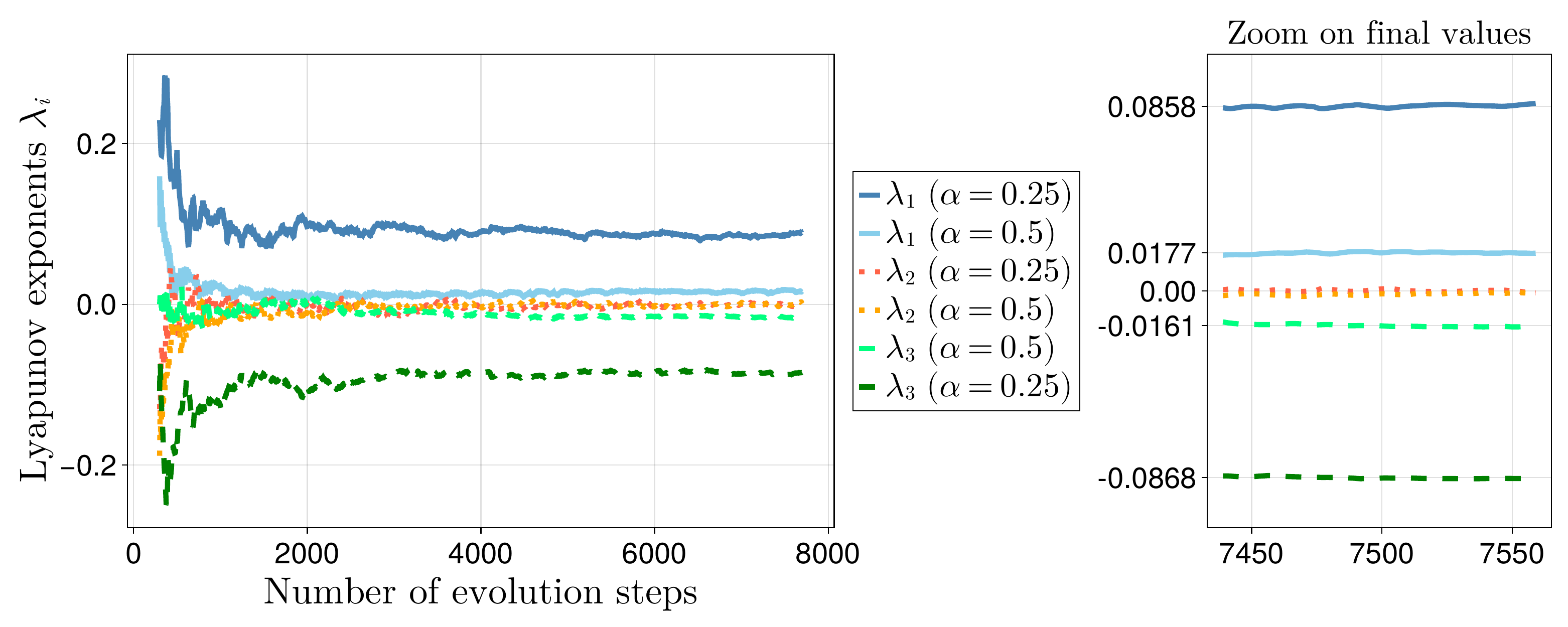}
  \caption{\label{fig:expolyapvstime}Evolution of Lyapunov exponents   versus the number of computation steps, for the two previous cases ($\alpha = 0.25$ and $0.5$, ${\bf x_0} = (0.15,0.15,0.15)$).}
\end{figure}

This method can easily be called up using the Julia package mentioned previously \cite{Datseris2018}. We have modified the default implementation to preserve the history of terms calculated during the process. For our starting point ${\bf x_0} = (0.15,0.15,0.15)$, and for the two values of $\alpha$ considered in \figurename\,\ref{fig:tori}, we plot the Lyapunov exponents $\lambda_i$ versus the number of evolution steps $N$ in \figurename\,\ref{fig:expolyapvstime}.    

The long-term shape of these curves agrees with Eq.\ (\ref{eq:lambda_chaos}), and the calculation converges fairly quickly, with a final value reached after nearly 7,000 steps. This confirms our choice of the ODE solver and associated time step. 
In both cases, we find a chaotic behavior with $\lambda_1 > 0$, but it is much more pronounced for $\alpha = 0.25$.

\begin{figure}
  \centering 
  \includegraphics[width=8.6cm]{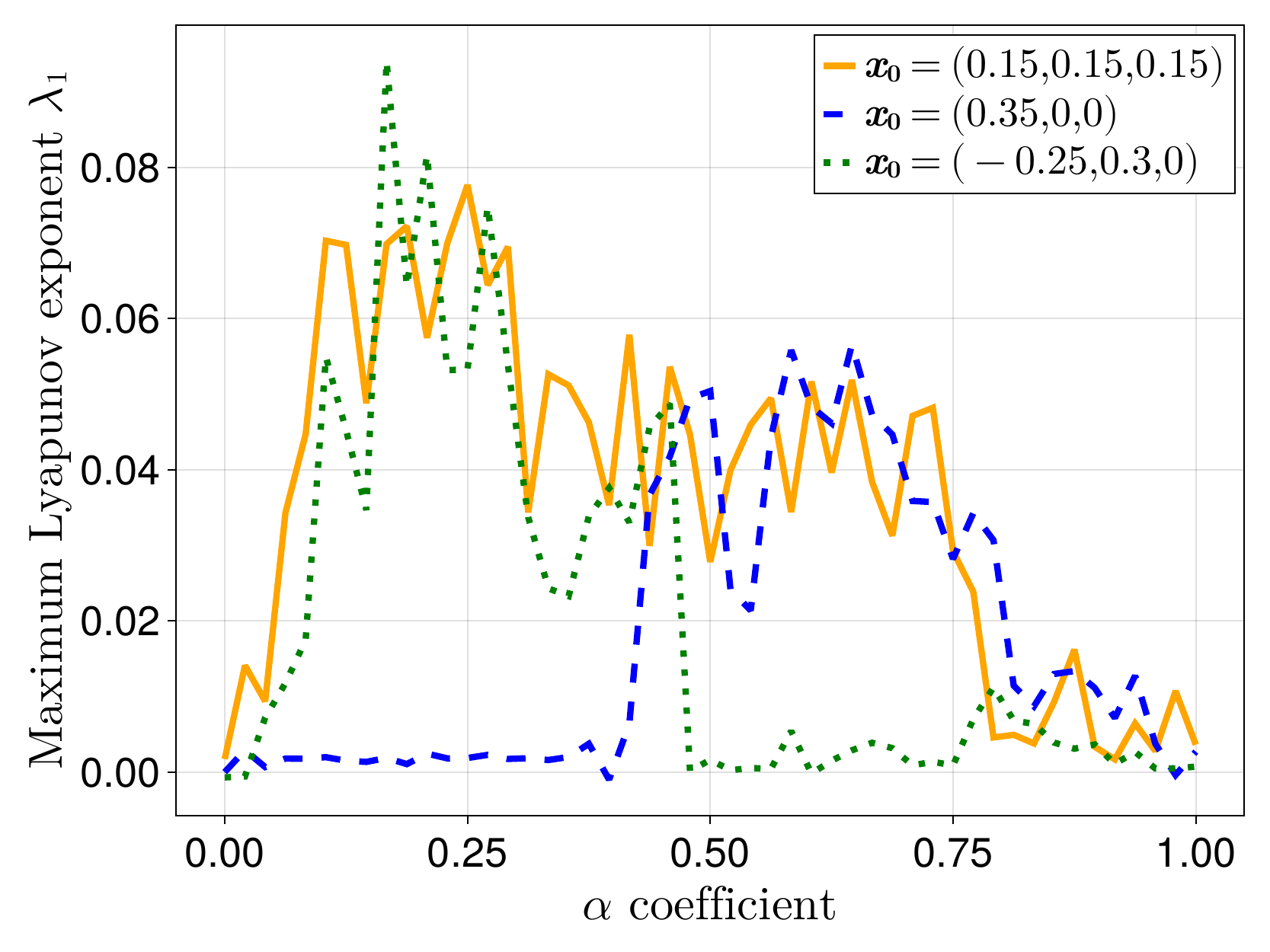}
  \caption{\label{fig:expolyaptori} Lyapunov exponents versus $\alpha$, for the trajectories shown in \figurename\,\ref{fig:tori}.
  }
\end{figure}

{The maximal Lyapunov exponent $\lambda_1$ also brings valuable information for starting points associated with the regular points observed in section \ref{subsec:Poinca}. It is shown in \figurename\,\ref{fig:expolyaptori} as a function of $\alpha$, for the points $x_0 = (0.35,0,0)$ (blue dashed line) and $x_0 = (-0.25,0.3,0)$ (green dotted line), corresponding respectively to the cases of one torus and two tori in  \figurename\,\ref{fig:tori}. 
In addition, the maximal Lyapunov exponent of point $x_0 = (0.15,0.15,0.15)$ has been plotted for comparison  in \figurename\,\ref{fig:expolyaptori} (orange solid line).
These three curves confirm the qualitative observations of \figurename\,\ref{fig:tori}.  Indeed, the blue dashed curve clearly shows a regular periodic trajectory for low values of $\alpha$, becoming chaotic as $\alpha$ increases. The green dotted curve follows the opposite evolution, and the switch between the two behaviors occurs around $\alpha = 0.45$.}

\medskip
To find a more global criterion that reflects the chaotic behavior of the entire flow, we have computed expansion entropies. These are shown in the next section.

\subsection{Expansion entropy}

The behavior of Lyapunov exponents is directly linked to the choice of the initial position of the trajectory. A more satisfactory and comprehensive way to characterize chaos without being too dependent on initial conditions is to consider numerical quantities such as topological entropy or expansion entropy. The topological entropy, with its standard definitions \cite{AdlerKonheimMcAndrew1965}, is however quite difficult to compute numerically. This is the reason why we prefer to deal with the expansion entropy whose definition is better suited to numerical evaluation. In particular, expansion entropy does not require to identify a compact invariant set, in contrast with topological entropy. We refer to \cite{HuntOtt2015} for a review and a discussion on various chaos indicators and also for the definition and properties of the expansion entropy.
An important result states that for autonomous systems on a compact manifold, the topological entropy coincides with the expansion entropy for a regular flow \cite{Kozlovski1998}.

\smallskip
The expansion entropy is defined as
\begin{equation}\label{eq:H0}
H_0(S) = \lim_{T\to+\infty} \frac{1}{T} \ln E_T(\varphi_T,S)
\end{equation}
with
\begin{equation}
E_T(\varphi_T,S) = \frac{1}{\text{vol}(S)} \int_{S_T} G(\nabla \varphi_T(\bf x))\,d{\bf x}
\end{equation}
where $\varphi_t$ is the flow associated with dynamical system \eqref{dxdt} i.e. ${\bf x}(t) = \varphi_t({\bf x}_0)$, $S$ is a subset of the fluid tank $\Omega_f$ and $S_T$ is the set of ${\bf x}$ such that $\varphi_t({\bf x}) \in S$ for all $t\in [0,T]$ (the trajectories that remain in $S$ up to time $T$). Finally, $G(A)$ denotes the product of the singular values of a matrix $A$ that are greater than 1 (if none of the singular values are $>1$ then we set $G(A)=1$).
We consider that chaos occurs in $S$ if $H_0(S) > 0$. The computations of the expansion entropy are performed in the whole fluid domain, that is with $S=S_T=\Omega_f$.

\smallskip 

There is also a link between the Lyapunov exponents and $G(\nabla \varphi_T)$. Indeed, due to the Oseledec's multiplicative ergodic theorem (see \cite{EckmannRuelle1985}) it can be shown \cite{KatokHasselblatt1995} that the exponential growth rate of $G(\nabla \varphi_T(\x))$ as $T\to+\infty$ is the sum of the positive Lyapunov exponents of the trajectory starting from $\x$ at time $t=0$. In a chaotic region, thanks to \eqref{eq:lambda_chaos}, we have
\begin{equation}
\lim_{T \to \infty} \frac{1}{T} \ln G(\nabla \varphi_T(\x)) = \lambda_1.
\end{equation}
As a consequence, the limit in \eqref{eq:H0} can be seen as a space average on $S$ of the sum of the positive Lyapunov exponents of trajectories that remain in $S$.

\smallskip
Numerically, the expansion entropy is computed as follows.
First, we randomly generate a uniform distribution of $N$ points $\{\x_1, \x_2, \cdots, \x_N\}$ in $S=\Omega_f$, then we compute
\begin{equation}\label{eq:ET}
\hat{E}_T(\varphi_T) =  \frac{1}{N} \sum_{i=1}^N G(\nabla \varphi_T(\x_i))
\end{equation}
which is an estimate of $E_T(\varphi_T,\Omega_f)$ given by \eqref{eq:ET}. Then, we compute 
the approximate expansion entropy 
\begin{equation}
\hat{H}_0 = \frac{1}{T} \ln \hat{E}_T(\varphi_T)
\end{equation}
which is an approximation of $H_0(\Omega_f)$ for large $N$ and $T$. In practice, we choose a number $q$ of different batches of $N$ points and we take the average of the approximate expansion entropy computed with each batch. We use the Julia package {\tt ChaosTools.jl} with the function {\tt expansionentropy} which computes $\hat{H}_0$ from $q$ batches.

\begin{figure}
  \centering
  \includegraphics[width=8.6cm]{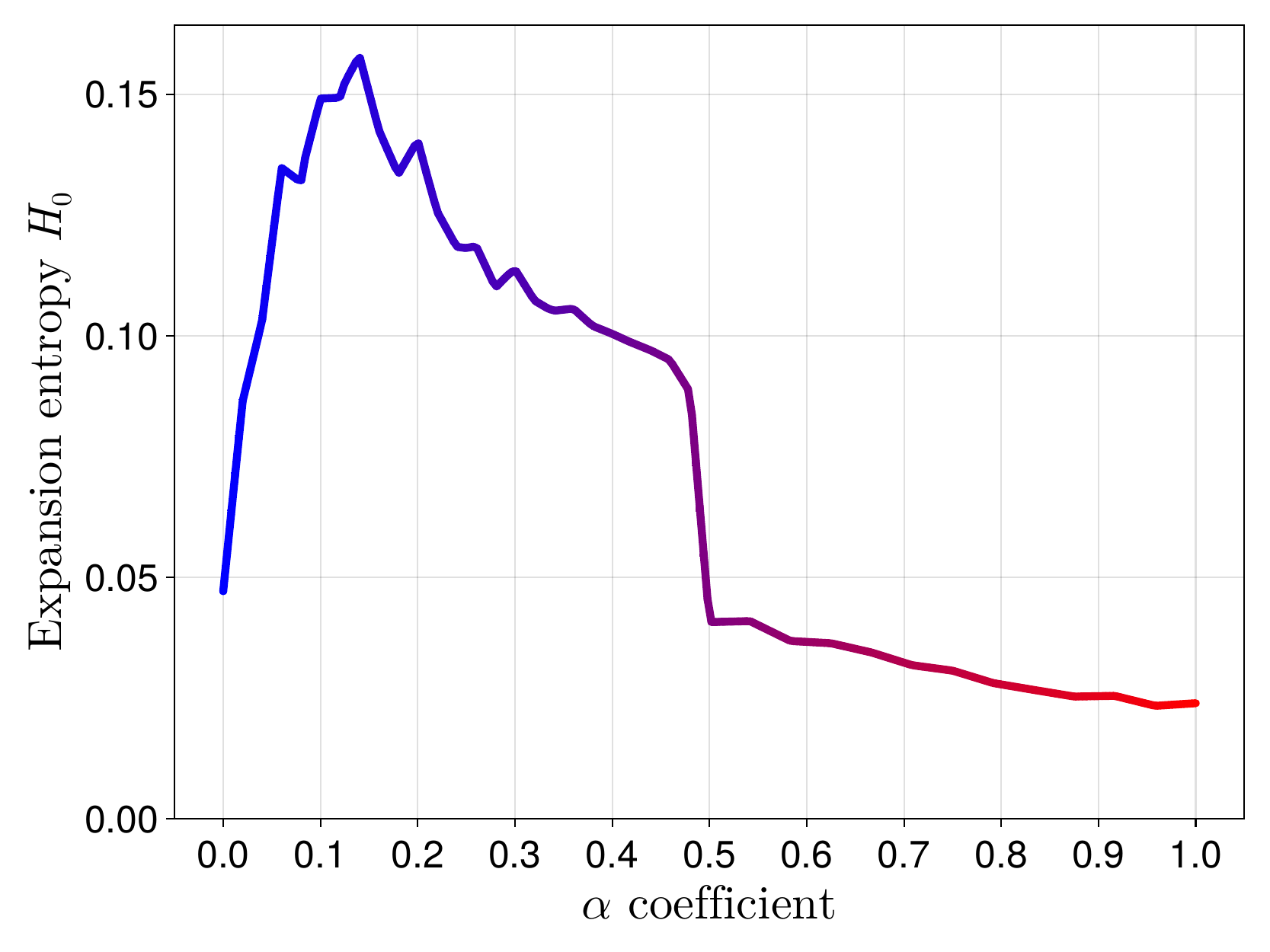}
  \caption{\label{fig:entropy} Expansion entropy of the entire flow  versus $\alpha$.}
\end{figure}

\smallskip
In \figurename\,\ref{fig:entropy}, the computed expansion entropy $\hat{H}_0$ is depicted with respect to the parameter $\alpha$. We chose $N=1000$ points uniformly distributed in $\Omega_f$ and $q=10$ batches.  
We see that the expansion entropy is always positive and higher for values of $\alpha$ less than 0.5. A peak with a maximum of $\hat{H}_0$ appears around  $\alpha=0.14$. 
The chaotic behavior of the system is therefore more pronounced around this value of $\alpha$. 
We also point out that for some initial conditions, the Lyapunov exponents behave like the expansion entropy. For instance, this is the case for the initial condition $\x_0=(-0.25, 0.3, 0)$ as can be seen with the green dotted curve in \figurename\ref{fig:expolyaptori}.

\subsection{Chaotic mixing analysis}\label{sec:mixing}

The  Poincar\'e sections, Lyapunov exponents and expansion entropies studied in the previous sections show that the dynamics of many fluid points is chaotic. The system is therefore expected to mix efficiently. To study the efficiency of mixing in a practical manner, we have computed contamination rates, homogeneities and mixing times. These quantities are widely used in chemical engineering or biology, and are presented in the following lines.

First, we simulate the dispersion of clouds of $N_p = 5000$ particles initially located within a small cube (width $\frac{a}{100}$) near the center of the cube,  by solving Eq.\ (\ref{dxdt}). The final time of the simulation of the cloud of particles is taken to be of a few convective times
\begin{equation}
  \label{eq:tc}
  t_c = \frac{a}{U}  
\end{equation}
where $U \simeq 0.1$ is an order of magnitude of the fluid velocity ${\bf v}$ in the MHD computations. This corresponds to $t_c= 10$ non-dimensional units (as a reminder, $a=1$ in our case study).
Note that, as explained in the previous section, the contribution of each magnet  has been parameterized by means of the parameter $\alpha$ (Eq.\ (\ref{defalpha})):
$\alpha=1$ corresponds to stirring by the side magnet alone (magnet  \#1, blue colors in the following figures), whereas $\alpha=0$ corresponds to the central magnet alone (magnet \#2, red). 

\begin{figure}[b!] 
  \centering
  \centering\renewcommand{\arraystretch}{1.5}
  \begin{tabular}{c | c | c | c}
    $\alpha = 0$ &  $\alpha = 0.2$ &  $\alpha = 0.4$ & $\alpha = 0.5$\\[-0.25em]
    \includegraphics[width=0.22\textwidth]{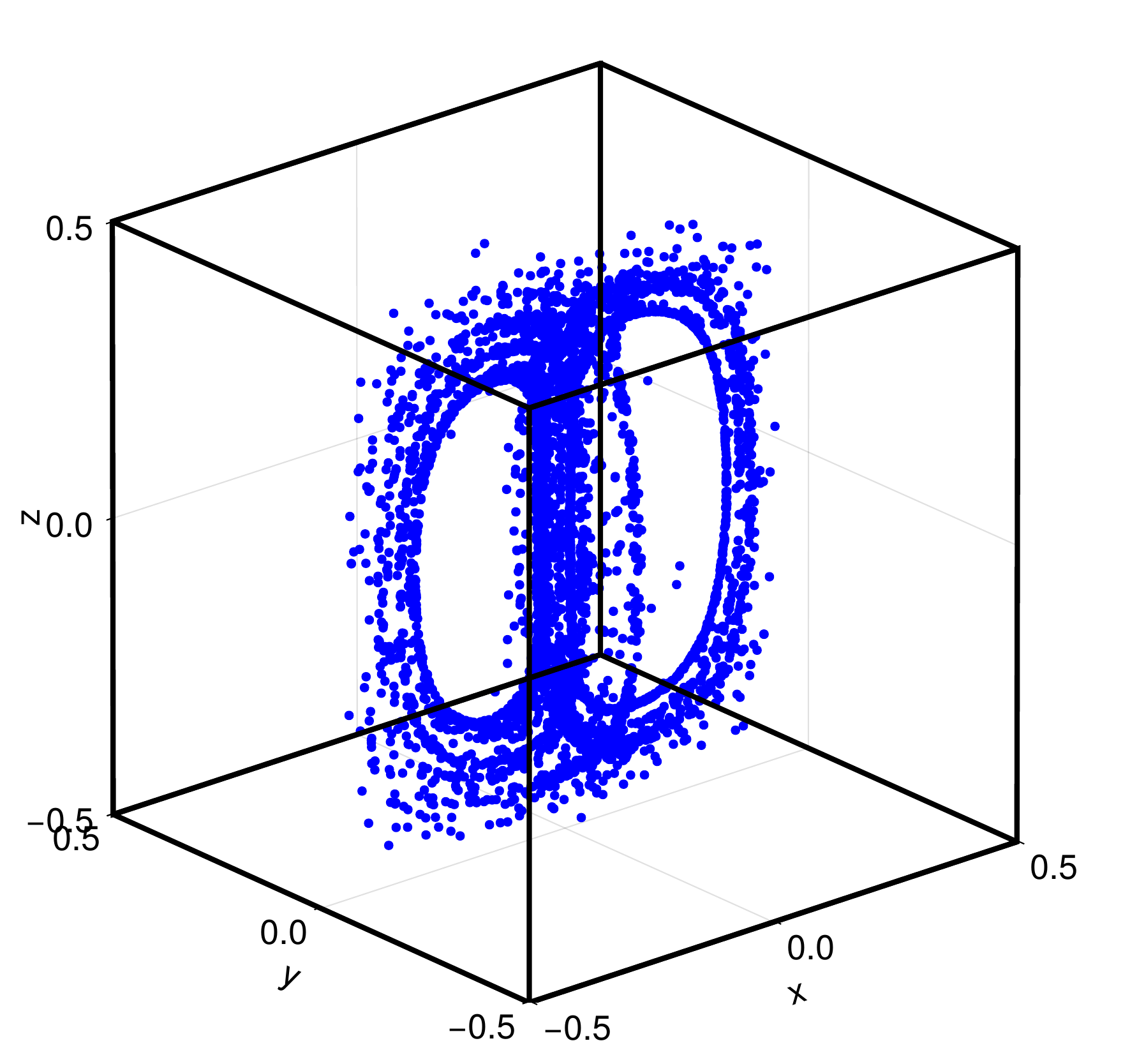} & \includegraphics[width=0.22\textwidth]{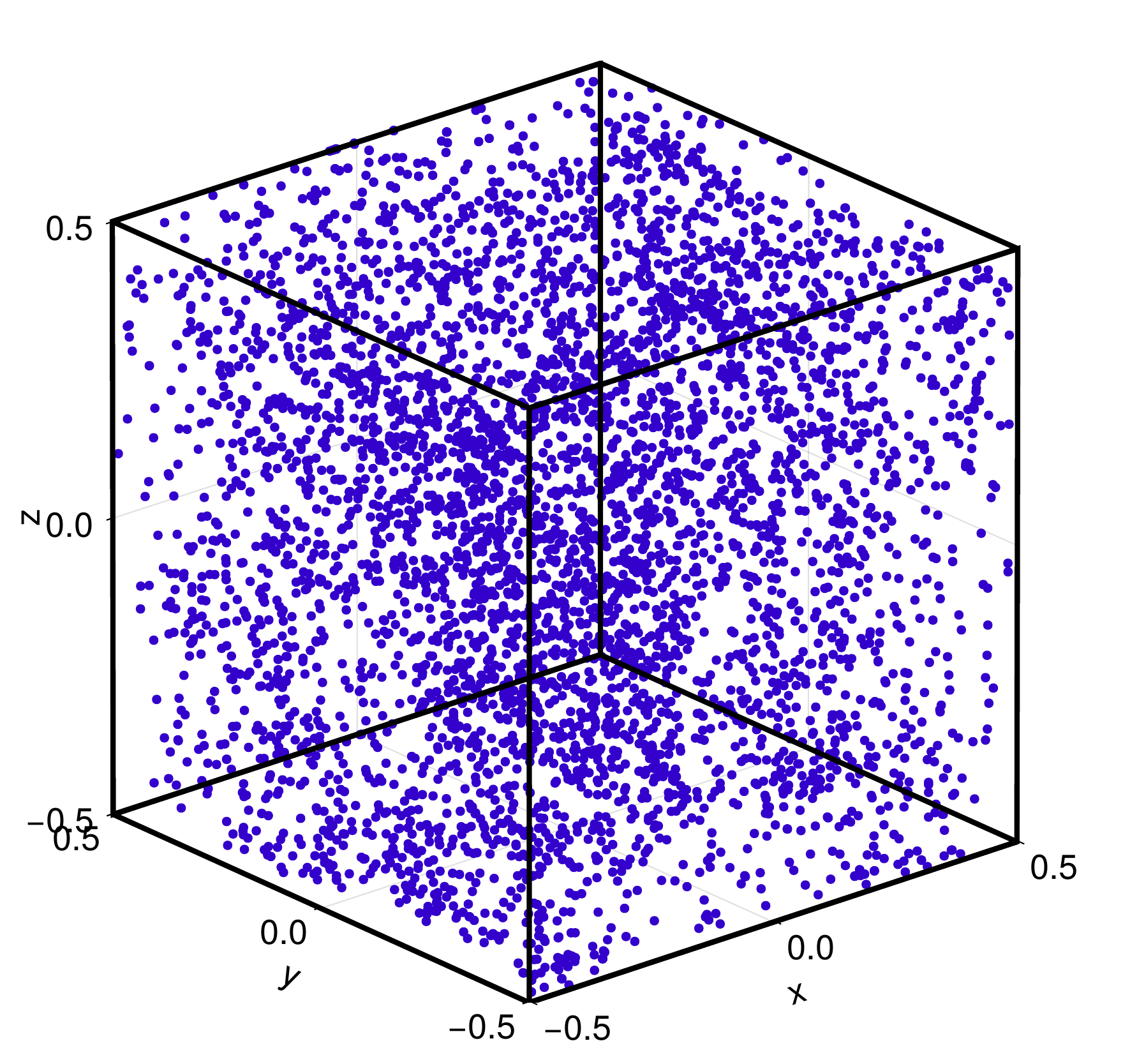} & \includegraphics[width=0.22\textwidth]{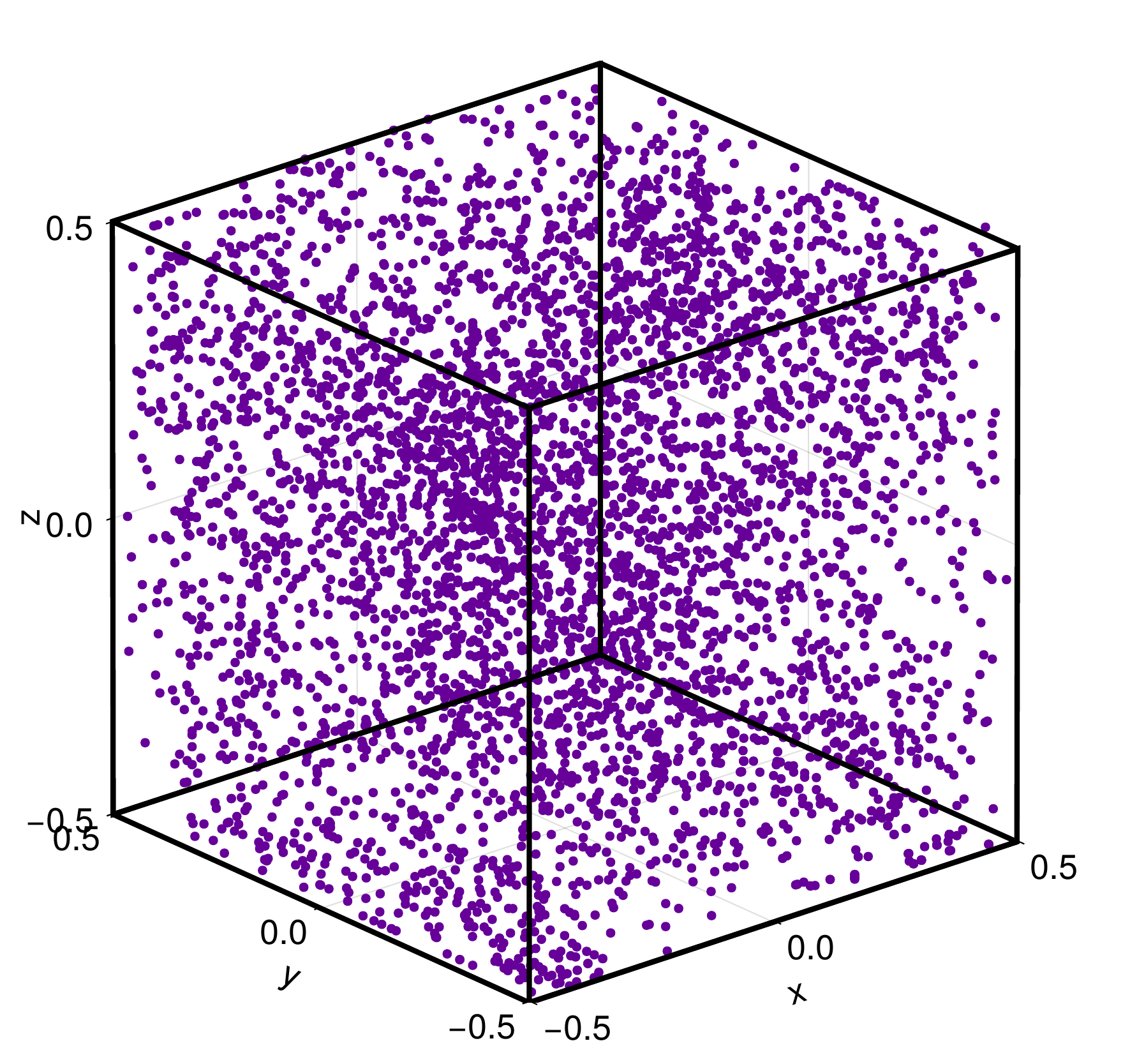} &
     \includegraphics[width=0.22\textwidth]{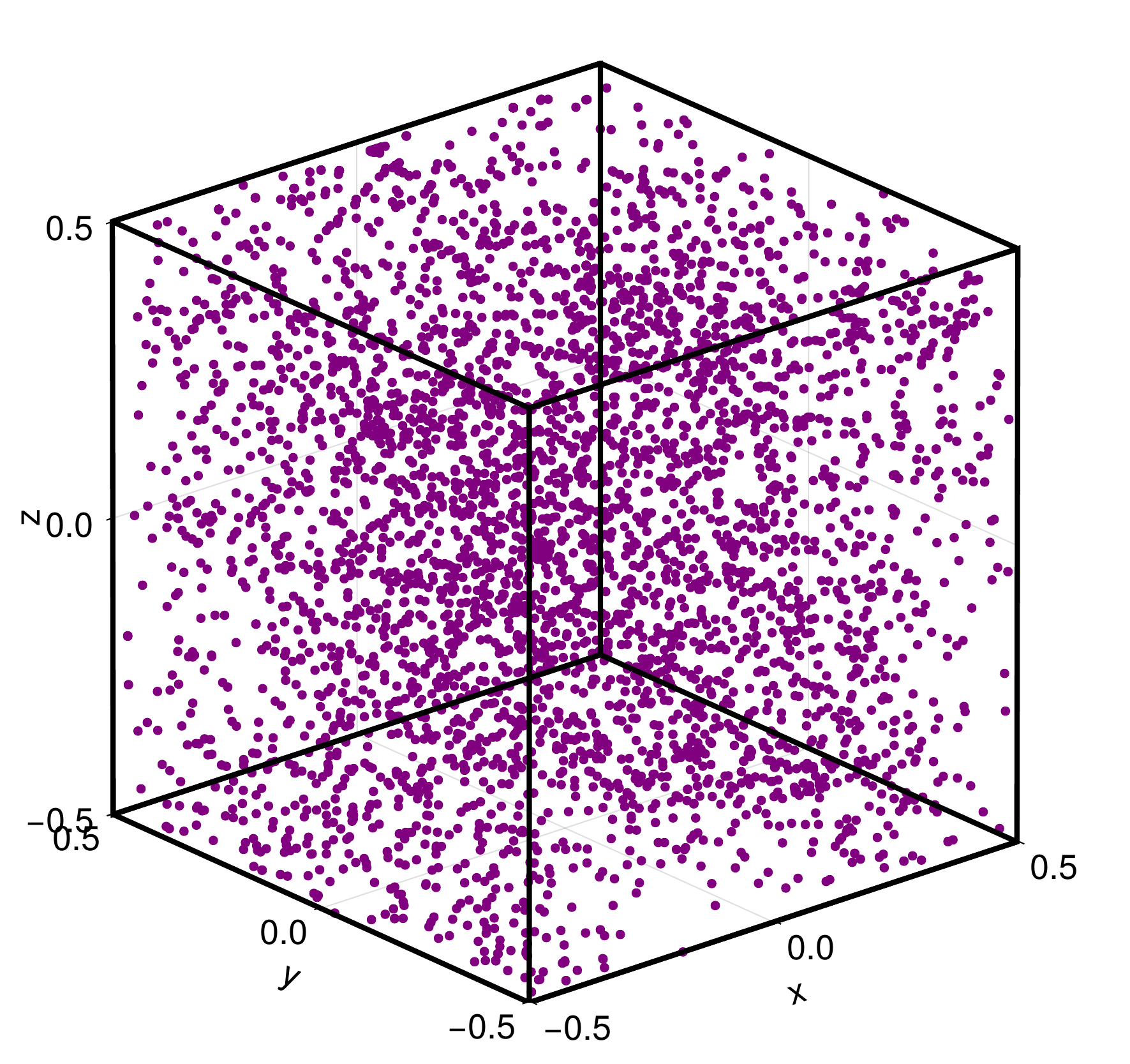} \\
    \hline
     $\alpha = 0.6$ &  $\alpha = 0.8$ & $\alpha =1$ & \\[-0.25em]
     \includegraphics[width=0.22\textwidth]{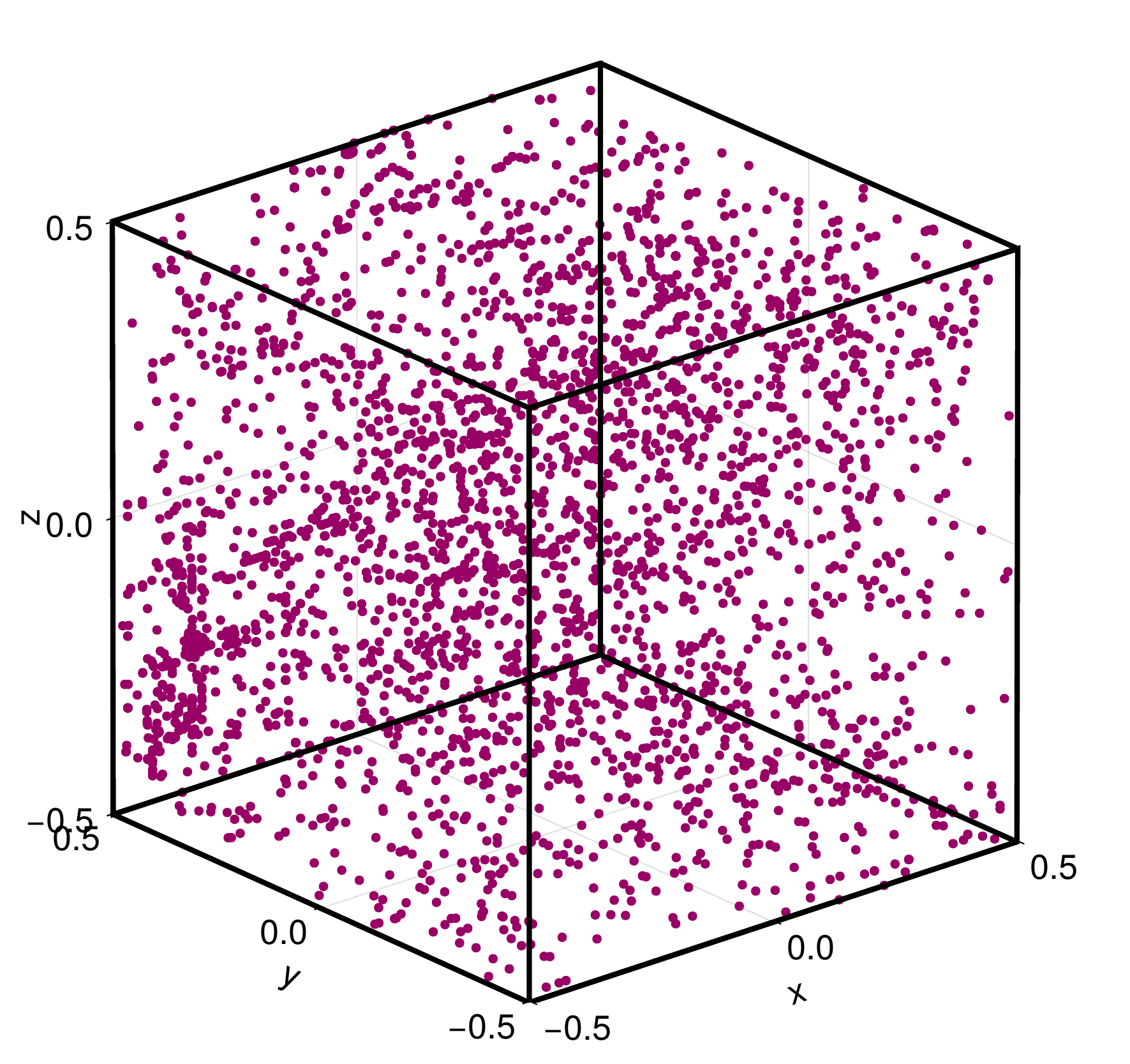} &
     \includegraphics[width=0.22\textwidth]{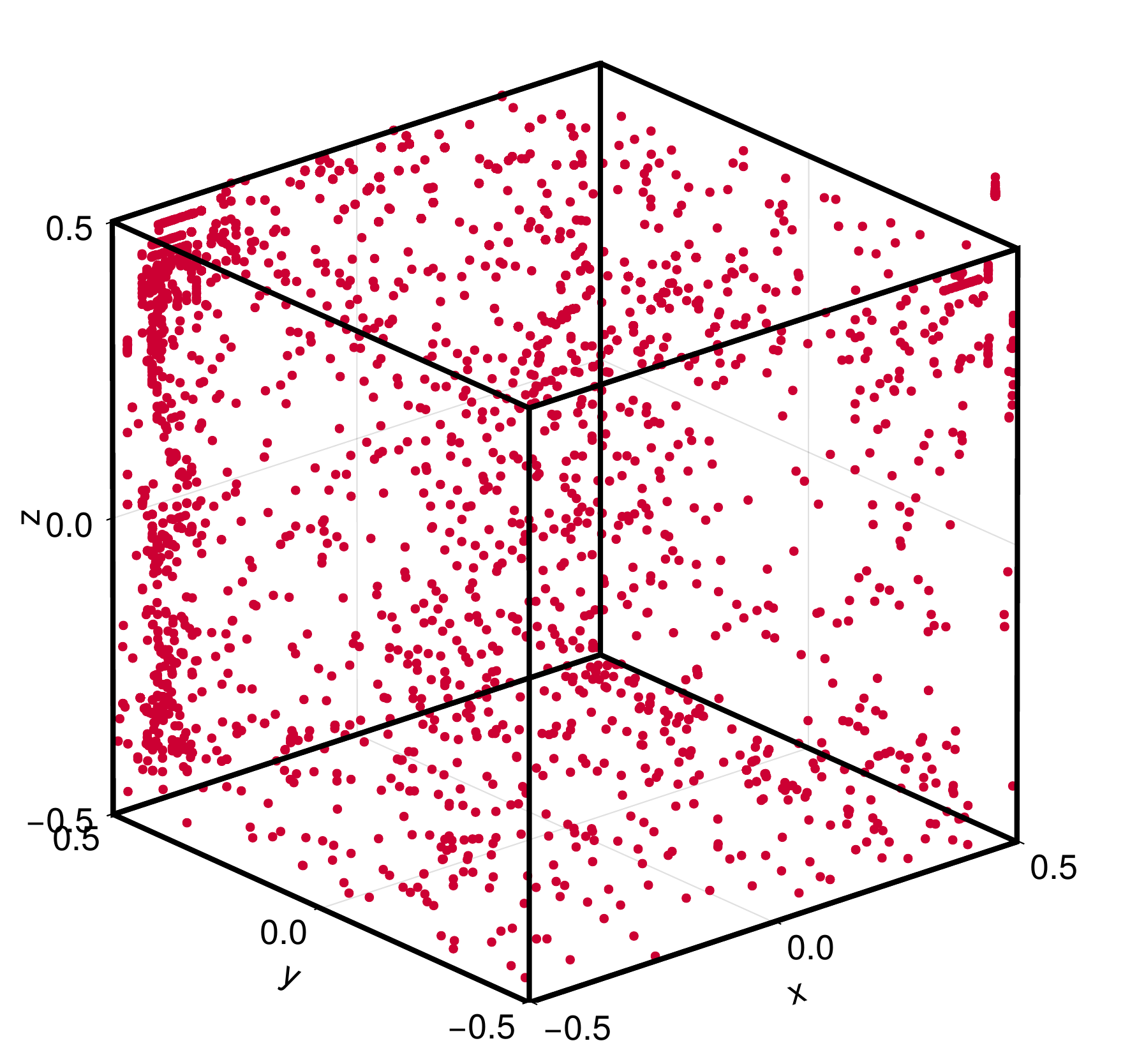} & \includegraphics[width=0.22\textwidth]{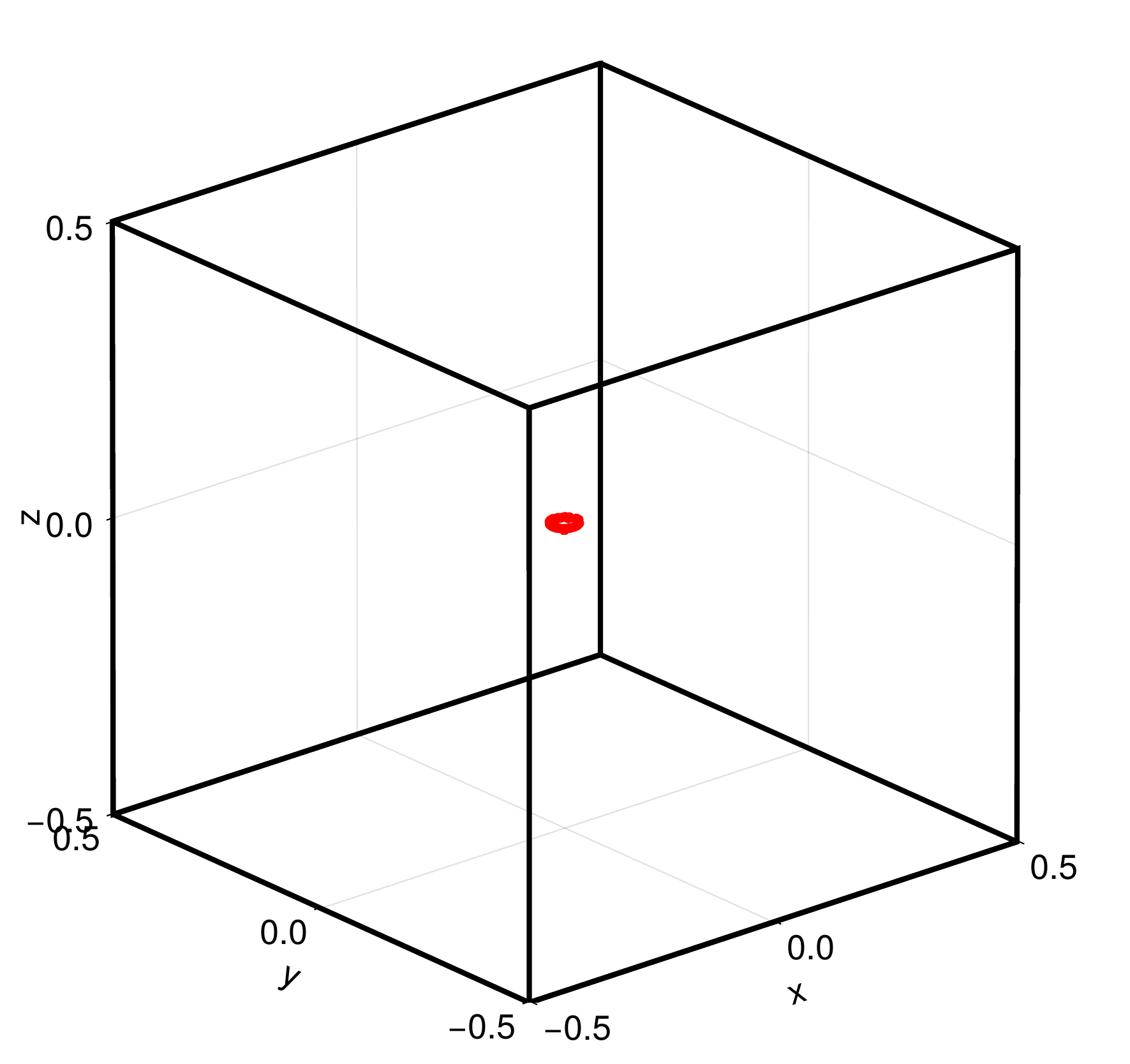} &
   \end{tabular}
  \caption{\label{fig:clouds}Clouds of 5000 particles after 250 convective times, for various values of $\alpha$. Particles were initially released in a cube with side ${1}/{100}$ located at the center of the domain. The poloidal flow $\mathbf v_2$ ($\alpha=0$) stretches the cloud but keeps it in a quasi-planar domain. The toroidal flow alone $\mathbf v_1$ ($\alpha=1$) has very little effect on the tiny initial cloud.}
\end{figure} 
 
\figurename\,\ref{fig:clouds} shows theses clouds at $t=250\, t_c$, for various values of $\alpha$. We observe that the shape of the cloud highly depends on $\alpha$ when $\alpha$ approaches 0 or 1, and mixing appears very poor there. Differences between transport by the approximately poloidal flow ($\mathbf v_2$ alone, $\alpha=0$) and the toroidal flow  ($\mathbf v_1$ alone, $\alpha=1$) are clearly visible. The poloidal flow alone highly deforms the cloud of particles within a quasi-planar domain. In contrast, the sole toroidal flow has little effect on the initial spot of particles, which rotates near the box center without being stretched.

\medskip
In a second time, we divide the flow domain into $N_c= 125,000 \,(= 50^3)$ cubic sub-domains ("cells"), then simulate the dispersion of a much bigger cloud of $N_p = 1,000,000$ particles initially located, as above, within one of these cells near the center of the cube. We then derive two quantities of great interest for practical applications: the contamination rate $C(t)$ and the final homogeneity $H_\infty$. The former is the proportion of cells which were visited by at least one particle during the time interval $[0,t]$. The latter is based on the final particle density $n_l$, that is the number of particles in  cell $l \in [1,N_c]$ at the final time of the simulation (see \cite{engler2004} for example). At that time, we calculate the standard deviation $\sigma$ of $n_l$, then the final homogeneity of the distribution defined as 
\begin{equation}
  \label{eq:Hdet} 
  H_\infty = 1 - \frac{\sigma}{\sigma_{\max}}
\end{equation}
where $\sigma_{max}$ is the maximum value of the standard deviation, corresponding to the case where all particles gather in a single cell: $\sigma_{max} = N_p \sqrt{N_c - 1}/{N_c} \simeq N_p/\sqrt{N_c}$.
Note that some authors choose slightly different definitions (e.g.  $H_\infty = 1 -  ({\sigma}/{\sigma_{\max}})^2$)  \cite{danckwerts1952}  \cite{berthiaux2006}. In any case, poor mixing corresponds to $H_\infty = 0$, whereas efficient mixing corresponds to $H_\infty \simeq 1$.

\begin{figure} 
  \centering
  \begin{tabular}{c c}
    \includegraphics[width=0.475\textwidth]{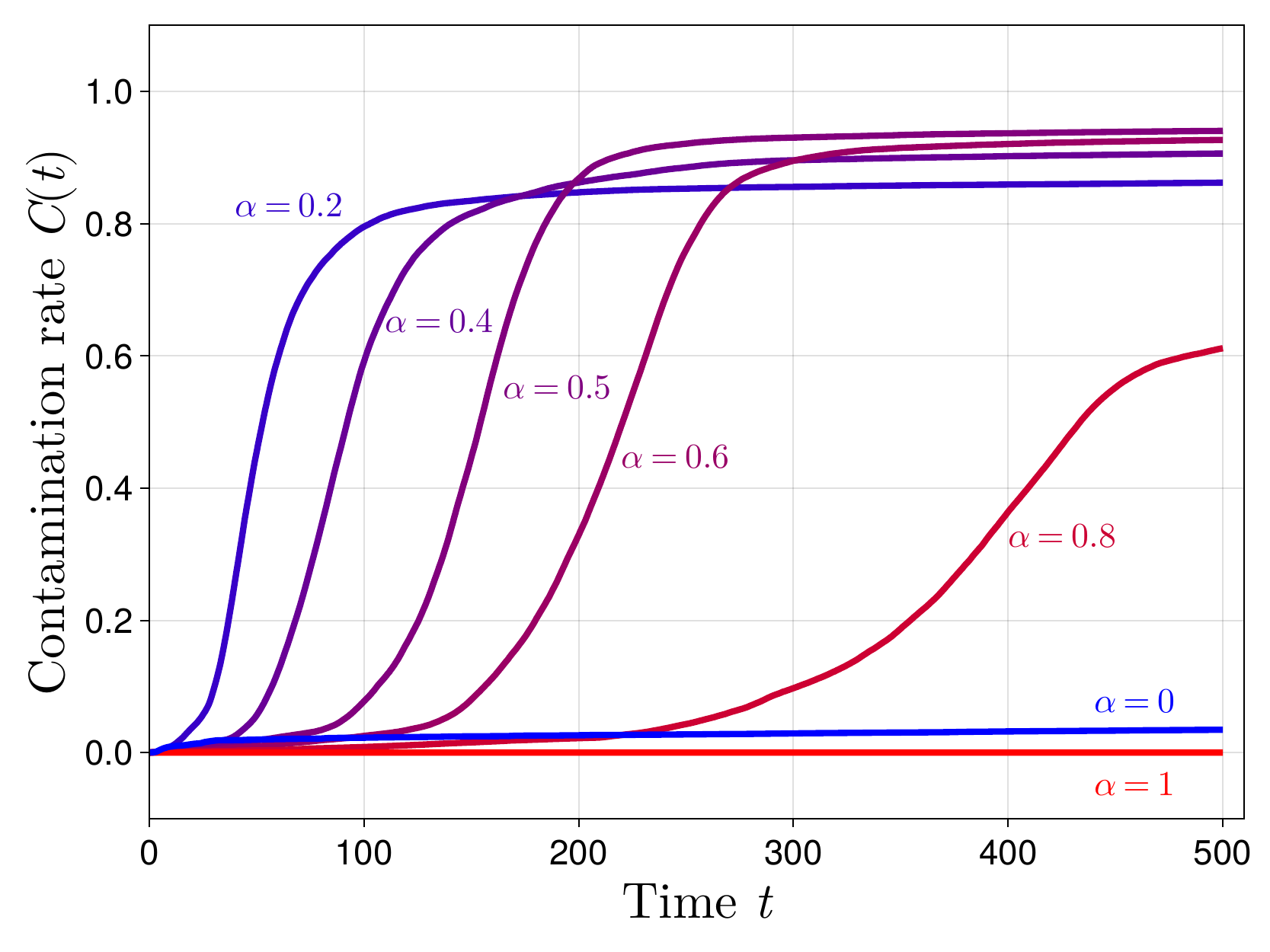} & 
    \includegraphics[width=0.475\textwidth]{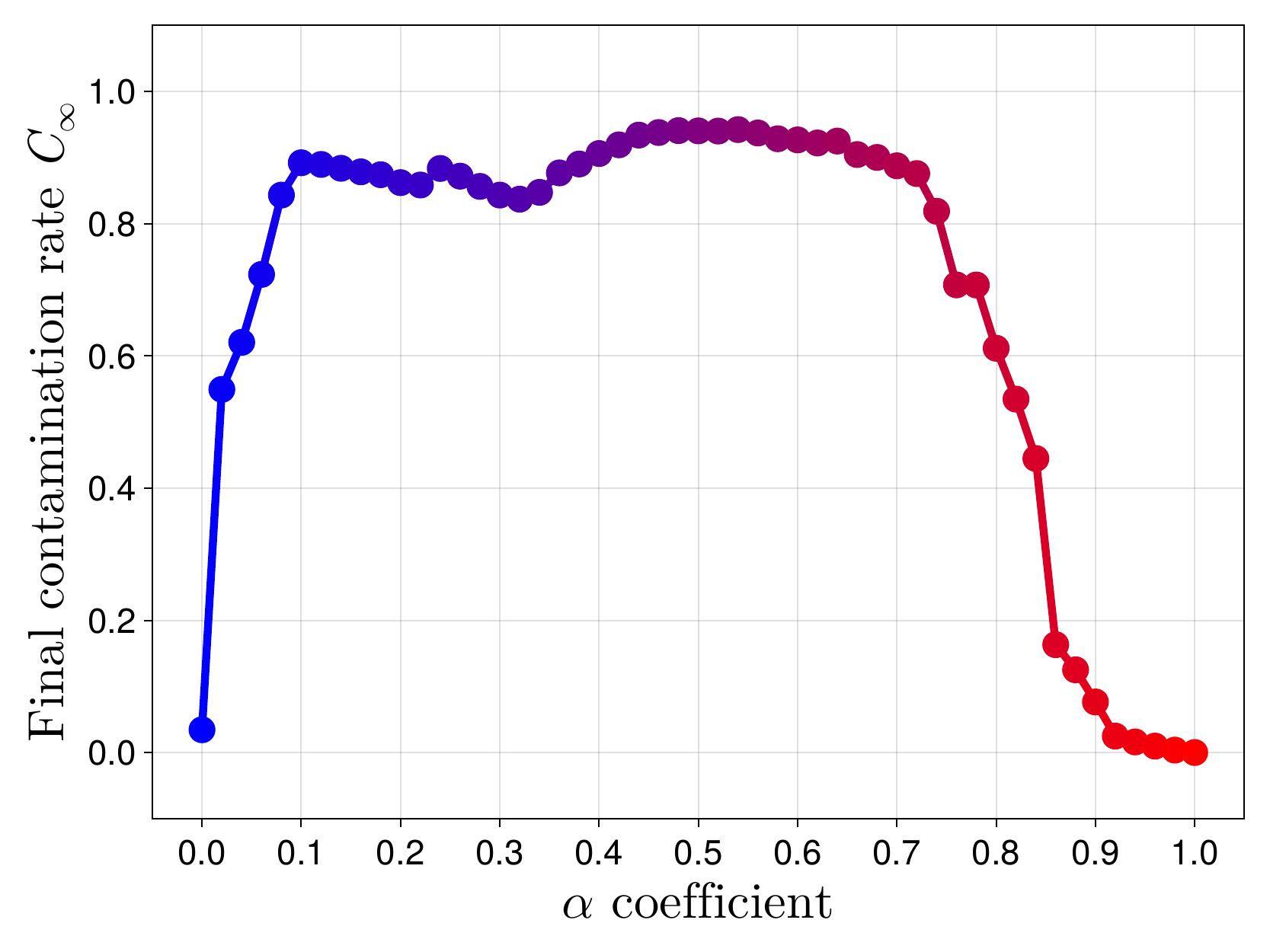}\\ 
    \multicolumn{2}{c}{\includegraphics[width=0.475\textwidth]{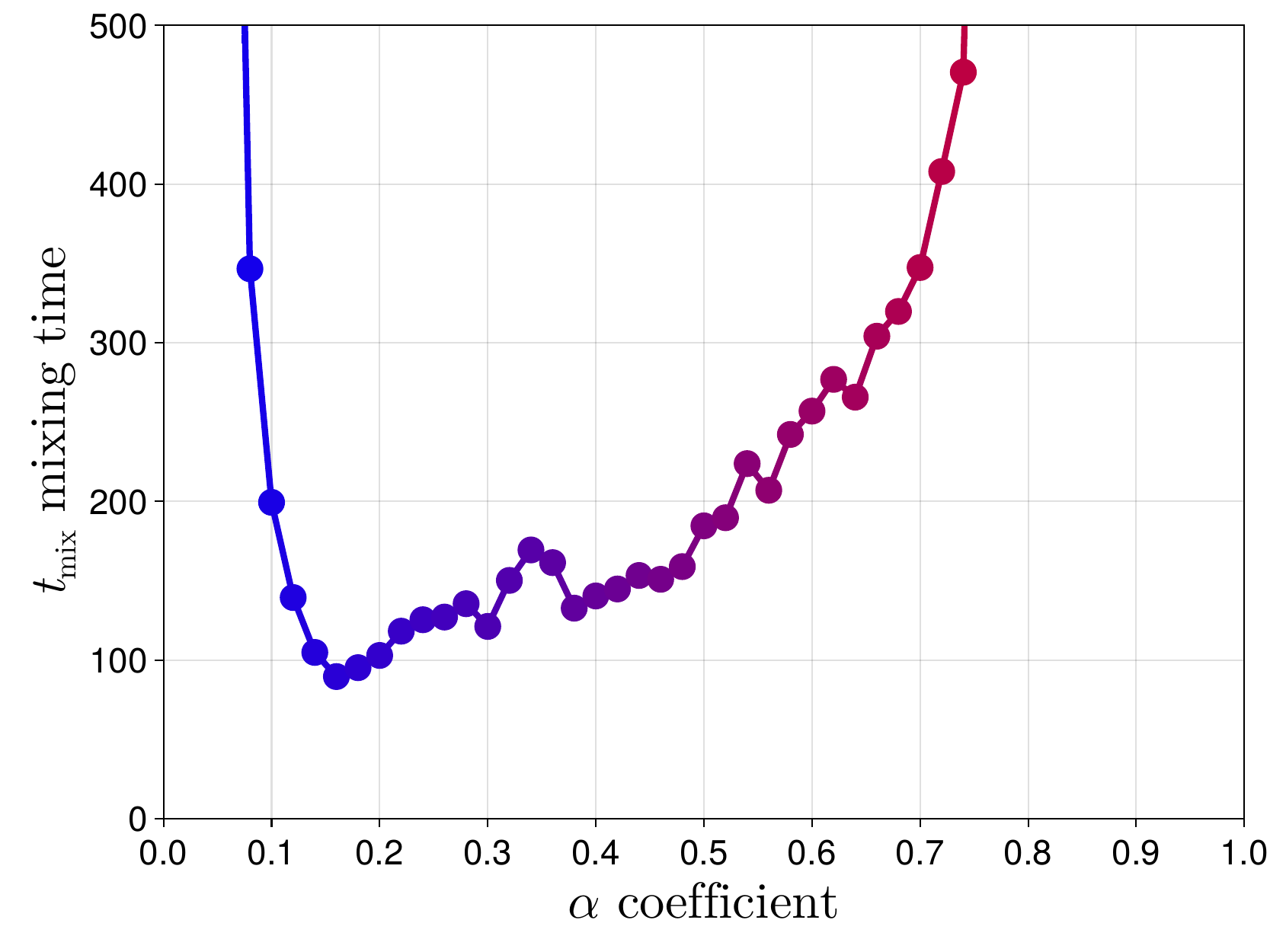}}
  \end{tabular}
  \caption{\label{fig:contamination} Upper graph, left: contamination rate $C(t)$ versus time for various values of $\alpha$ from 0 (blue) to 1 (red). Upper graph, right:  final values of $C_{\infty}$, at $t = 500 = 50\,t_c$. Lower graph: time of mixing (defined by $C(t_{mix}) = 80\%$) versus $\alpha$.}
\end{figure}

\medskip
\figurename\,\ref{fig:clouds} suggests that there is a wide range of $\alpha$'s where mixing seems efficient.  This is confirmed in \figurename\,\ref{fig:contamination}, as the contamination rate remains small for the extremal values of $\alpha$ (single magnet), whereas $C(t)$ increases quickly for intermediate values of $\alpha$ (two magnets). 
\figurename\,\ref{fig:contamination} shows the growth of contamination rates $C(t)$ for various $\alpha$'s (upper left graph), as well as the final contamination rate $C_\infty$ (upper right graph). For $\alpha$ in the range $[0.1,0.7]$, mixing is efficient  with a contamination rate $C_\infty \geq 80 \%$. It is slightly more efficient when $0.4 < \alpha < 0.65$; for these values, the largest final contamination rate $C_\infty \geq 90 \%$ is reached.

\medskip
The time of mixing is also an important parameter to quantify the quality of the device, and is sensitive to the parameter $\alpha$. It is defined here as the time $t_{mix}$ such that $C(t_{\text{mix}}) = 0.8$ (i.e. 80\% of the cells have been visited at least once).
It is plotted on the lower graph of \figurename\,\ref{fig:contamination} which shows that mixing is faster for $0.1 < \alpha < 0.6$ with a mixing time $t_{\text{mix}}\leq 300$ (i.e $\leq 30\,t_c$). The minimal $t_{\text{mix}}$ value is reached for $\alpha = 0.14$, which corresponds to the maximum of the expansion entropy $H_0$ plotted in \figurename\,\ref{fig:entropy}.

\begin{figure} 
  \centering
  \begin{tabular}{c c}
    \includegraphics[width=0.475\textwidth]{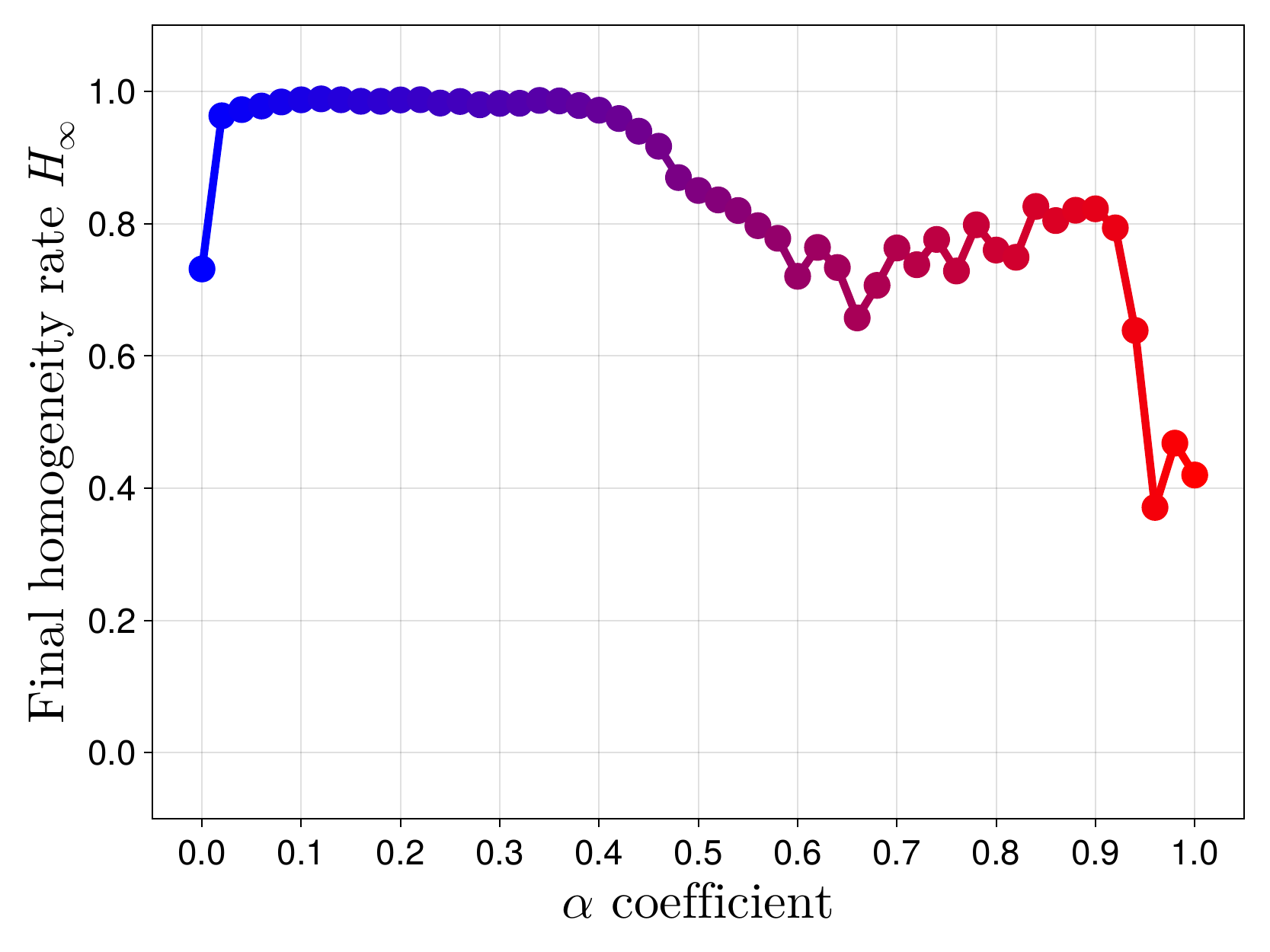} & 
    \includegraphics[width=0.475\textwidth]{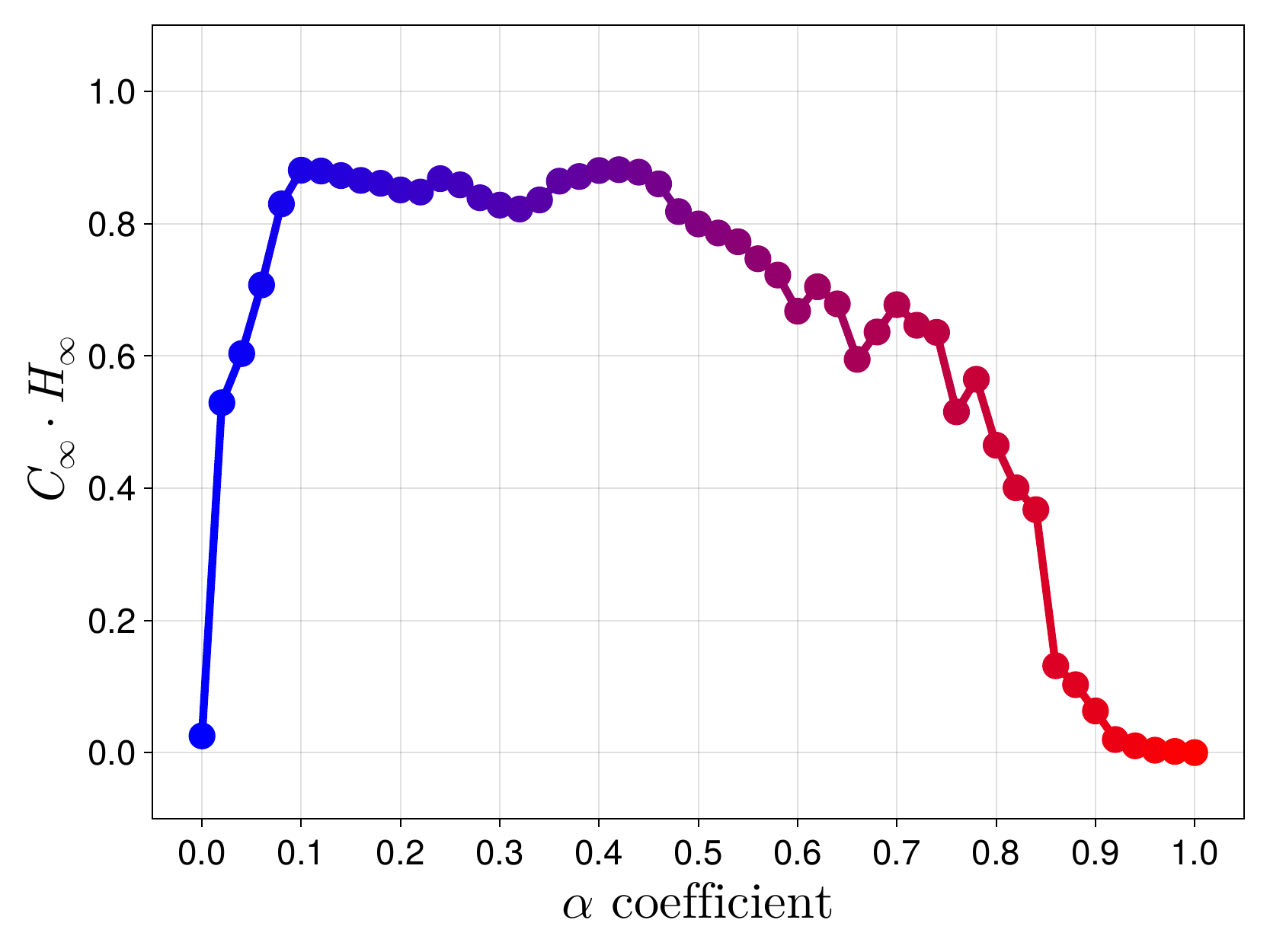}
  \end{tabular}
  \caption{\label{fig:homogeneity} Left: Final homogeneity $H_{\infty}$ versus $\alpha$, right: product of final contamination and homogeneity ($C_{\infty}\times H_{\infty}$) versus $\alpha$.}
\end{figure}

\medskip 
The spatial quality of final mixing is quantified by means of the homogeneity coefficient $H_\infty$ defined above. We observe that mixing is excellent for $\alpha$ between 0.02 and 0.4 (\figurename\,\ref{fig:homogeneity} left) with $H_\infty$ values close to 1, and is a bit poorer when $0.5 < \alpha < 0.9$, though it remains close to 80 \% there. And for $\alpha > 0.95$, it is clearly insufficient with $H_\infty$ values under 50\%.

Finally, by plotting the product of our two quantity, final contamination rate and homogeneity $C\infty\times H\infty$ as a function of $\alpha$, we can determine a range of good compromise in terms of contamination rate, homogeneity and time of mixing. According to \figurename\,\ref{fig:homogeneity} (right), mixing is particularly efficient for $\alpha \in [0.08, 0.5]$. These values correspond perfectly with the range where the expansion entropy plotted in \figurename\ref{fig:entropy} is the highest, which confirms our results.

\section{Concluding remarks}

The steady MHD device presented in this work has been shown to mix efficiently.  It generates velocity fields which effectively produce a three-dimensional Stokes flow structure corresponding to a single recirculation cell superposed to an orthogonal double cell. Both elementary flows are created by means of magnets placed in an appropriate manner: the first magnet  is located in an asymetric manner and pushes the flow along the side of the tank, thus creating a large recirculation cell. The second magnet is placed along a mid-plane of the tank, and creates a double recirculation cell. Due to the linearity of the Stokes equation, the flow resulting from the action of both magnets is a linear combination of these elementary flows.

A weak vorticity -- streamfunction formulation, based on the mixed finite element method, has been developed and implemented to calculate these velocity fields. 
Then, by computing Poincaré sections, Lyapunov exponents and expansion entropies, it was shown that most combinations of these two velocity fields produce fluid points that are indeed chaotic.  We also observed that, for a wide range of $\alpha$, mixing appears to be particularly efficient in terms of contamination and homogeneity rates. Following this theoretical study, the authors recommend using a value of $\alpha$ between 0.1 and 0.5 for best efficiency. 

Lyapunov exponents, as well as expansion entropies, suggest that the structure of the flow is different according to whether the first magnet or the second magnet dominate (i.e. $\alpha$ is nearest to 1 or 0). A detailed analysis of the model flow of Toussaint {\it et al.} \cite{articleVToussaint} shows that a Shilnikov chaos might be at work there, but the stagnation points involved in this Shilnikov chaos are different whether $\alpha$ is nearest to 1 or 0 \cite{holm1991,gonchenko2022}. Clearly, this model flow differs from ours in that it has periodic boundary conditions and an explicit solution. However, not too close to the walls, the structure of the MHD flow shares some common features with the flow of \cite{articleVToussaint}. This point will be analyzed in the near future. 

The results obtained in the present analysis are based exclusively on simulations. A demonstrator will be set up soon, and an experimental bench will be created.
This device will be designed with the help of the present study,  and measurements will be performed to validate experimentally our numerical results.


\appendix*
\section{\label{myappendix}Implementation of numerical fields computation}
 
In this part, we describe the main steps involved in the numerical resolution of the systems described in section \ref{model_num}. The first step consists in generating the meshing of the geometry. For this purpose, the free and open-source software Gmsh is used \cite{gmsh}. The mesh size in the whole domain is controlled by the number $n_e$ of elements on the edges of the cubic tank. We therefore impose a structured mesh in $\Omega_f$; and outside, the characteristic length of the elements is then gradually scaled up as the distance from the tank increases, until the boundary $\Gamma_{\infty}$ is reached. The mesh of the domain with the chosen control parameter value ($n_e = 15$) is displayed on \figurename\,\ref{fig:mesh}, and some characteristics are provided in \tablename\,\ref{tab:mesh}.

\begin{table}[hb]
  \caption{\label{tab:mesh}Some statistics of the chosen mesh}  
  \renewcommand{\arraystretch}{1.2}  
  \centering
  \begin{tabular}{ c || c | c | c }
    \hline\hline
    Domain & Nodes & Triangles & Tetrahedra \\
    \hline\hline
    $\Omega$ & 18,350 & 6,866 & 107,674 \\
    \hline
    $\Omega_f$ & 3,423 & 3,256 & 15,964 \\
    \hline\hline
  \end{tabular}
\end{table}

Afterwards, the formulations detailed in the previous section were then implemented in C\texttt{++} language, using the free and open-source GmshFEM library \cite{gmshfem,TheseARoyer}. Thanks to this library, it is relatively easy to define the approximate subspaces based on our mesh of the global needed function spaces $\text{H}_0(\Grad,\Omega)$, $\text{H}(\Grad,\Omega_f)$, $\text{H}_0(\Grad,\Omega_f)$, $\textbf{H}(\Curl,\Omega_f)$, and $\textbf{H}_0(\Curl,\Omega_f)$. 
These finite dimensional function spaces are respectively called $\text{W}_0^0(\Omega)$, $\text{W}^0(\Omega_f)$, $\text{W}_0^0(\Omega_f)$, $\text{W}^1(\Omega_f)$ and $\text{W}_0^1(\Omega_f)$, and are defined thanks to hierarchical basis functions. The three first spaces are generated using 0-form nodal basis functions which are the classical nodal basis functions of Lagrange elements, their order are noted $p_n$. 
The two other spaces $\text{W}^1(\Omega_f)$ and $\text{W}_0^1(\Omega_f)$ are generated by 1-form edge basis functions of order $p_e$ built with the same functions as the previous ones, see \cite{ren,ThesisGeuzaine,hierarchicalBF} for details.

The numerical implementation and runs were performed on a dedicated workstation with 48 threads @3.5\,GHz and 192\,GiB of RAM. For the best possible accuracy of results, we choose to use high order elements with an interpolation order of $p_n = 4$ for the nodal basis functions of $\text{W}_0^0(\Omega)$, $\text{W}^0(\Omega_f)$ and $\text{W}_0^0(\Omega_f)$, and $p_e = 3$ for the edge basis functions of $\text{W}^1(\Omega_f)$ and $\text{W}_0^1(\Omega_f)$. These values combined with a mesh parameterized with the chosen value of $n_e$ allow us to make the best use of all our  computing resources. The synthesis of the systems we solve (using the MUMPS direct solver \cite{mumps}) is given by (\ref{eq:syst_discr_compl}), and some details upon the numerical runs are reported in \tablename\,\ref{tab:numsyst}.

\begin{figure*}[ht!]
\begin{equation}
  \label{eq:syst_discr_compl}
  \renewcommand{\arraystretch}{1.5}
  \hspace{-1.15em}
  \begin{array}{l}
    \left(\Sigma_{hm,i}\right) \left\{
    \begin{array}{l}
        \text{Find~} \phi_{hi} \in \text{W}_0^0(\Omega) \text{~such that:}  \\
        \displaystyle \int_{\Omega_i} \Grad {\phi_{hi}} \cdot \Grad {\xi_{hi}}~\text{d}{\bf x} - \int_{\Omega_{mi}} {\bf {\bf m_i}} \cdot \Grad {\xi_{hi}}~\text{d}{\bf x} \\ \displaystyle \qquad\quad
         + \int_{\Omega_{\infty}} {\bf J_{\mathcal{F}}^{-1}} \Grad {\phi_{hi}} \cdot {\bf J_{\mathcal{F}}^{-1}} \Grad {\xi_{hi}}~|\det \mathbf{J}_{\mathcal{F}}|~\text{d}{\bf X} = 0 ~,~~\forall \xi_{hi} \in \text{W}_0^0(\Omega)
    \end{array}\right.\\
    \quad\big\downarrow\phi_{hi}\\
    \left(\Sigma_{hf,i}^{(1)}\right) \left\{
    \begin{array}{l}
      \text{Find~} \left(\boldsymbol{\omega_{hi}^0},\lambda_{hi}^0\right) \in \text{W}_0^1(\Omega_f) \times \text{W}^0(\Omega_f) \text{~such that:~}  \\
      \displaystyle \int_{\Omega_f} \Curl \boldsymbol{\omega_{hi}^0}\cdot\Curl \boldsymbol{\varphi_{hi}}~\text{d}{\bf x} + \int_{\Omega_f} \Grad \lambda_{hi}^0 \cdot \boldsymbol{\varphi_{hi}}~\text{d}{\bf x} \\ \displaystyle \qquad\qquad
      + \int_{\Omega_f}  \left({\bf j_0}\times\Grad {\phi_{hi}}\right)\cdot\Curl \boldsymbol{\varphi_{hi}}~\text{d}{\bf x}  = 0 ~,~~ \forall \boldsymbol{\varphi_{hi}} \in \text{W}^1_0(\Omega_f) \\
      \displaystyle \int_{\Omega_f} \boldsymbol{\omega_{hi}^0} \cdot \Grad \mu_{hi}~\text{d}{\bf x} = 0 ~,~~ \forall \mu_{hi} \in \text{W}_0^0(\Omega_f)\\
    \end{array}
    \right.\\
    \quad\big\downarrow\boldsymbol{\omega_{hi}^0}\\
    \left(\Sigma_{hf,i}^{(2)}\right)\left\{
    \begin{array}{l}
      \text{Find} \left(\boldsymbol{\omega_{hi}^*},\lambda_{hi}^*, \boldsymbol{\psi_{hi}},\eta_{hi}\right) \in \text{W}^1(\Omega_f) \times \text{W}^0(\Omega_f) \times \text{W}_0^1(\Omega_f) \times \text{W}^0(\Omega_f)  \text{~such that:}   \\
      \displaystyle \int_{\Omega_f} \boldsymbol{\omega_{hi}^*} \cdot \boldsymbol{\varphi_{hi}}~\text{d}{\bf x} - \int_{\Omega_f} \Curl \boldsymbol{\psi_{hi}} \cdot \Curl \boldsymbol{\varphi_{hi}}~\text{d}{\bf x} + \int_{\Omega_f} \Grad \eta_{hi} \cdot \boldsymbol{\varphi_{hi}}~\text{d}{\bf x} = \\ \hfill \displaystyle
      - \int_{\Omega_f} \boldsymbol{\omega_{hi}^0} \cdot \boldsymbol{\varphi_{hi}}~\text{d}{\bf x}  ~,~~ \forall \boldsymbol{\varphi_{hi}} \in \text{W}^1(\Omega_f) \\
      \displaystyle \int_{\Omega_f} \Curl \boldsymbol{\omega_{hi}^*} \cdot \Curl \boldsymbol{\xi_{hi}}~\text{d}{\bf x} + \int_{\Omega_f} \Grad \lambda_{hi}^* \cdot \boldsymbol{\xi_{hi}}~\text{d}{\bf x} = 0 ~,~~  \forall \boldsymbol{\xi_{hi}} \in \text{W}_0^1(\Omega_f)\\
      \displaystyle \int_{\Omega_f} \boldsymbol{\psi_{hi}} \cdot \Grad \chi_{hi}~\text{d}{\bf x} = 0 ~,~~ \forall \chi_{hi} \in \text{W}_0^0(\Omega_f)\\
      \displaystyle \int_{\Omega_f} ( \boldsymbol{\omega_{hi}^0} + \boldsymbol{\omega_{hi}^*} ) \cdot \Grad \mu_{hi}~\text{d}{\bf x} = 0 ~,~~ \forall \mu_{hi} \in \text{W}^0(\Omega_f)\\
    \end{array}
    \right.    
  \end{array}
\end{equation}
\end{figure*}  

\begin{table}[h!]
   \caption{\label{tab:numsyst}Details of the numerical resolution ($n_e = 15$, $p_n = 4$, $p_e = 3$)}
  \renewcommand{\arraystretch}{1.5}  
  \centering
  \begin{tabular}{ c || c | c | c }
    \hline\hline
    System & $\left(\Sigma_{hm,i}\right)$ & $\left(\Sigma_{hf,i}^{(1)}\right)$ & $\left(\Sigma_{hf,i}^{(2)}\right)$ \\
    \hline\hline
    DOFs &  1,207,089 & 527,823 & 1,198,914 \\
    \hline
    Assembling time & 2.47 sec & 1.92 sec & 4.12 sec \\
    \hline
    Solving time & 208 sec & 92 sec & 3050 sec \\
    \hline\hline
    \end{tabular}
\end{table} 
 
The very low value of assembling time highlights the efficiency of the chosen solver. Moreover, even if the computation time of the last system $\left(\Sigma_{hf,i}^{(2)}\right)$ seems high compared to the others (due to a non-symmetry of the stiffness matrix), the duration of the whole resolution remains largely acceptable for such a quite big problem.


\FloatBarrier

\begin{acknowledgments}
  All the numerical tools used in this article are free and open-source software. Then, the authors would like to express their gratitude to the developers and contributors of the corresponding projects for providing and sharing their work to the community. We especially thank: Christophe Geuzaine and Jean-François Remacle for the mesh generator \texttt{Gmsh}; Anthony Royer, Eric Béchet and Christophe Geuzaine again for the \texttt{GmshFEM} library; and George Datseris for the \texttt{DynamicalSystems.jl} Julia package.
\end{acknowledgments}



%

\end{document}